\documentclass[letterpaper,11pt]{article}

\usepackage{anysize}
\marginsize{1in}{1in}{1in}{1in}

\usepackage[english]{babel} 
\usepackage[T1]{fontenc} 
\usepackage{textgreek} 
\usepackage{libertine} 
\usepackage{amsmath,amssymb}
\usepackage[round]{natbib} 
\usepackage{multicol} 

\usepackage{graphicx} 
\usepackage{float} 
\usepackage{wrapfig} 
\usepackage[font=normalsize,format=plain,labelfont=bf,up,textfont=it,up]{caption}

\usepackage{mdwlist} 
\usepackage{titlesec} 
\titlespacing{\section}{0pt}{8pt}{4pt}
\titlespacing{\subsection}{0pt}{6pt}{2pt}
\usepackage[hidelinks]{hyperref} 
\usepackage[dvipscolors]{xcolor}
\usepackage[font=small]{caption}
\usepackage{siunitx}
\usepackage{tocloft}

\usepackage{multirow}

\usepackage[final]{pdfpages}

\usepackage[modulo]{lineno}

\graphicspath{{Figures/}}

\begin{document}

\begin{center}
	{\LARGE Segmented coronagraph design and analysis (SCDA): \\an initial design study of apodized vortex coronagraphs}
\end{center}

\begin{center}
	{\large G. Ruane$^1$, J. Jewell$^2$, D. Mawet$^{1,2}$, S. Shaklan$^2$, and C. Stark$^3$}
\end{center}
\begin{center}
	{ $^1$ Department of Astronomy, California Institute of Technology\\
	$^2$ Jet Propulsion Laboratory, California Institute of Technology\\
	$^3$ Space Telescope Science Institute}
\end{center}

\smallskip

\renewcommand{\cftaftertoctitle}{\thispagestyle{empty}}
\setcounter{secnumdepth}{3}
\setcounter{tocdepth}{3}
\tableofcontents

\clearpage
\setcounter{page}{1}


\section{Summary}\label{sec:summary}

The segmented coronagraph design and analysis (SCDA) study is a coordinated effort, led by Stuart Shaklan (JPL) and supported by NASA's Exoplanet Exploration Program (ExEP), to provide efficient coronagraph design concepts for exoplanet imaging with future segmented aperture space telescopes \citep{SCDA1,SCDA2}. This document serves as an update on the apodized vortex coronagraph designs devised by the Caltech/JPL SCDA team. Apodized vortex coronagraphs come in two flavors, where the apodization is achieved either by use of 1) a gray-scale semi-transparent pupil mask or 2) a pair of deformable mirrors in series. Each approach has attractive benefits. This document presents a comprehensive review of the former type. Future theoretical investigations will further explore the use of deformable mirrors for apodization. 

\section{Introduction and background}\label{sec:intro}
Future NASA exoplanet imaging mission concepts, such as HabEx \citep{Mennesson2016} and LUVIOR \citep{Bolcar2015}, would have sufficiently large apertures (4m-16m) to enable direct spectroscopic analysis of light from a variety of substellar objects ranging from gas giants to habitable rocky planets, drastically improving our understanding of planetary formation and potentially leading to the first unambiguous detection of signs-of-life on another world. However, such observations pose a formidable technical challenge owing to the extremely high brightness contrast and small angular separation between a planet and its host star. The HabEx and LUVIOR space telescopes would therefore require imaging instruments capable of precisely rejecting light from the bright parent star while also maintain high sensitivity to the precious few planet photons. Coronagraphs are designed to accomplish this through fine control of the optical wavefront and suppression of unwanted stellar radiation that otherwise overwhelms the weak planet signal. 

For structural reasons, the primary mirror of large telescopes tend to have various discontinuities, such as a secondary mirror, spider support structures, and gaps between mirror segments, which limit the performance and drive the complexity of conventional coronagraph designs. The optical system needs to be specially tailored to each telescope aperture shape in order to compensate for the complicated stellar diffraction patterns that contaminate the field of view.

\begin{figure}[h]
    \centering
    \includegraphics[width=\linewidth]{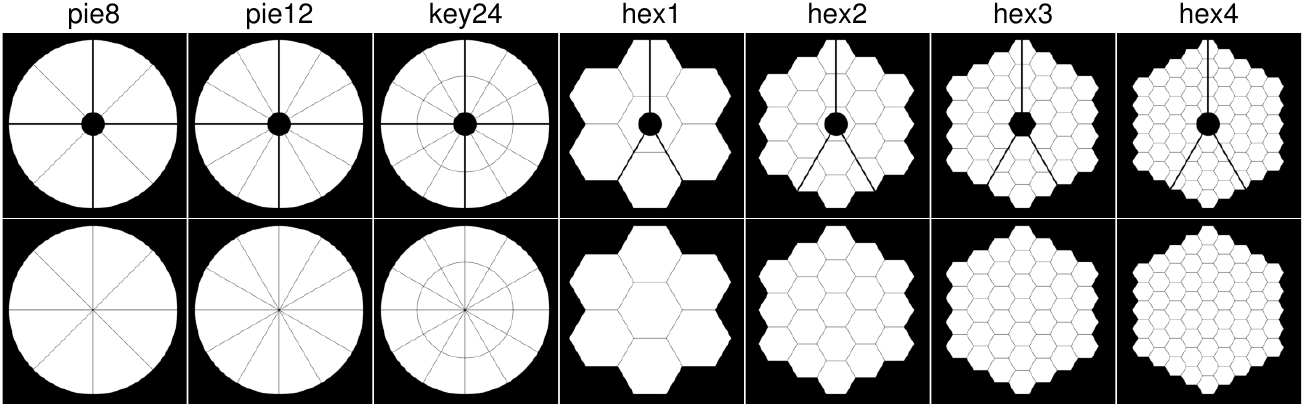}
    \caption{Apertures considered as part of the SCDA study, including (top row) obscured and (bottom row) unobscured varieties. The flat-to-flat diameters are 12~m, the secondary supports are 10~cm thick, and the segment gaps are $\sim$20~mm.}
    \label{fig:SCDAapertures}
\end{figure}

The Segmented Coronagraph Design and Analysis (SCDA) study is a coordinated effort to develop highly efficient optical design concepts that passively reject starlight on realistic segmented aperture telescopes. The apertures considered by SCDA are shown in Fig. \ref{fig:SCDAapertures}. Several segment shapes are represented, including piewedges (pie8 and pie12), keystones (key24), and hexagons (hex1, hex2, hex3, and hex4). In each case, we compare performance of on- and off-axis designs; i.e. with and without a central obscuration and spider support structures. 

The main coronagraph architectures studied by the SCDA teams are the apodized vortex coronagraph, hybrid Lyot coronagraph (HLC), apodized pupil Lyot coronagraph (APLC), and phase induced amplitude apodized complex mask coronagraph (PIAACMC). This document provides an overview of a promising family of apodized vortex coronagraph designs. 

\section{Vortex coronagraphs for segmented aperture space telescopes}\label{sec:approach}

\begin{figure}[h]
    \centering
    \includegraphics[width=0.85\linewidth]{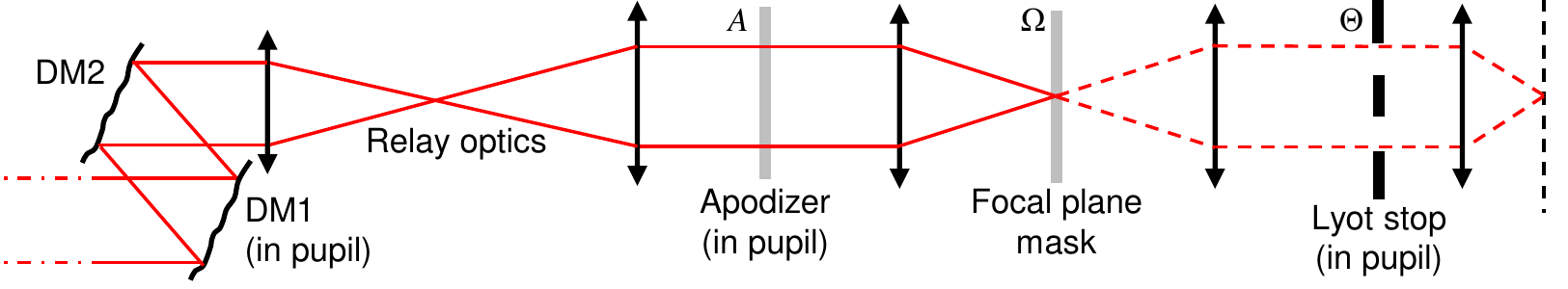}
    \caption{Schematic of a coronagraph instrument with deformable mirrors DM1 and DM2, a pupil-plane apodizer~$A$, focal plane mask~$\Omega$, and a Lyot stop~$\Theta$ in the downstream pupil. The black arrows represent powered optics.}
    \label{fig:diagram}
\end{figure}

A high-contrast imaging coronagraph instrument is usually made up of a wavefront control unit with one or two deformable mirrors (DMs) and a series of coronagraphic masks arranged between powered optics. Fig. \ref{fig:diagram} shows an example system with two DMs, a pupil plane apodizer $A$, a focal plane mask $\Omega$, and a Lyot stop $\Theta$ in the downstream pupil. The jointly optimized masks in a coronagraph manipulate the amplitude and phase of the incoming beam to passively suppress starlight while maintaining maximal planet throughput. 

Several coronagraph throughput definitions may be found in the literature. In this work, we use two measures: the total energy throughput (i.e. the fraction of energy from a point source that transmits through the Lyot stop) and the fraction of energy within 0.7$\lambda/D$ of the source position. The latter definition is the relevant quantity for detecting point sources in noisy images using aperture photometry, as is the case in common post-processing approaches for exoplanet detection. We note however that this may be a conservative estimate for the effective throughput since advanced matched-filtering and local deconvolution techniques can theoretically make use of the planet light outside of the PSF core.

The vortex coronagraph \citep[VC; ][]{Mawet2005} has a focal plane mask with complex transmittance $\exp(il\phi)$, where $l$ is an even, nonzero integer known as the charge and $\phi$ is the azimuthal angle, which acts to relocate starlight outside of a downstream Lyot stop. The VC has been demonstrated to provide high sensitivity to planets at small angular separations \citep{Serabyn2010}. However, complicated aperture shapes limit the performance of the conventional VCs \citep{Mawet2010b} and thereby drive the complexity of the optical design \citep{Mawet2011_improved,Mawet2013_ringapod,Carlotti2014,Ruane2015_SPIE,Ruane2015_LPM} and/or requirements for wavefront control \citep{Pueyo2013,Mazoyer2015}. Here, we present an approach to overcome this technical challenge in the case of segmented aperture space telescopes by introducing optimized gray-scale pupil apodizers that provide polychromatic ($\Delta\lambda/\lambda \ge 0.1$) suppression of diffracted starlight, at angular separations $<10\lambda/D$, potentially down to the $10^{-10}$ level on a segmented aperture telescope \citep{Ruane2016}. We simulate the performance for imaging stars of finite size and in the presence of aberrations, including low-order Zernike phase modes in the entrance pupil and pupil field distortions induced in the coronagraph (modeled as a shifted or magnified Lyot stop). 

\subsection{Optimization algorithm: Auxiliary field optimization}\label{sec:approach}
We use an iterative numerical method, known as Auxiliary Field Optimization \citep{Jewell2017_SPIE}, to determine the optimal coronagraph design that efficiently minimizes starlight at the final image plane by use of a gray-scale semi-transparent pupil mask.

The problem to find a the optimal gray-scale apodizer is written as 
\begin{equation}
\underset{w}\min\left(||QCw||^2 + b||w - P A||^2\right),
\end{equation}
where $Q$ is a matrix that represents the desired ``dark hole'' region in the image plane, $C$ is the coronagraph operator that propagates the field from the apodizer plane to the final image plane, $P$ is the original telescope pupil function, $A$ is the apodizer function, and $w$ is the so-called auxiliary field in the apodizer plane which strikes a balance (as regulated by the weight $b$) between the field needed to generate a zero-valued dark hole and the physical field in the apodizer plane. Assuming the pupil and focal planes are related by Fourier transform propagation operators $F$, the coronagraph operator may be written $C=F \Theta F^{-1} \Omega F$, where $\Omega$ and $\Theta$ represent the focal plane mask and Lyot stop transmittance, respectively. The auxiliary field $w$ is calculated by
\begin{equation}
w=(b I + C^\dagger Q C)^{-1} b A P.
\end{equation}
To reduce computation time, the dimensionality of the inverted (square) matrix is reduced from the number of samples in the pupil plane to the number of samples in the dark hole region, by use of the Woodbury matrix identity:
\begin{equation}
w = \left[ I - C^\dagger Q(b Q + Q C C^\dagger Q)^{-1}Q C\right] P A.
\end{equation}
Since $w$ may be a complex function with infinite support, the physical apodizer is taken to be $A = |w|$ and $A$ is thresholded such that samples where $A>1$ are set to one and non-zero values outside of the original telescope pupil $P$ support are set to zero. A new auxiliary field is calculated based on the updated pupil field, and the process is repeated. The matrix $C C^\dagger$ is calculated once for each choice of focal plane mask and Lyot stop as follows:
\begin{equation}
C C^\dagger = (F \Theta F^{-1} \Omega F) (F^{-1} \Omega^\dagger F \Theta^\dagger F^{-1}) = F \Theta F^{-1}  |\Omega|^2 F \Theta^\dagger F^{-1}.
\end{equation}
Since the focal plane mask has phase-only transmittance (i.e. $|\Omega|^2=I$), this matrix simplifies to $C C^\dagger = F |\Theta|^2 F^{-1}$ and only depends on the squared modulus of Lyot stop function. 

The solutions for gray-scale apodizers are calculated for a single wavelength. Theoretically, since the focal plane mask has no radial features and the coronagraph masks reside in pupil and focal planes, the monochromatic result applies to all wavelengths. These assumptions are valid over a bandpass where achromatic apodizers and phase masks may be fabricated to high fidelity. 

The coronagraph optimization procedure typically creates a dark hole in the light from the star at the cost of planet throughput. In order to balance the cost of throughput losses, we adopt performance metrics outlined in section \ref{sec:perfmetrics}. 

\subsection{Apodized vortex coronagraph designs}\label{sec:VCdesigns}

\begin{figure}[p]
    \centering
    \begin{center}\textbf{Apertures with central obscurations}\end{center}
    \includegraphics[width=0.7\linewidth]{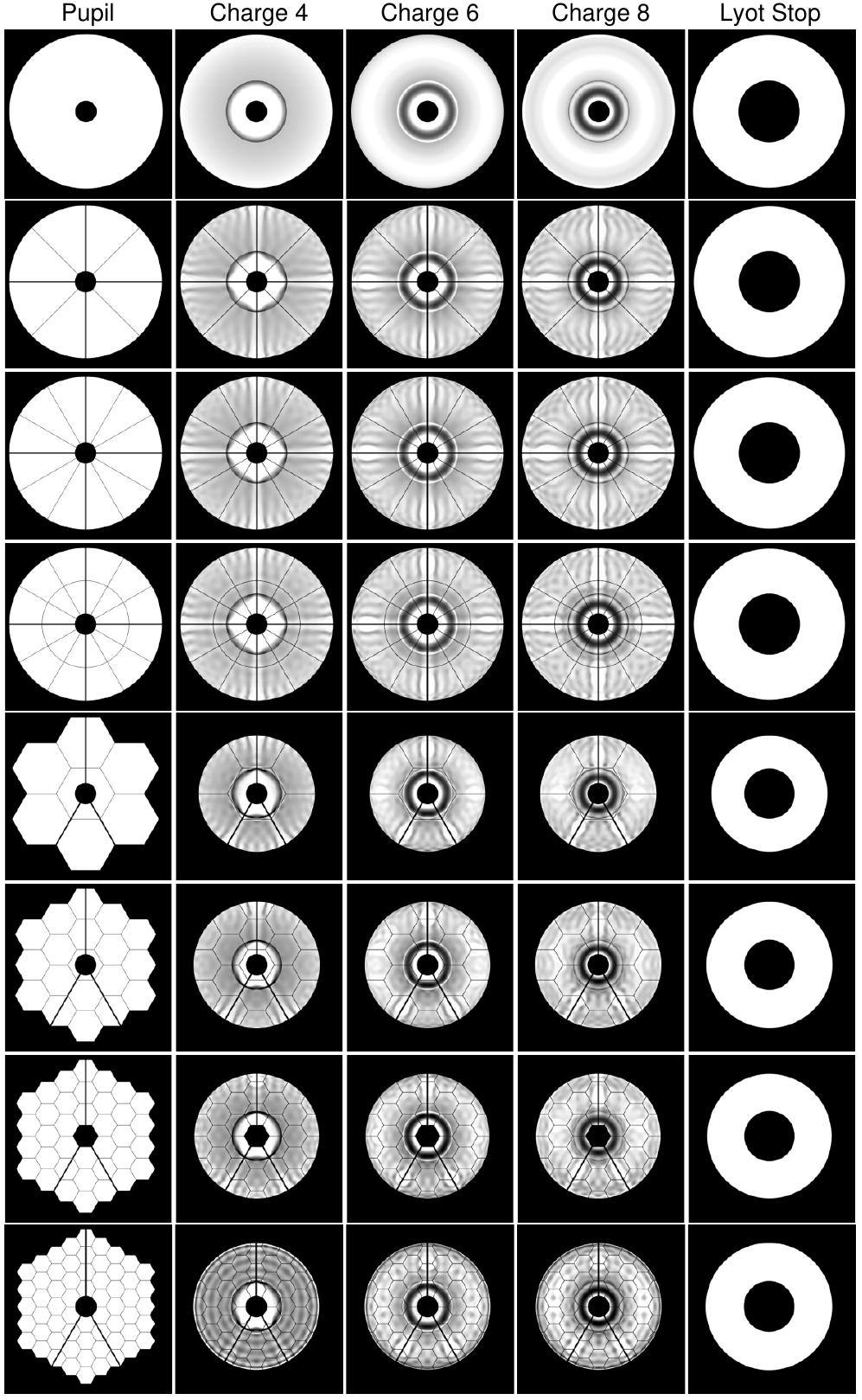}
    \caption{Apodizers for obscured, segmented apertures.}
    \label{fig:apodizers_obs}
\end{figure}

\begin{figure}[p]
    \centering
    \begin{center}\textbf{Apertures without central obscurations}\end{center}
    \includegraphics[width=0.7\linewidth]{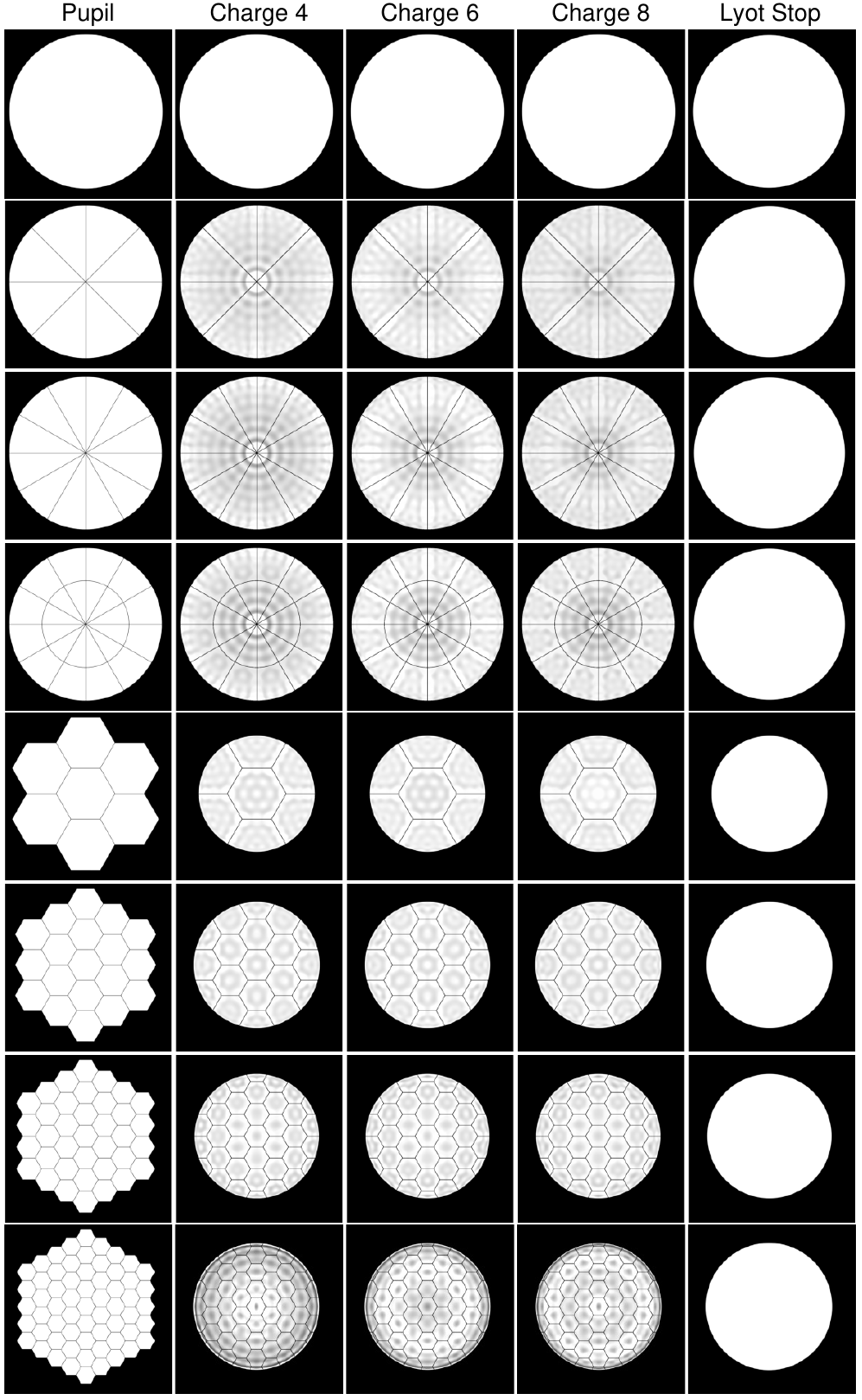}
    \caption{Apodizers for unobscured, segmented apertures.}
    \label{fig:apodizers_unobs}
\end{figure}

Vortex coronagraph designs that make use of gray-scale, apodizing pupil masks to compensate for diffraction owing to mirror segmentation are shown in Figs. \ref{fig:apodizers_obs} and \ref{fig:apodizers_unobs} for obscured and unobscured apertures, respectively. The first column is the field magnitude incident on the apodizer in each case. The second, third, and fourth columns show the optimized apodization function for charge 4, 6, and 8 coronagraphs. The final column is the corresponding Lyot stop. Each apodization function is optimized to generate a dark hole in the stellar field in annulus from angular coordinate 3$\lambda/D$ to 10$\lambda/D$. Additionally, in the case of unobscured apertures, the apodizers are also compatible with higher charge focal plane masks.

The solutions for the annular pupils are akin to those theorized by \citet{Mawet2013_ringapod} and \citet{Fogarty2017_SPIE}. However, the auxiliary field optimization approach allows us to generate apodizers for arbitrary pupils, for the first time, that would otherwise be too difficult to describe analytically. In all cases, the Lyot stops are chosen to be an annuli based on prior knowledge of the properties of apodized vortex coronagraphs for simpler apertures. The inner and outer radii of the Lyot stops are indicated in Table \ref{tab:LSproperties} and in the case of unobscured apertures, the inner radii are set to zero.

\begin{table}[h]
\small
\centering
\begin{tabular}{ *{9}{c} }
 & annulus & pie8 & pie12 & key24 & hex1 & hex2 & hex3 & hex4 \\
\hline
Inner radius (\%) & 40 & 40 & 40 & 40 & 33 & 33 & 33 & 33 \\
Outer radius (\%) & 99 & 99 & 99 & 99 & 76 & 82.5 & 81.5 & 82.5 \\
\hline
\end{tabular}
\caption{Inner and outer radii of the Lyot stops in Figs. \ref{fig:apodizers_obs} and \ref{fig:apodizers_unobs}, given in percent of the full geometric pupil radius.}
\label{tab:LSproperties}
\end{table}

The apodizers for the hex apertures clip the outer regions of the original pupil to create a circular boundary. The outer circle is chosen to be the largest circle that fits inside the pupil and the inner Lyot stop radius is tuned to find the optimal trade off between starlight suppression and coronagraph throughput. Each apodizer is designed to suppress the starlight to the $\sim10^{-10}$ level within a radius of $10\lambda/D$ about the star. In practice, the outer radius of the dark hole will limited by the maximum spatial frequency represented to high precision in the fabricated masks. The simulated performance of these designs is discussed further in section \ref{sec:perf}.

\subsection{Performance}\label{sec:perf}

\subsubsection{Performance metrics}\label{sec:perfmetrics}

We adopt two metrics for quantifying the performance of the coronagraph designs. The first is based on finding the design parameters that minimize the estimated exposure time for a typical observation. In the stellar photon noise limited regime, the exposure time scales as $\eta_s/\eta_p^2$, where $\eta_s$ is the starlight suppression level and $\eta_p$ is the coronagraph throughput \citep{Ruane2016,Ruane2017_SPIE}. Therefore, we define our first performance metric as 
\begin{equation}
M = \frac{I_{avg}}{\eta_c^2},
\label{eqn:m1}
\end{equation}
where $I_{avg}$ is the mean irradiance in the dark hole and $\eta_c$ is the relative PSF core throughput of the coronagraph (i.e. energy within 0.7$\lambda/D$, normalized by the energy in the core of the telescope PSF).

The second metric is based on ExoEarth yield analyses developed by \citet{Stark2014,Stark2015}. Here, the exoEarth candidate yield is defined as the number of earth-like planets detected in V-band to signal-to-noise ratio (SNR) of 7 within 1 year of total integration time. The simulated mission images the nearest main sequence and sub-giant stars in the Hipparcos catalog, 10\% of which are assumed to host earth-like planets and each have 3 zodis of dust. A detection limit is set such that the signal from the planet must be greater than 10\% of the stellar signal. 

\textit{The optimal coronagraph minimizes $M$ and maximizes the exoEarth candidate yield}. The goal of this work is the estimate these performance metrics for each coronagraph, in the presence of a star of finite size and low order aberrations, in order to find optimal combinations of aperture shapes and coronagraph designs. 

\subsubsection{Ideal starlight suppression}
\begin{figure}[p]
    \begin{center}\textbf{Apertures with central obscurations}\end{center}
    \includegraphics[height=0.31\linewidth]{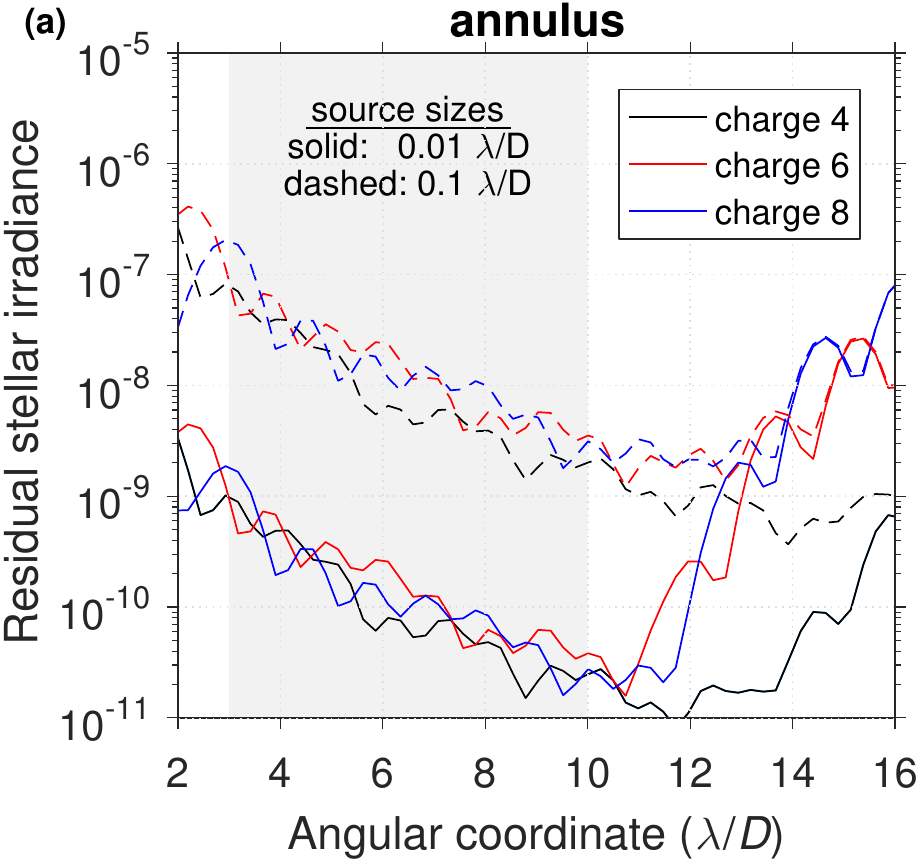}
    \includegraphics[height=0.31\linewidth]{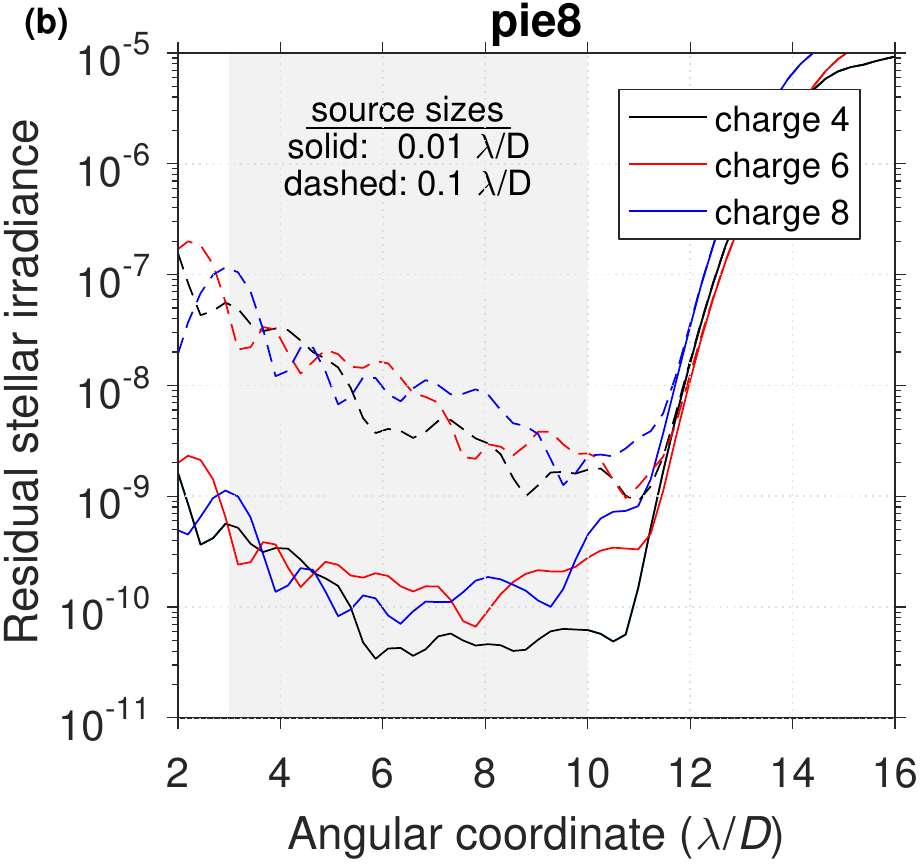}
    \includegraphics[height=0.31\linewidth]{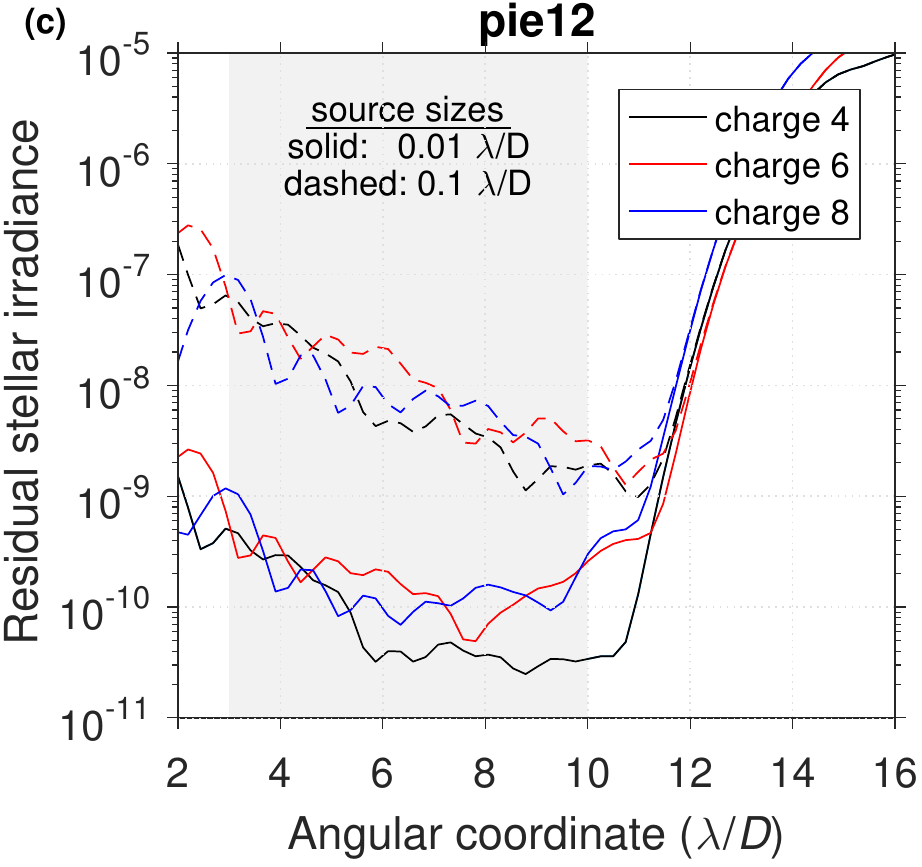}\\
    \includegraphics[height=0.31\linewidth]{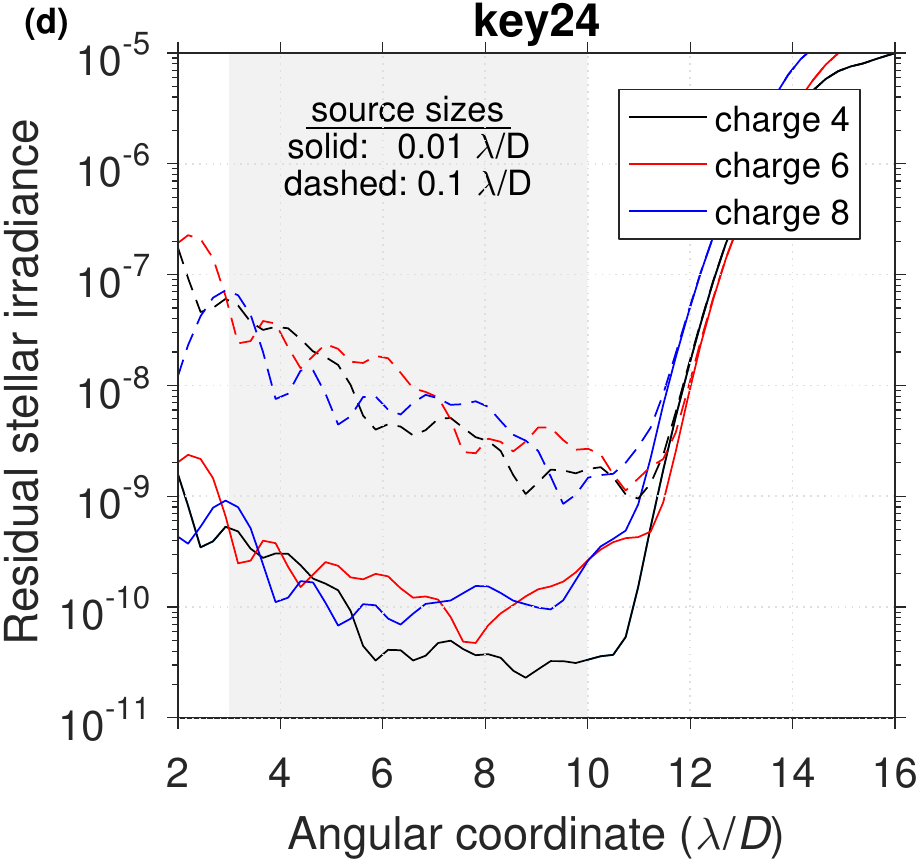}
    \includegraphics[height=0.31\linewidth]{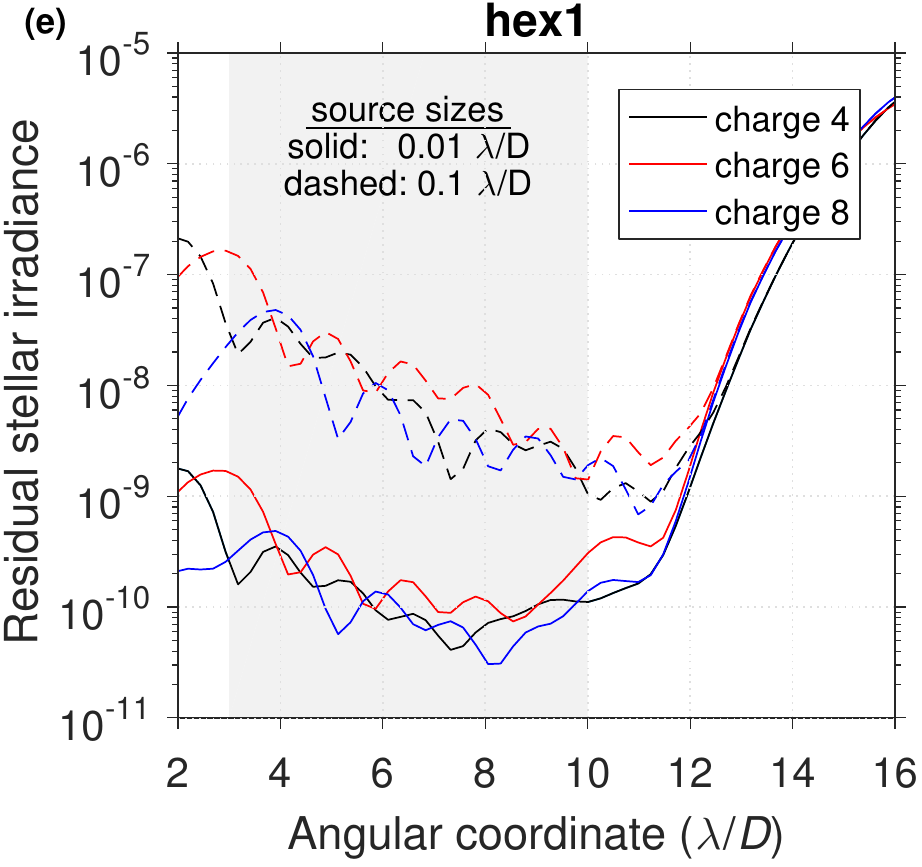}
    \includegraphics[height=0.31\linewidth]{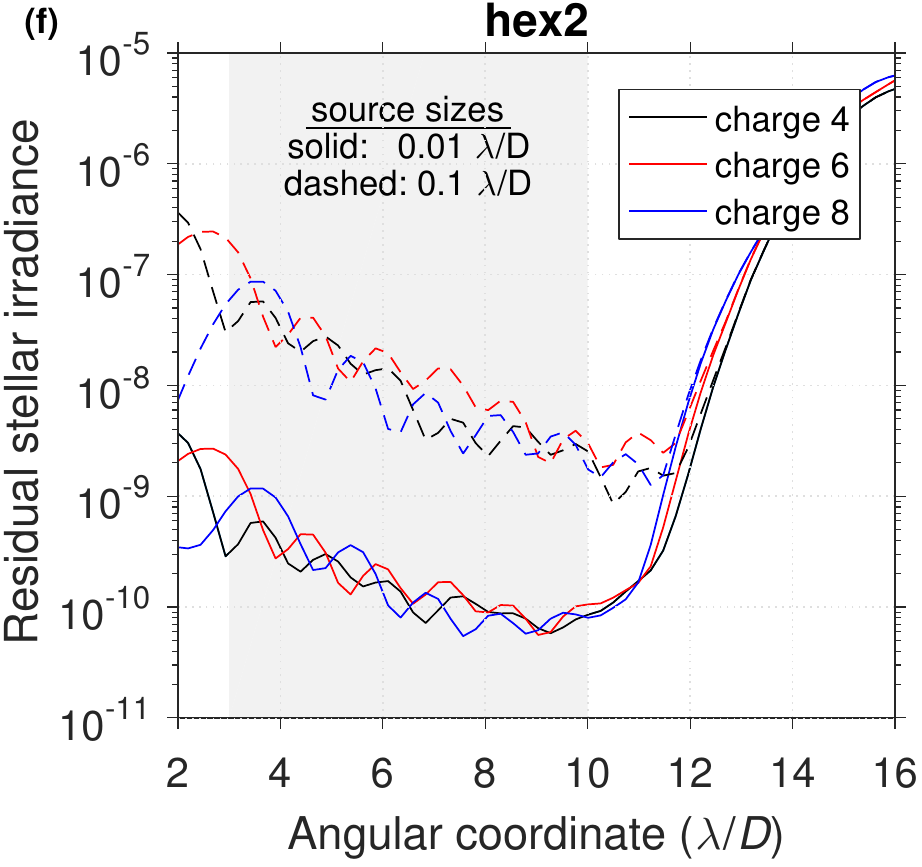}\\
    \includegraphics[height=0.31\linewidth]{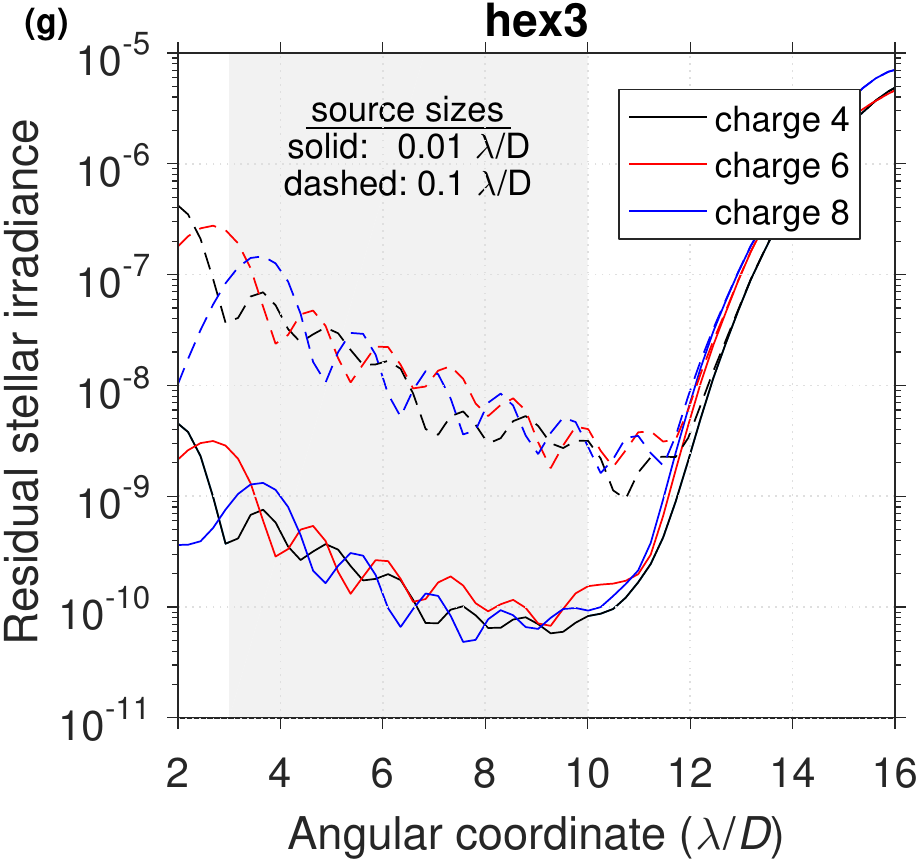}
    \includegraphics[height=0.31\linewidth]{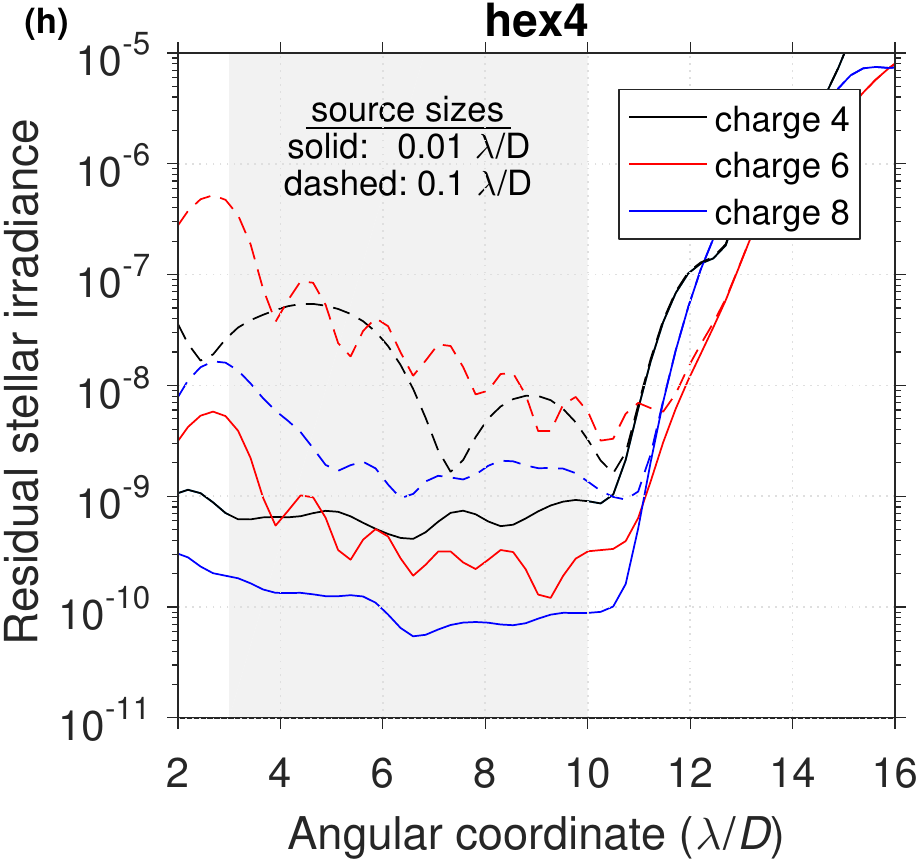}\\
    \caption{Residual starlight for the obscured pupils and source with angular diameter of (solid) 0.01$\lambda/D$ and (dashed) 0.1$\lambda/D$. PSF is normalized to the peak of the telescope PSF before the coronagraph. The gray box indicates the optimization region.}
    \label{fig:PSFS_obs}
\end{figure}

\begin{figure}[p]
    \begin{center}\textbf{Apertures without central obscurations}\end{center}
    \includegraphics[height=0.31\linewidth]{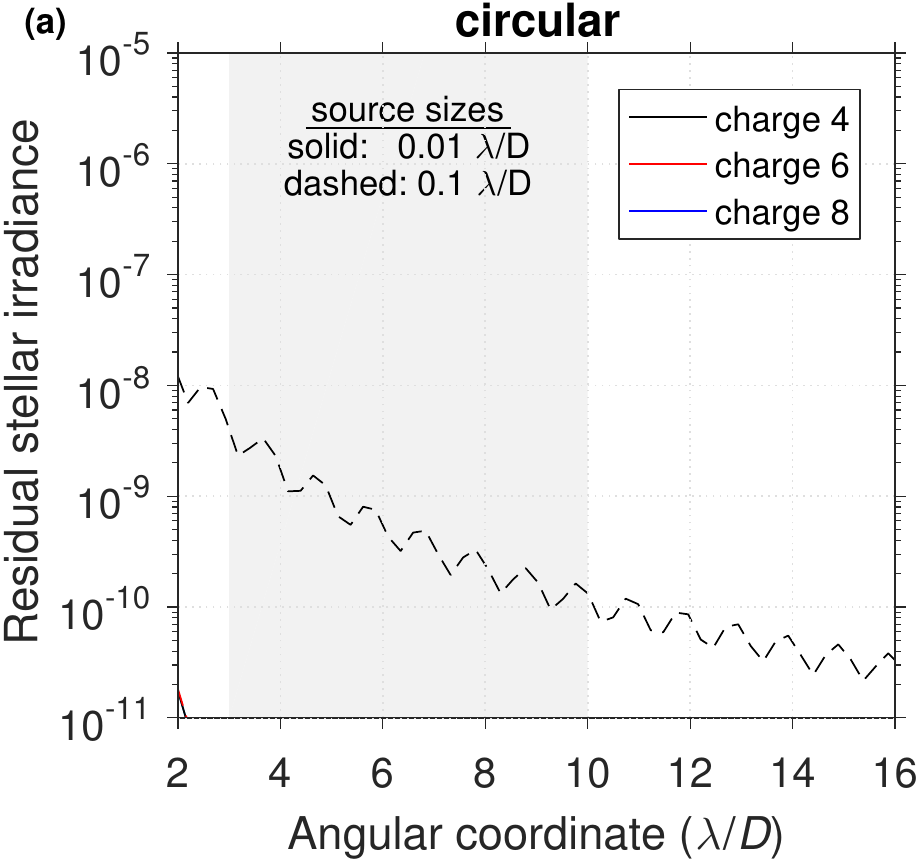}
    \includegraphics[height=0.31\linewidth]{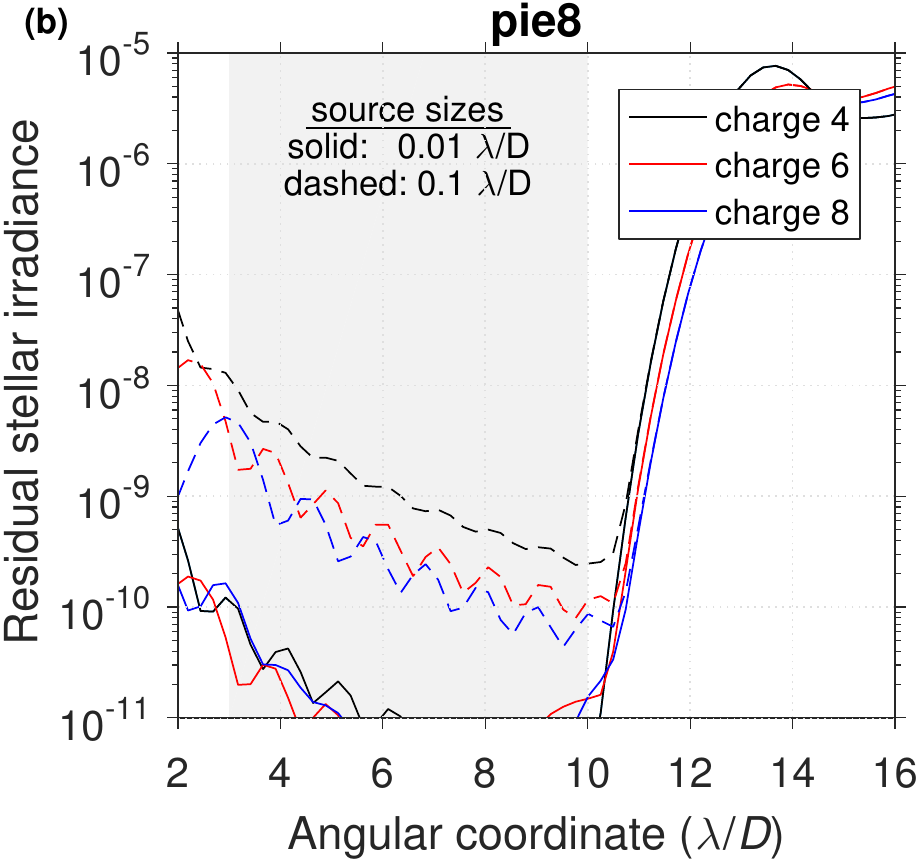}
    \includegraphics[height=0.31\linewidth]{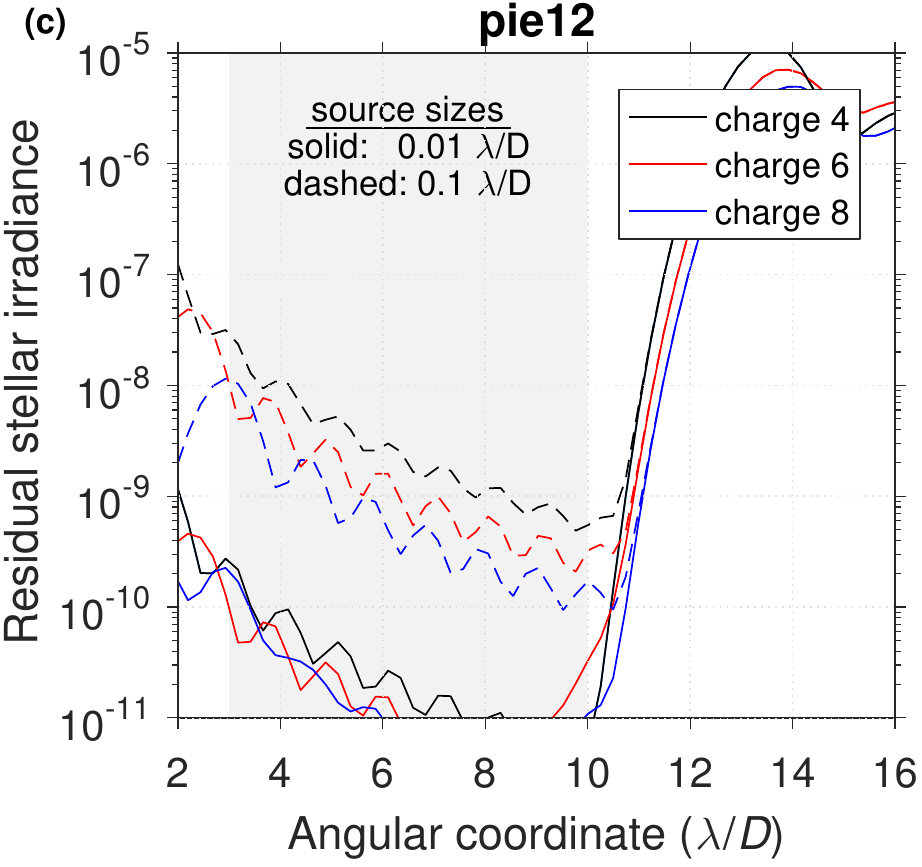}\\
    \includegraphics[height=0.31\linewidth]{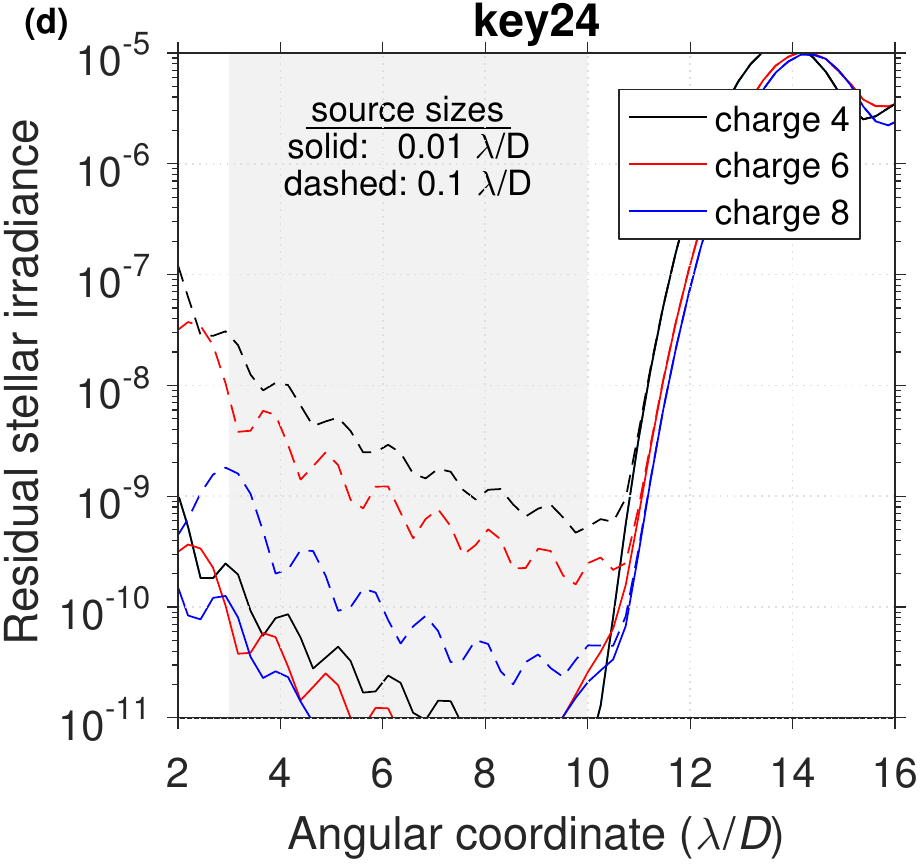}
    \includegraphics[height=0.31\linewidth]{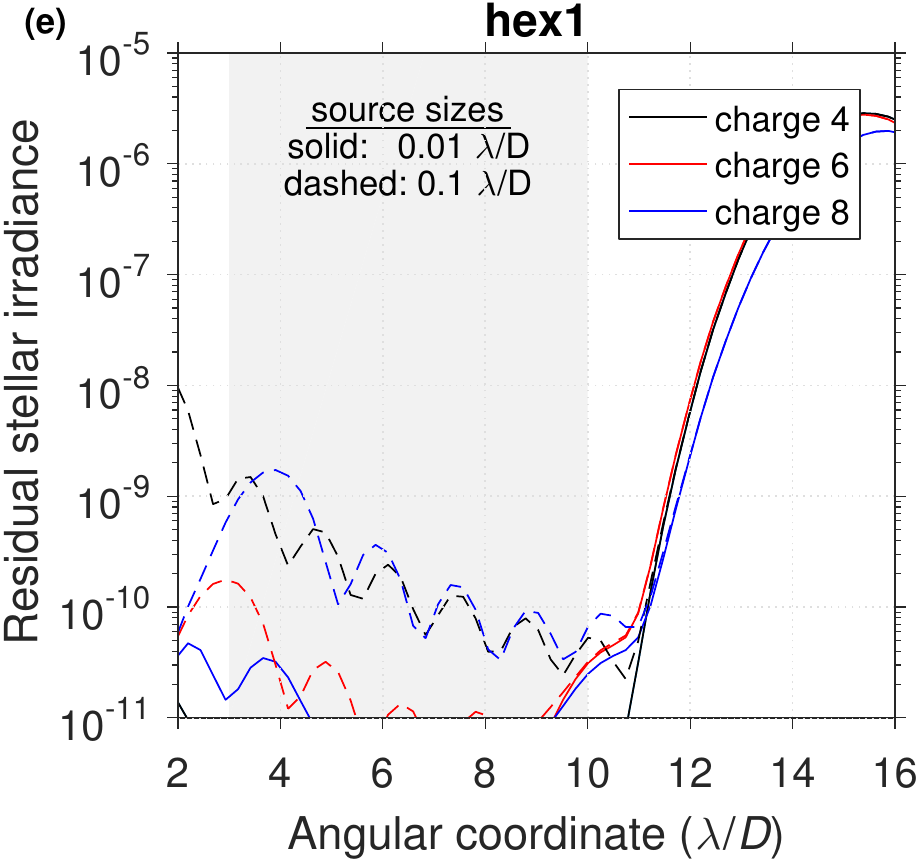}
    \includegraphics[height=0.31\linewidth]{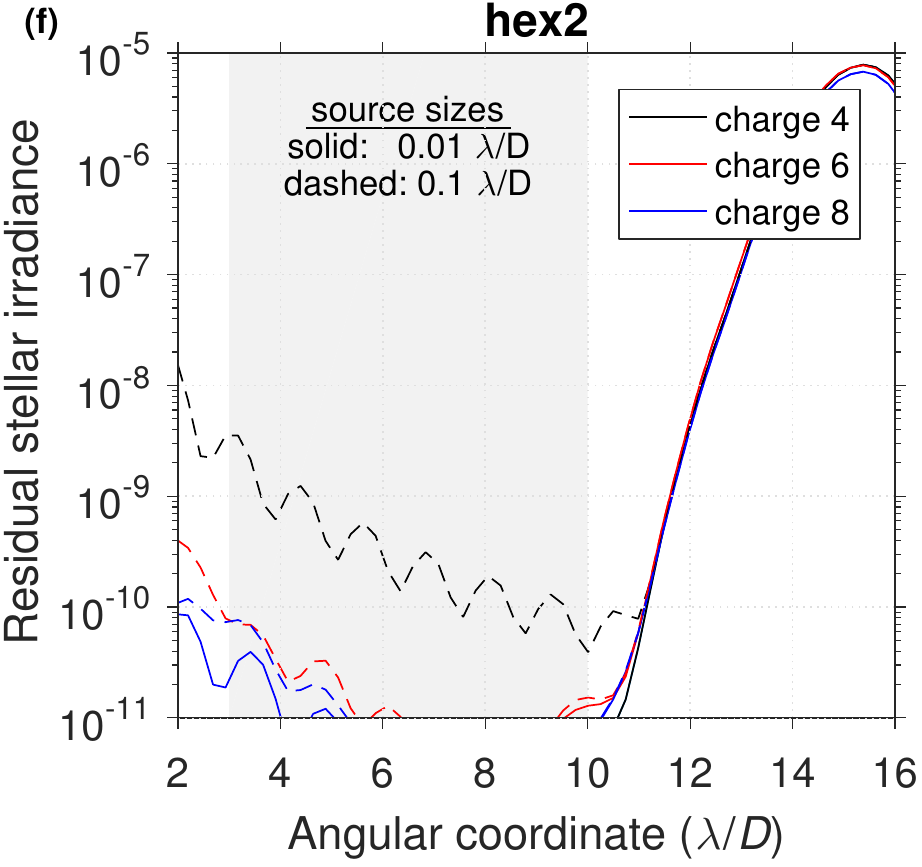}\\
    \includegraphics[height=0.31\linewidth]{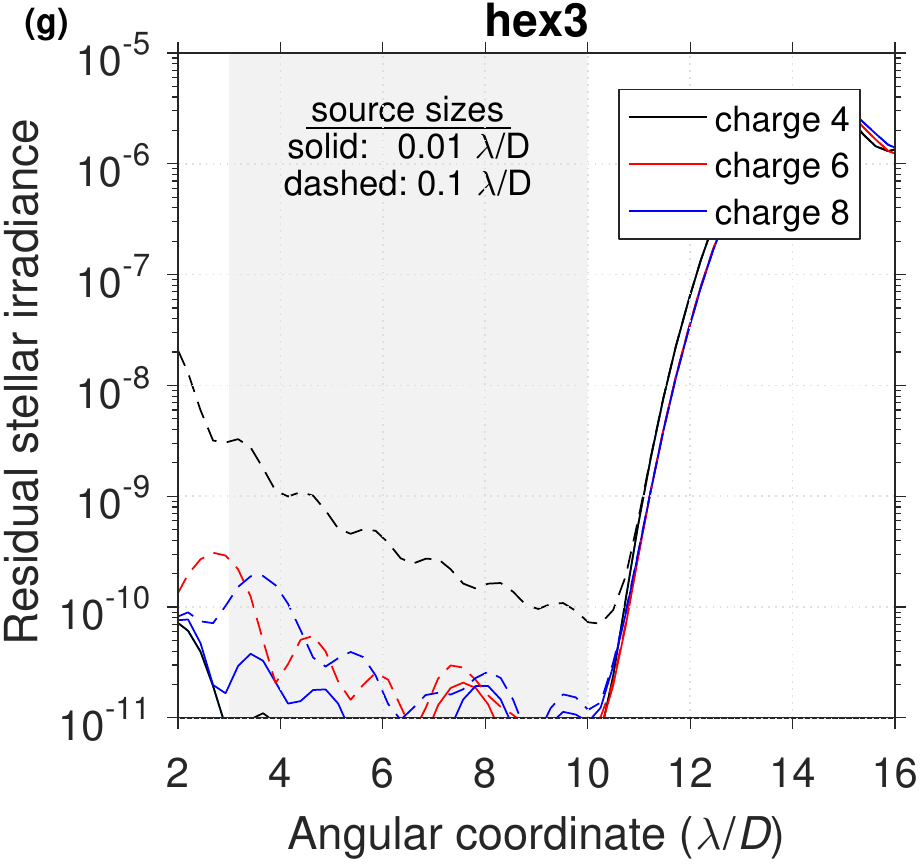}
    \includegraphics[height=0.31\linewidth]{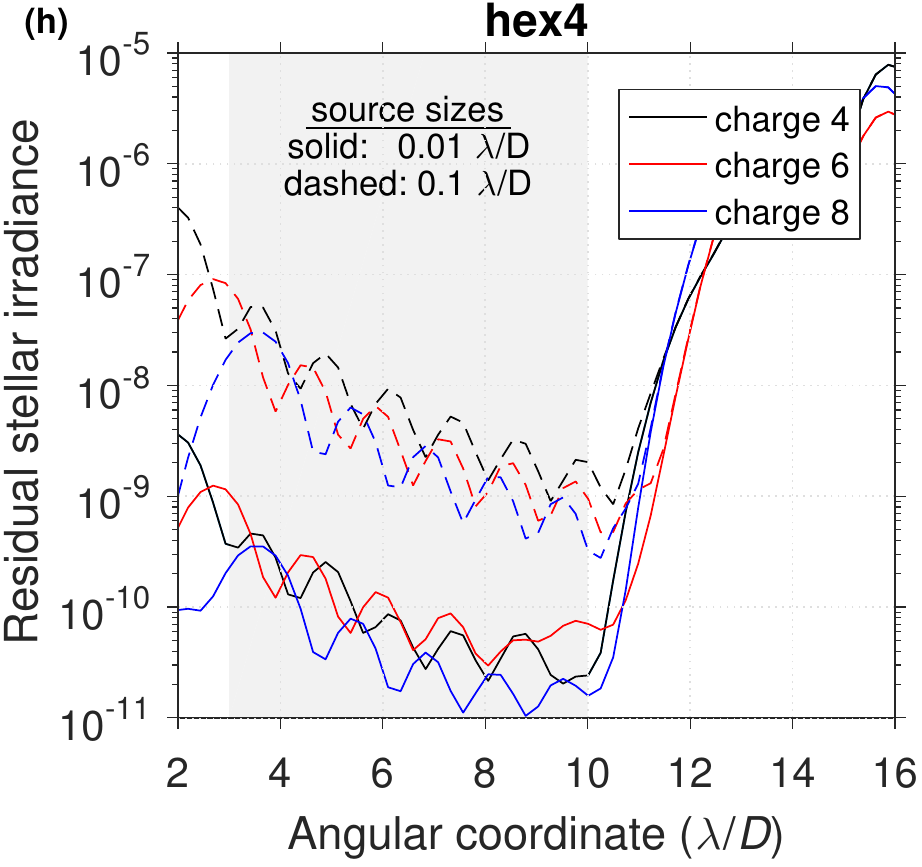}\\
    \caption{Same as Fig. \ref{fig:PSFS_obs}, but for the unobscured apertures.}
    \label{fig:PSFS_unobs}
\end{figure}

Figures \ref{fig:PSFS_obs} and \ref{fig:PSFS_unobs} show the residual stellar irradiance in the image plane after the coronagraph as a function of radial position in angular coordinates, for obscured and unobscured apertures, respectively. Two scenarios are included where the diameter of the star subtends 1\% and 10\% of $\lambda/D$, which for a 12~m space telescope roughly correspond to sun-like stars at 100~pc and 10~pc. We find that the vortex coronagraphs designed for unobscured apertures are generally more robust to the partial resolution of the star. However, there are some exceptions, including the case of a unobscured hex4, which we find leads to unusually high sensitivity with respect to other designs without the central obscuration. The best performance is achieved by charge 6 and 8 with an unobstructed circular aperture. 

The performance for a given stellar angular size is theoretically equivalent to the performance in the presence of jitter with amplitude on the order of the angular radius of the star. Therefore, the results are also representative of the effect of tip-tilt jitter with offsets on the order of 0.5\% and 5\% of $\lambda/D$ (or roughly 0.3\% and 3\% of $\lambda/D$ rms). 

\subsubsection{Coronagraph throughput}
\begin{figure}[p]
    \begin{center}\textbf{Apertures with central obscurations}\end{center}
    \includegraphics[height=0.31\linewidth]{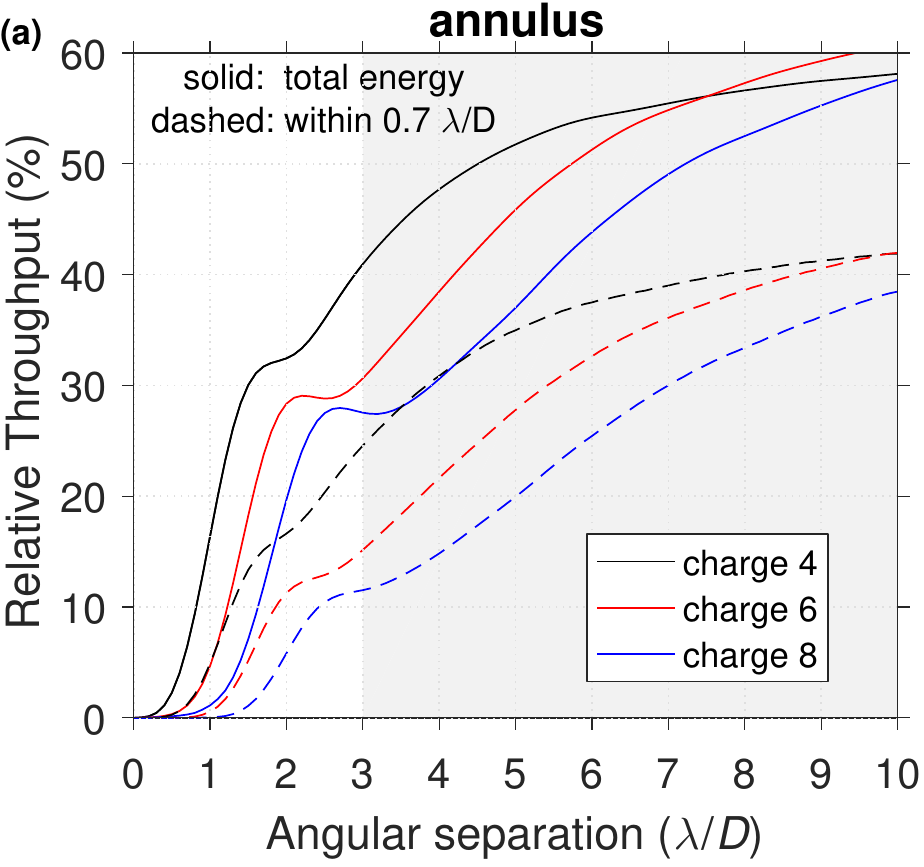}
    \includegraphics[height=0.31\linewidth]{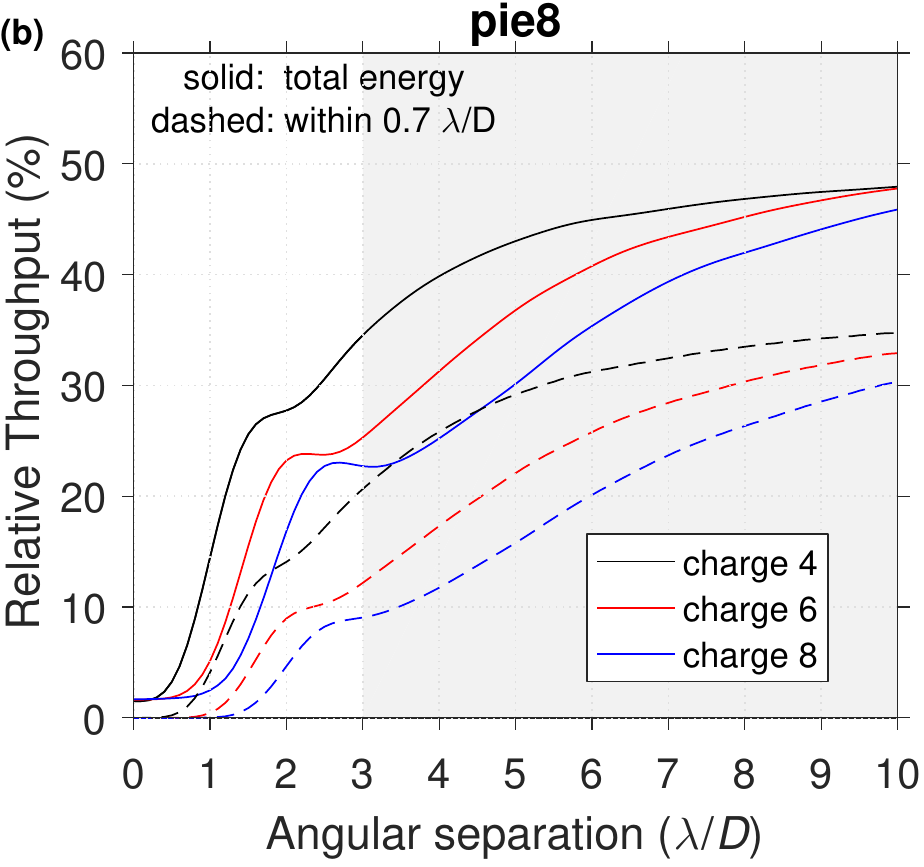}
    \includegraphics[height=0.31\linewidth]{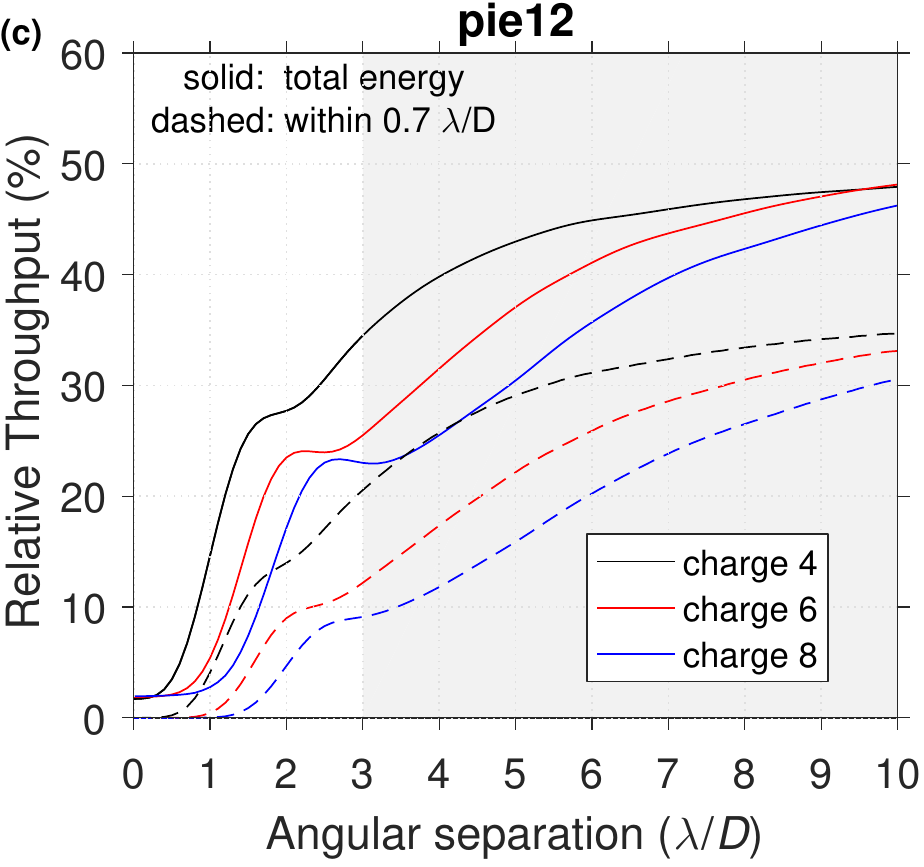}\\
    \includegraphics[height=0.31\linewidth]{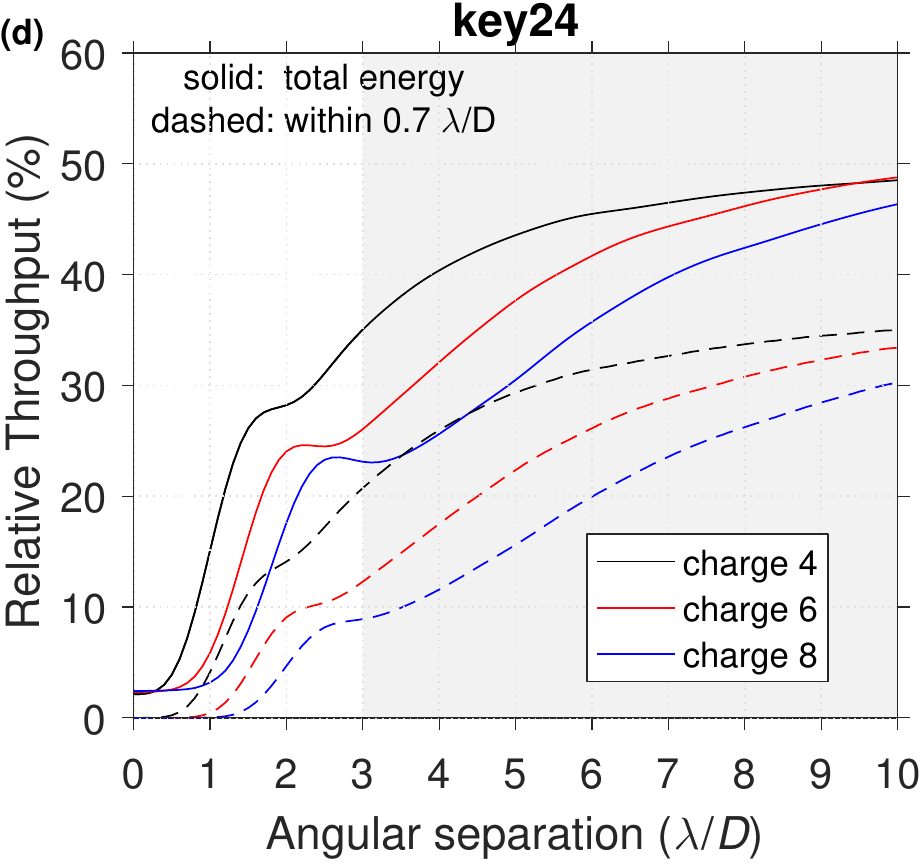}
    \includegraphics[height=0.31\linewidth]{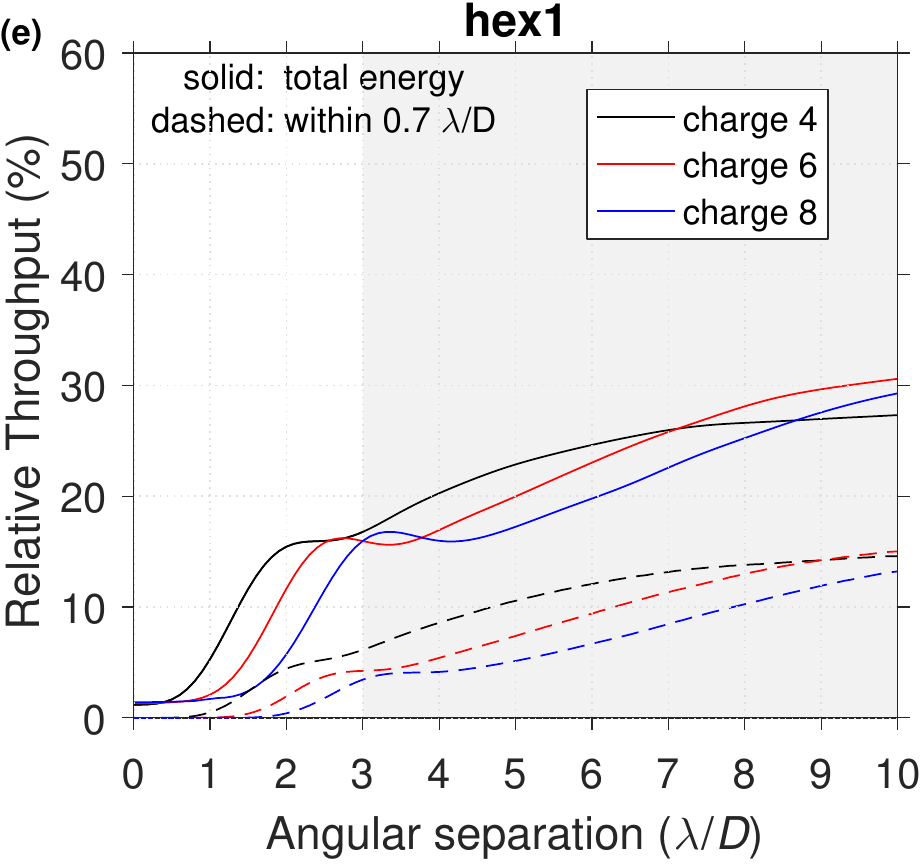}
    \includegraphics[height=0.31\linewidth]{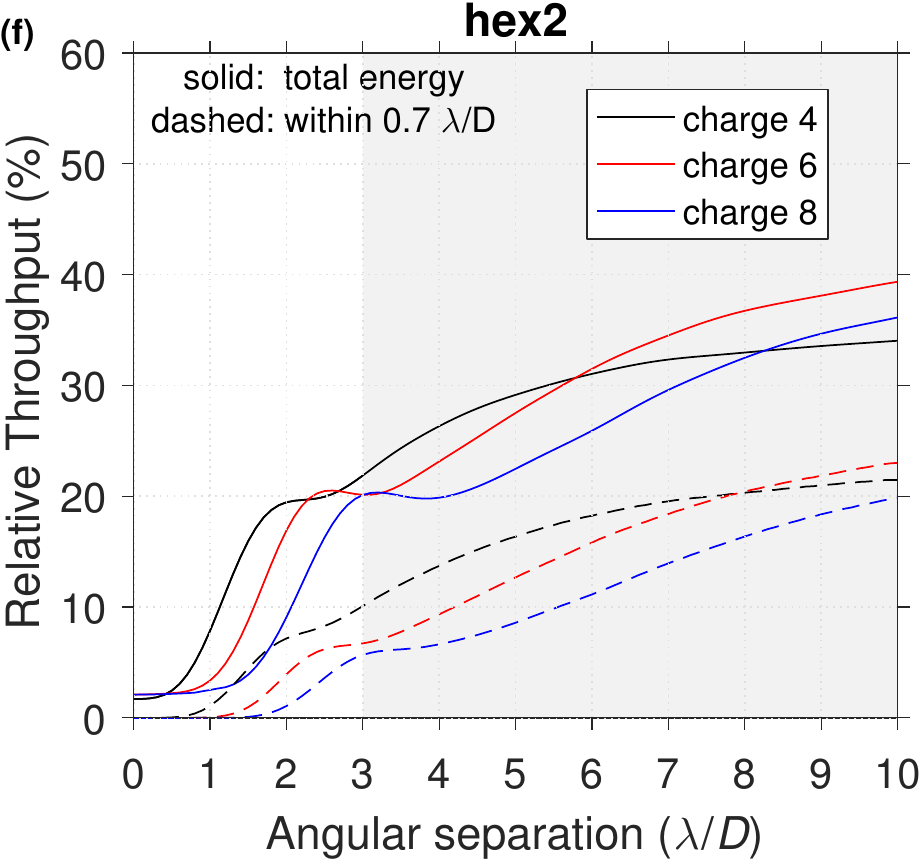}\\
    \includegraphics[height=0.31\linewidth]{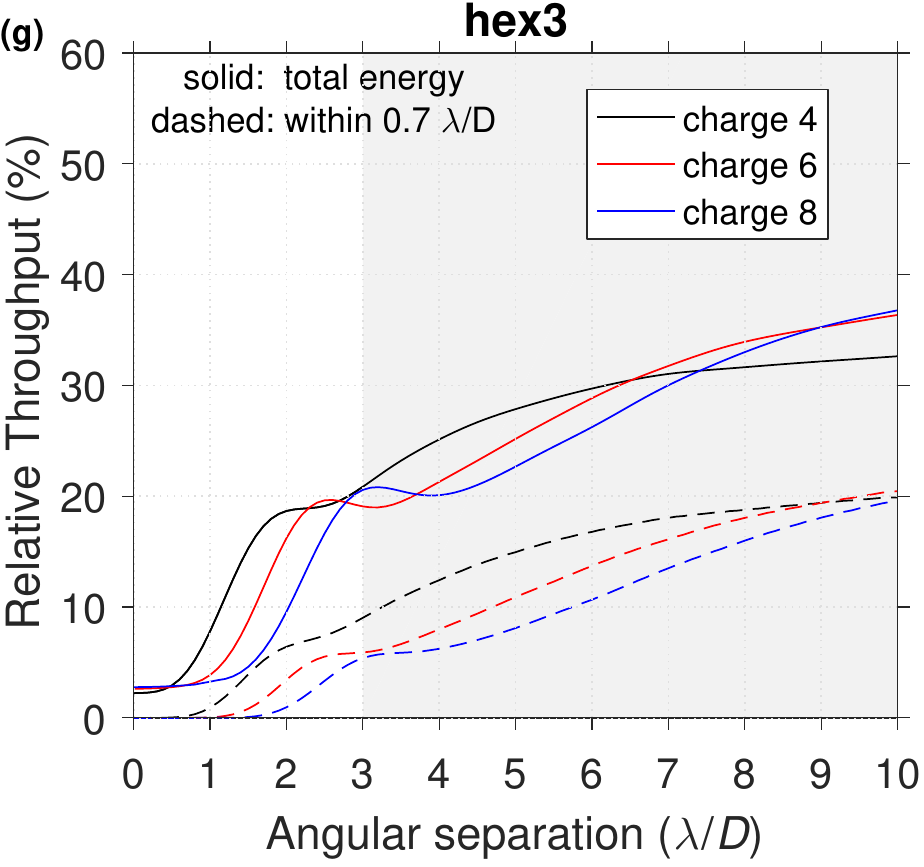}
    \includegraphics[height=0.31\linewidth]{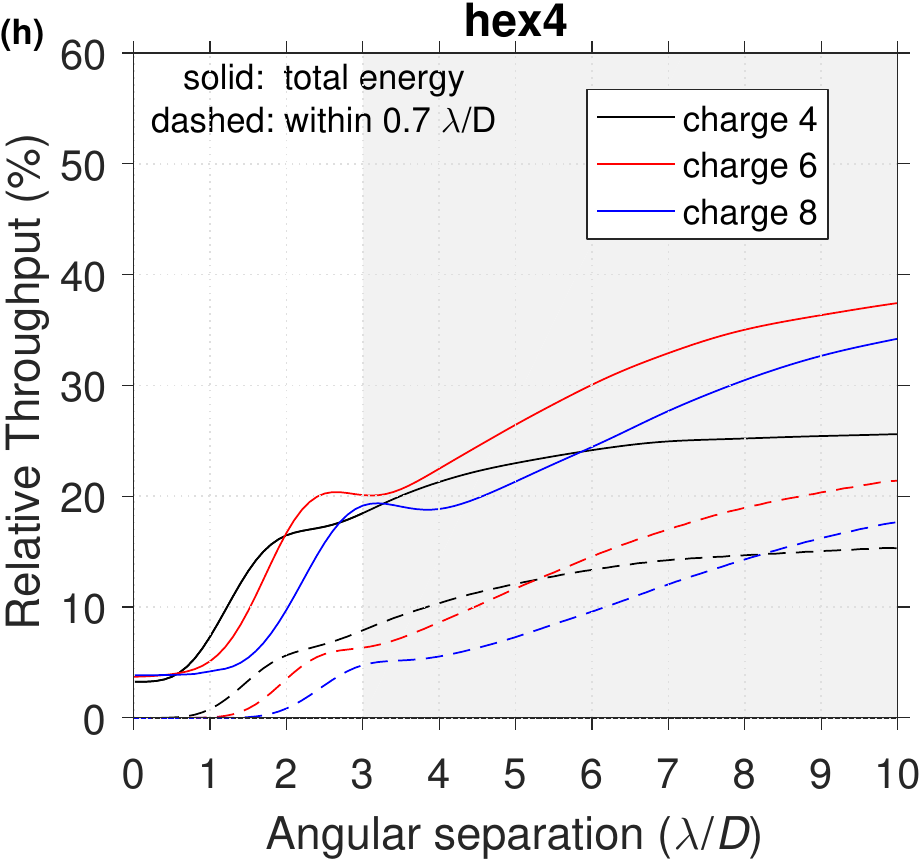}\\
    \caption{Relative throughput for each coronagraph as a function of angular separation. The (solid) total energy throughput and (dashed) PSF core throughput (encircled energy within 0.7~$\lambda/D$ of the source location) are shown. The gray box indicates the optimized starlight suppression region of the coronagraph.}
    \label{fig:Thpts_obs}
\end{figure}

\begin{figure}[p]
    \begin{center}\textbf{Apertures without central obscurations}\end{center}
    \includegraphics[height=0.31\linewidth]{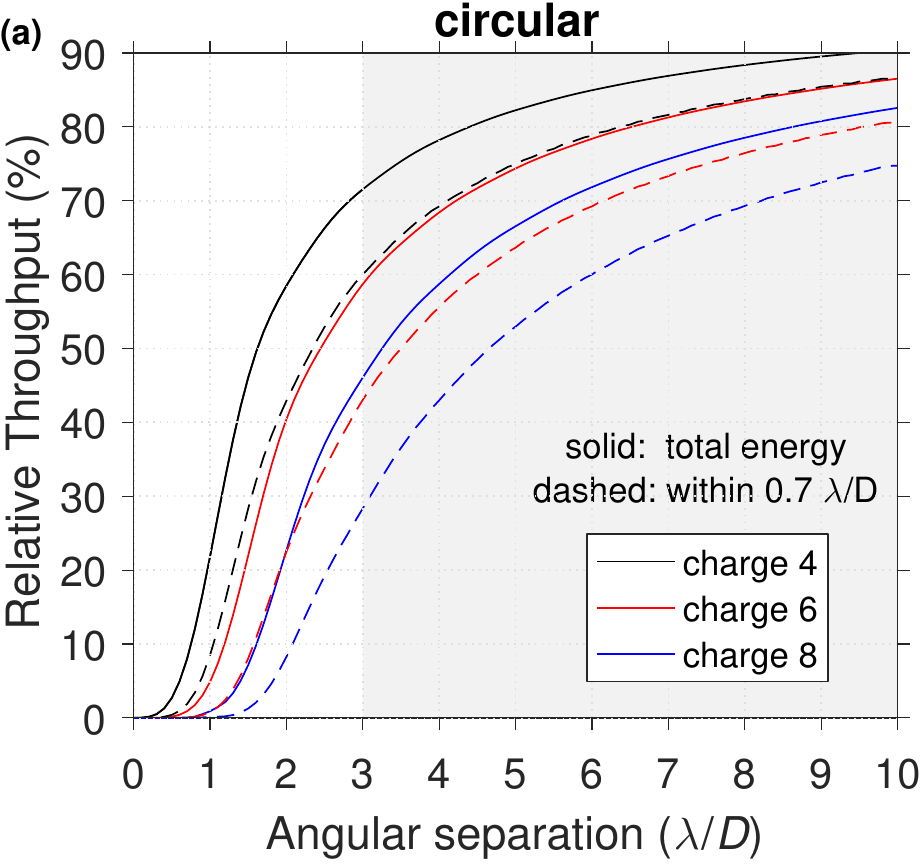}
    \includegraphics[height=0.31\linewidth]{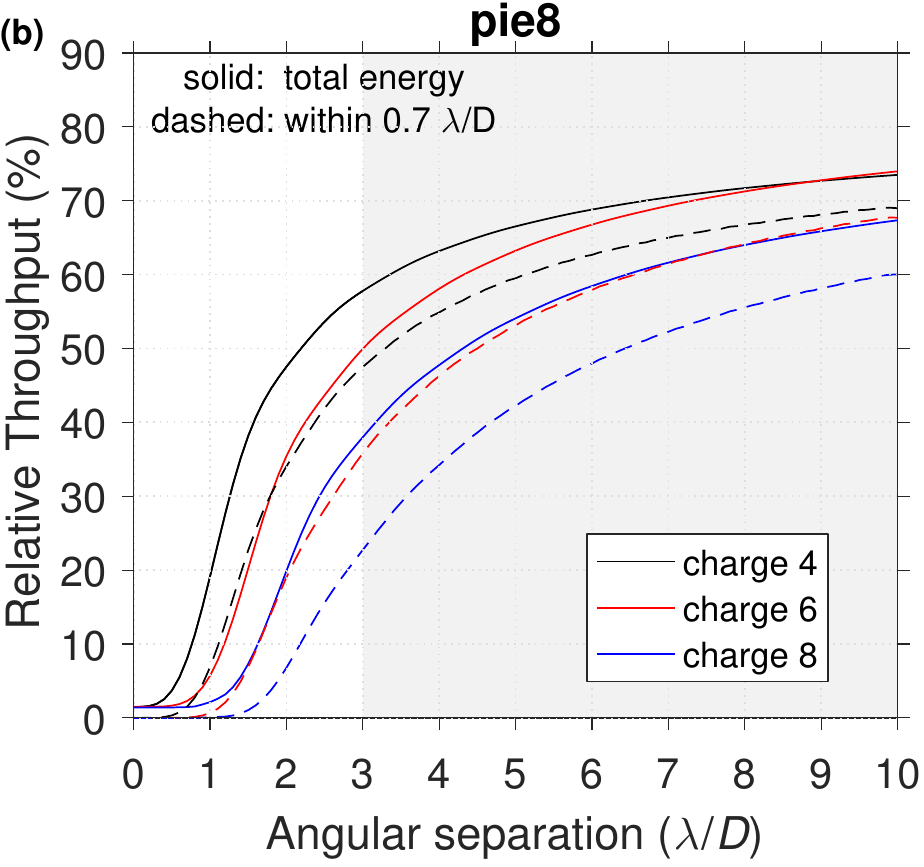}
    \includegraphics[height=0.31\linewidth]{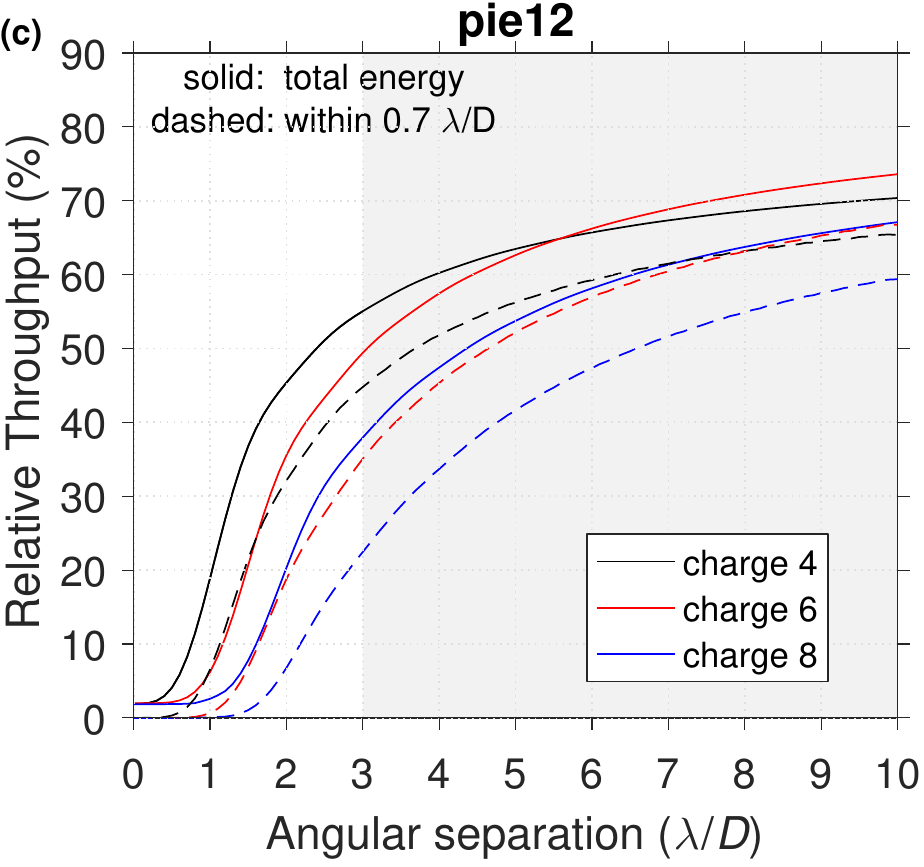}\\
    \includegraphics[height=0.31\linewidth]{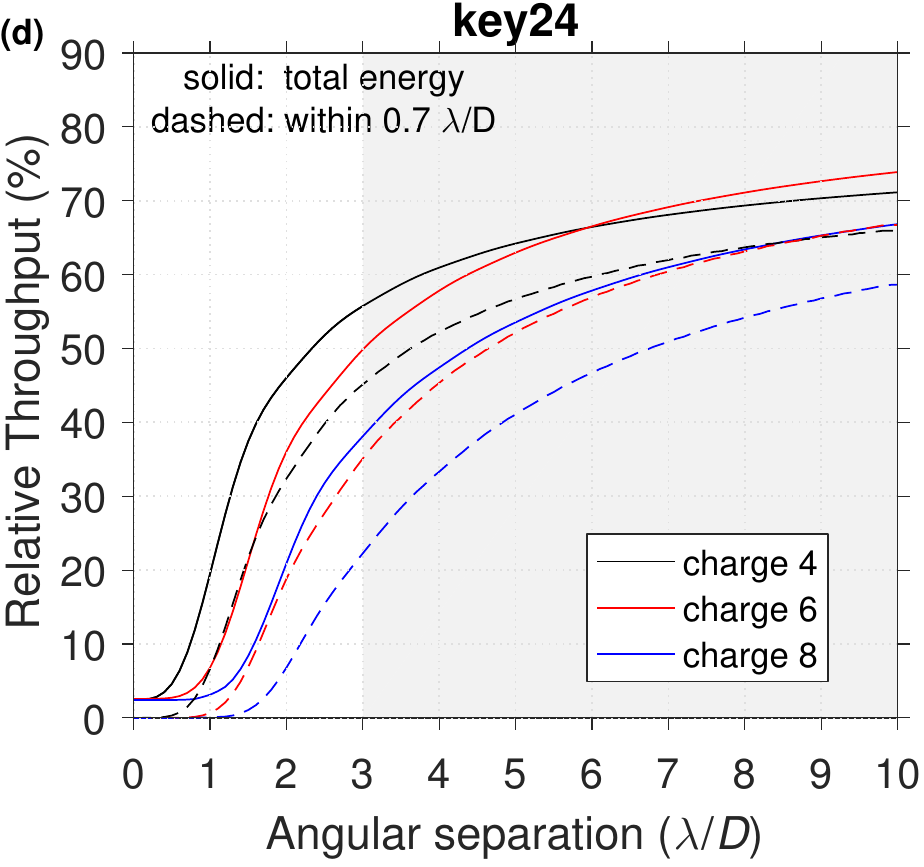}
    \includegraphics[height=0.31\linewidth]{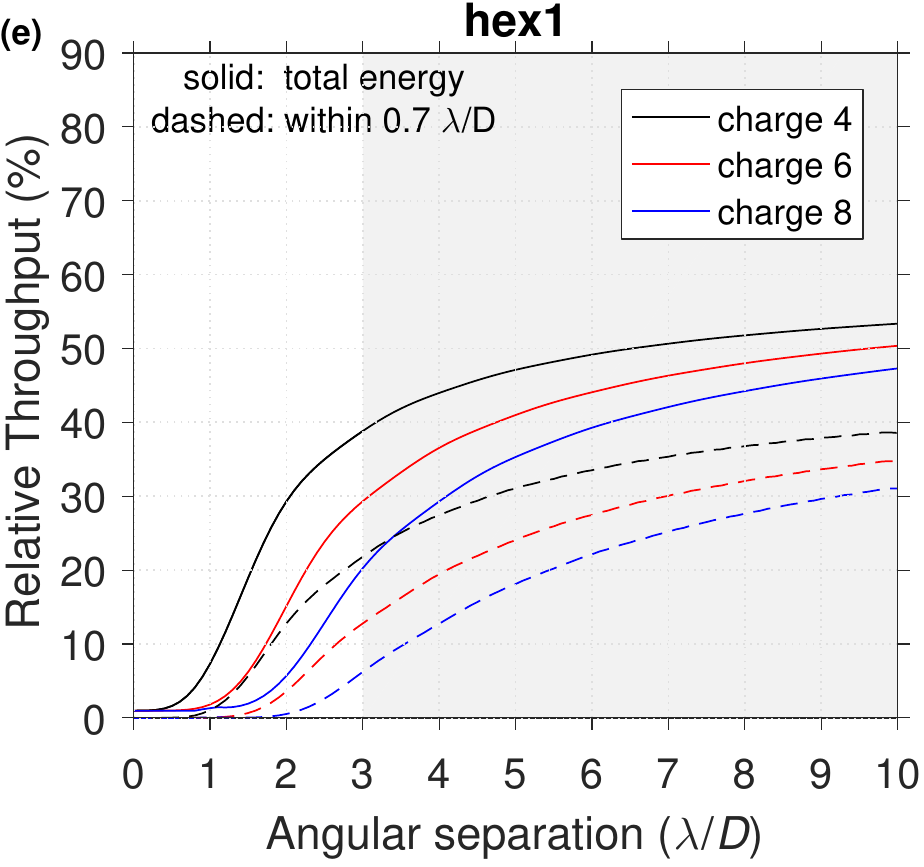}
    \includegraphics[height=0.31\linewidth]{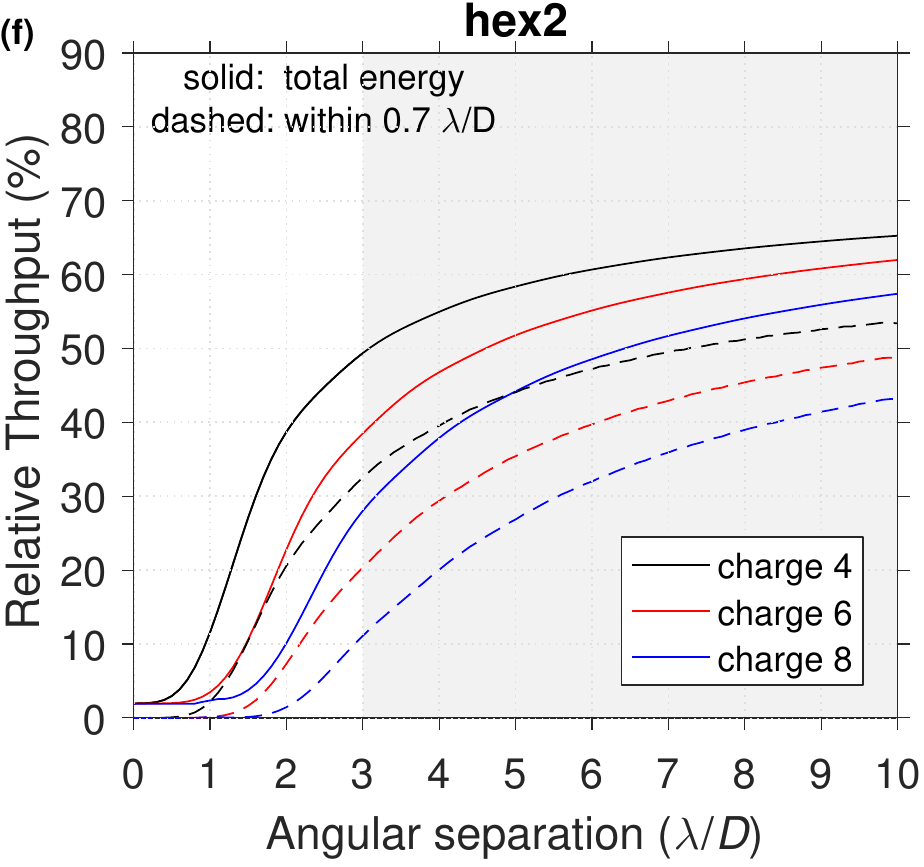}\\
    \includegraphics[height=0.31\linewidth]{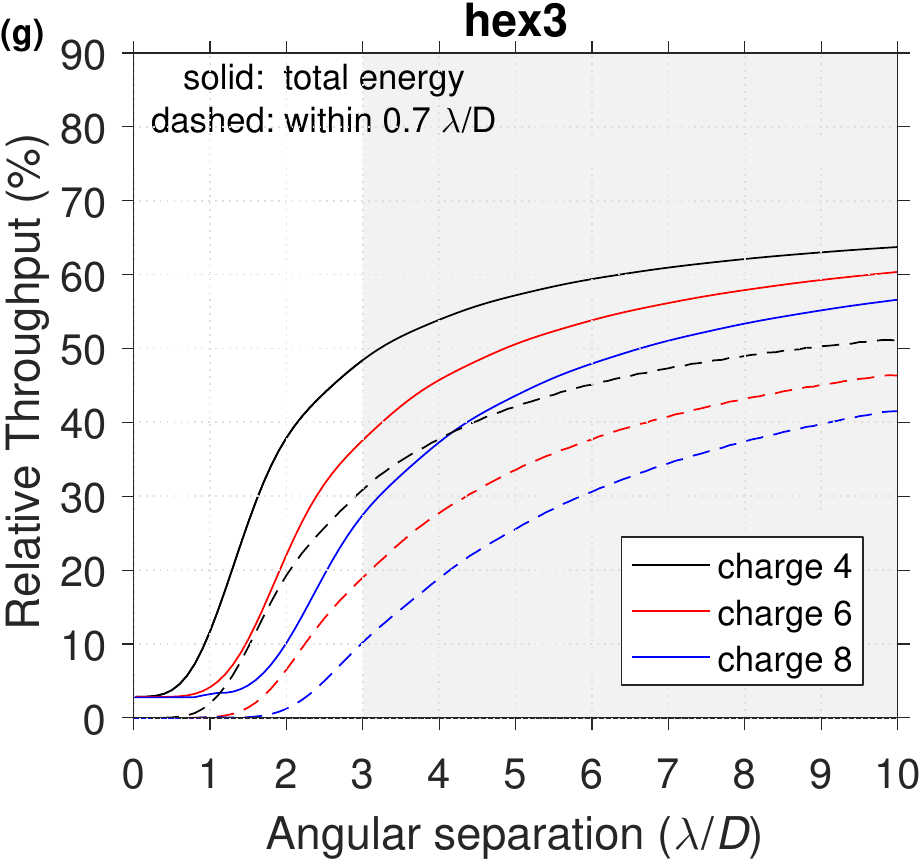}
    \includegraphics[height=0.31\linewidth]{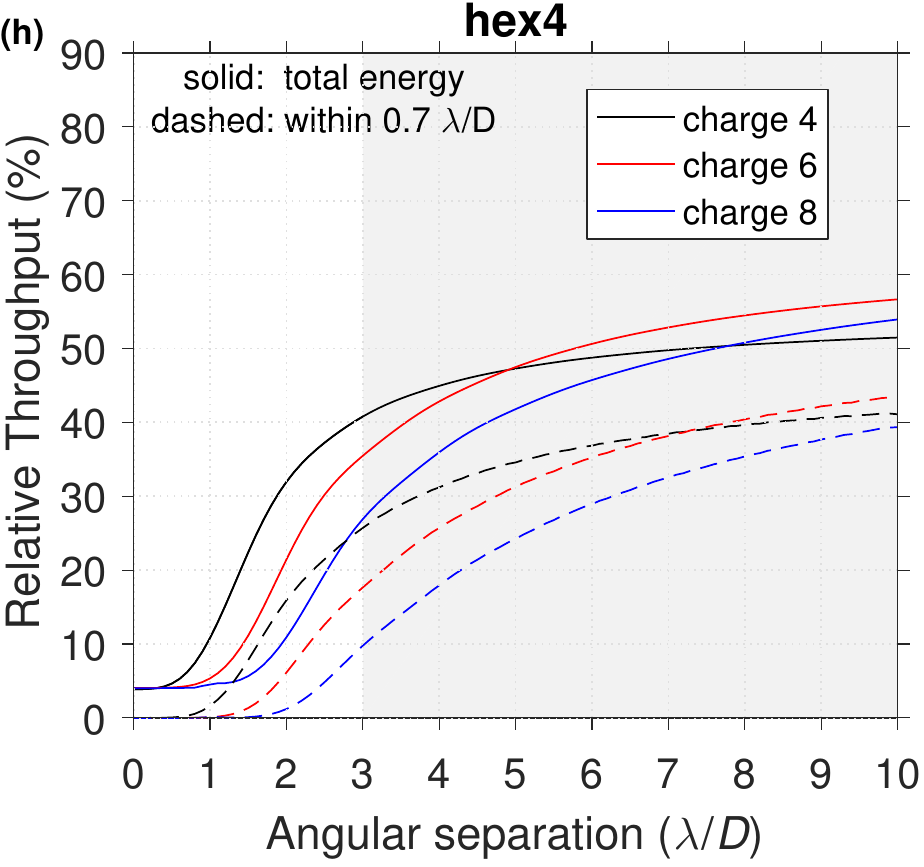}\\
    \caption{Same as Fig. \ref{fig:Thpts_obs}, but for the unobscured apertures.}
    \label{fig:Thpts_unobs}
\end{figure}

The throughput performance of each coronagraph design for obscured and unobscured apertures are shown in Figs. \ref{fig:Thpts_obs} and \ref{fig:Thpts_unobs}, respectively. As explained above, we use two measures: the total energy throughput (i.e. the fraction of energy from a point source that transmits through the Lyot stop) and the fraction of energy within 0.7$\lambda/D$ of the source position. 

Vortex coronagraphs have the unique property that planet signal is maintained at small angular separations (down to $\sim2~\lambda/D$). In all cases, the throughput gradually increases from a few $\lambda/D$ outward. The highest throughput is achieved with unobscured designs, mostly owing to the increased transmitting area of the Lyot stop and  encircled energy in the PSF core. A circular pupil offers the best throughput, followed by pie8, pie12, and key24 unobscured apertures. Using an obstructed aperture leads to a factor of $\sim$2 loss (see e.g. circle vs. annulus). This is one of the primary reasons to avoid a central obscuration for obtaining the best performance with vortex coronagraphs. By most measures, the overall performance of a high-contrast imaging instrument depends strongly on the coronagraph throughput. 

\subsubsection{Sensitivity to low order aberrations}
A major concern for high-contrast imaging with future space telescopes is that low order aberrations arise as a result of thermal changes, mechanical flexing, and polarization aberrations. The sensitivity of a vortex coronagraph with a circular pupil to such aberrations is well understood analytically \citep[see e.g.][]{Ruane2015_SPIE,Ruane2017_SPIE}. Specifically, a vortex coronagraph of charge $l$ is insensitive to Zernike phase modes $Z_n^m$ to first order (i.e. assuming $\exp(iZ_n^m)\approx1+iZ_n^m$) if $l$ is even and $|l|>n+|m|$. The minimum charge needed to sufficiently reject the lowest order Zernike modes are listed in Table \ref{tab:Zern}.

\begin{table}[h]
\small
\centering
\begin{tabular}{ *{8}{c} }
$Z_n^m$ & piston & tilt & defocus & astig. & coma & trefoil & spherical \\
\hline
$n$ & 0 & 1 & 2 & 2 & 3 & 3 & 4 \\
$m$ & 0 & $\pm$1 & 0 & $\pm$2 & $\pm$1 & $\pm$3 & 0 \\
Min. $l$ & 2 & 4 & 4 & 6 & 6 & 8 & 6\\
\hline
\end{tabular}
\caption{Minimum charge needed to reject each Zernike mode to first order.}
\label{tab:Zern}
\end{table}

\begin{figure}[p]
    \centering
    \includegraphics[height=0.3\linewidth]{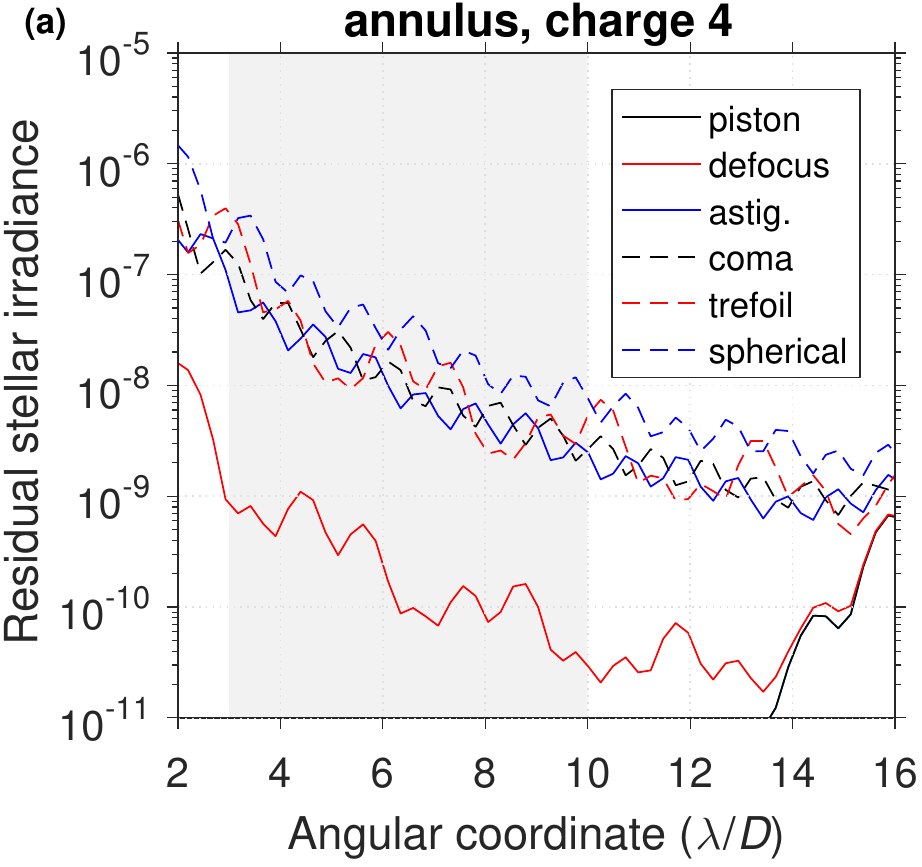}
    \includegraphics[height=0.3\linewidth]{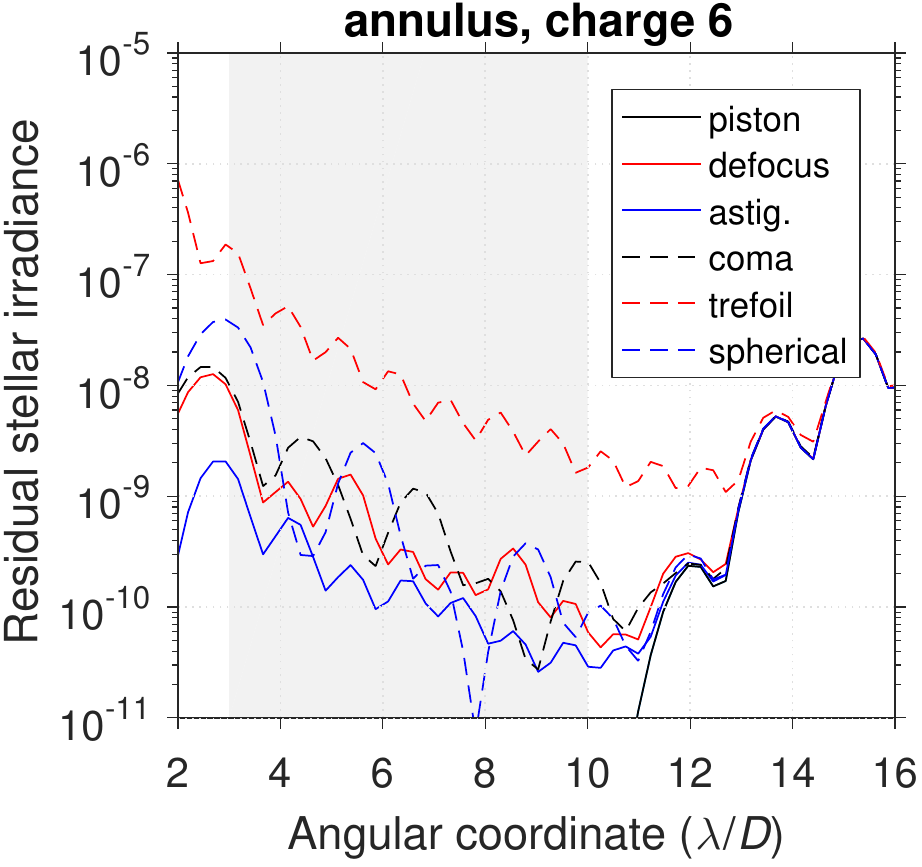}
    \includegraphics[height=0.3\linewidth]{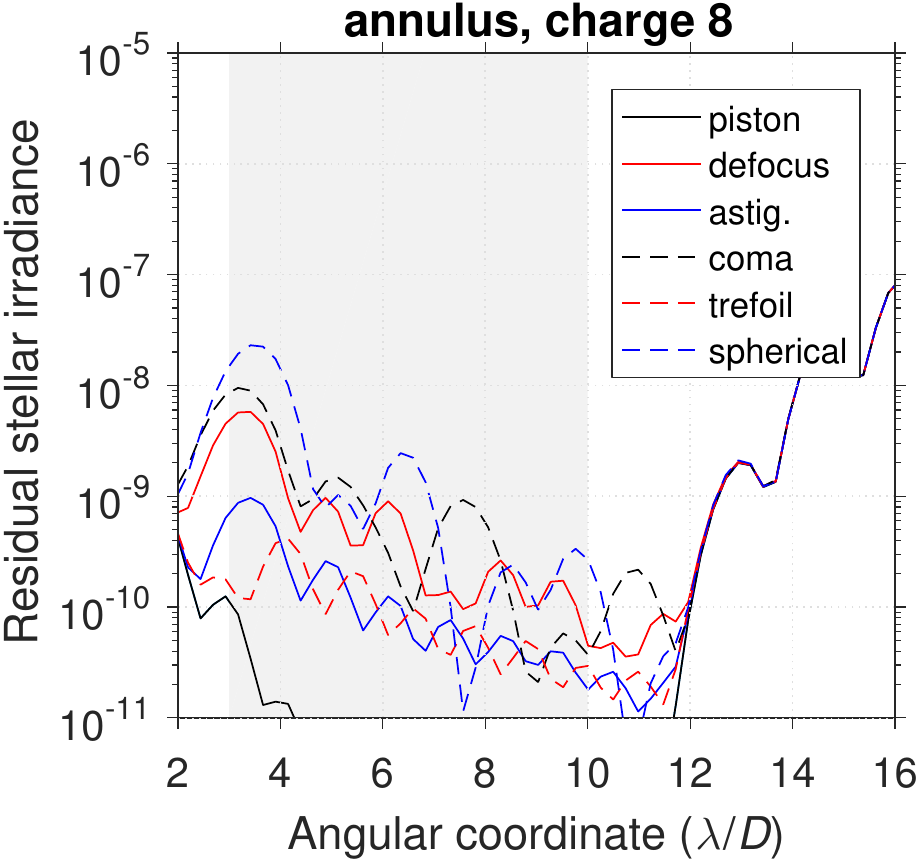}\\
    \includegraphics[height=0.3\linewidth]{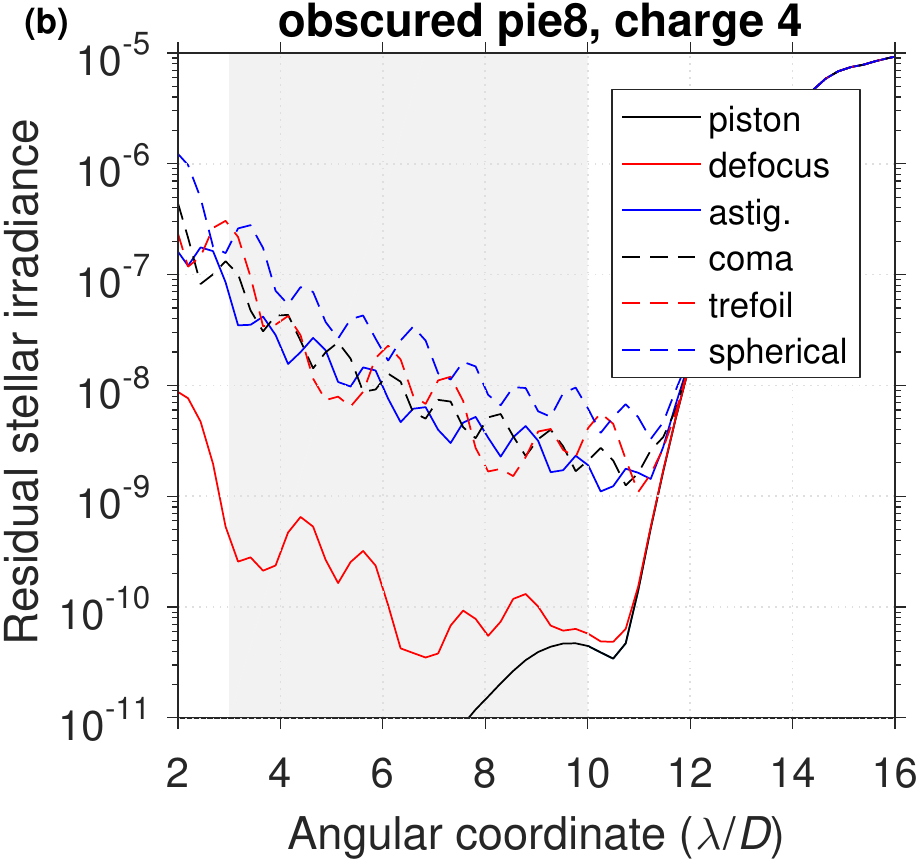}
    \includegraphics[height=0.3\linewidth]{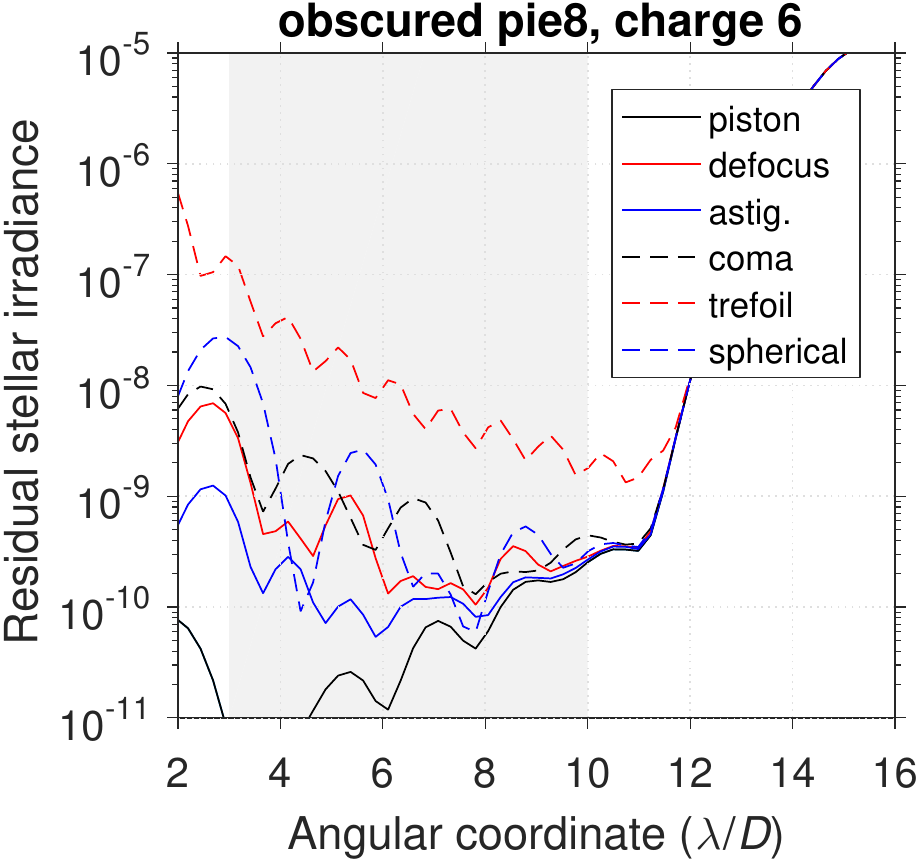}
    \includegraphics[height=0.3\linewidth]{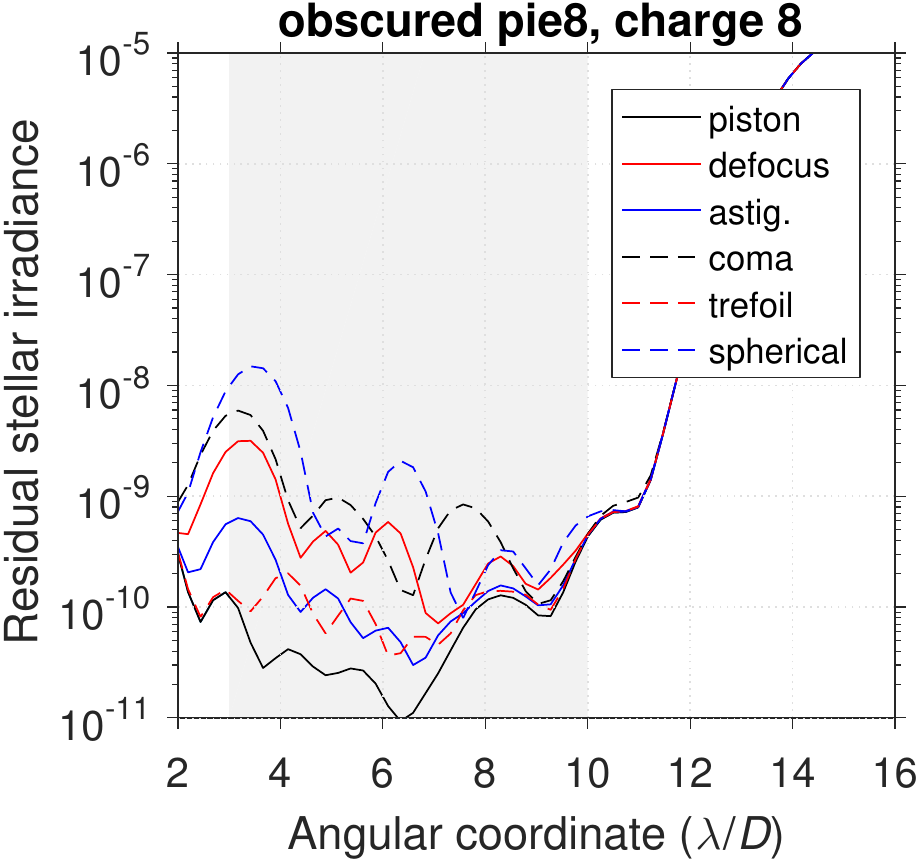}\\
    \includegraphics[height=0.3\linewidth]{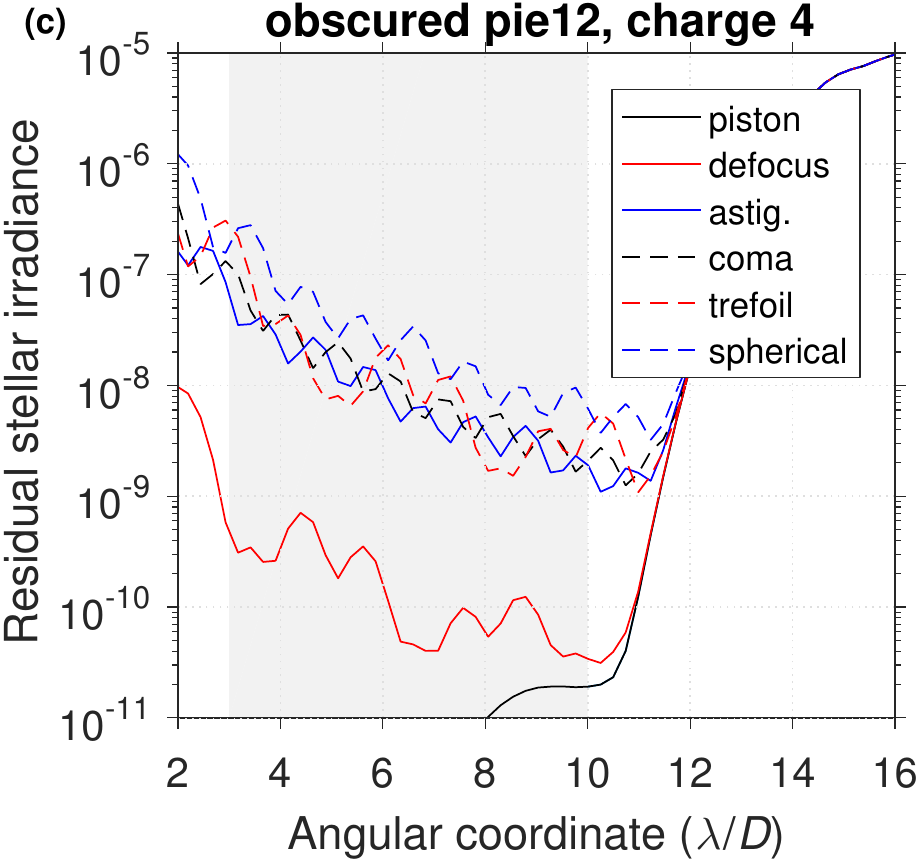}
    \includegraphics[height=0.3\linewidth]{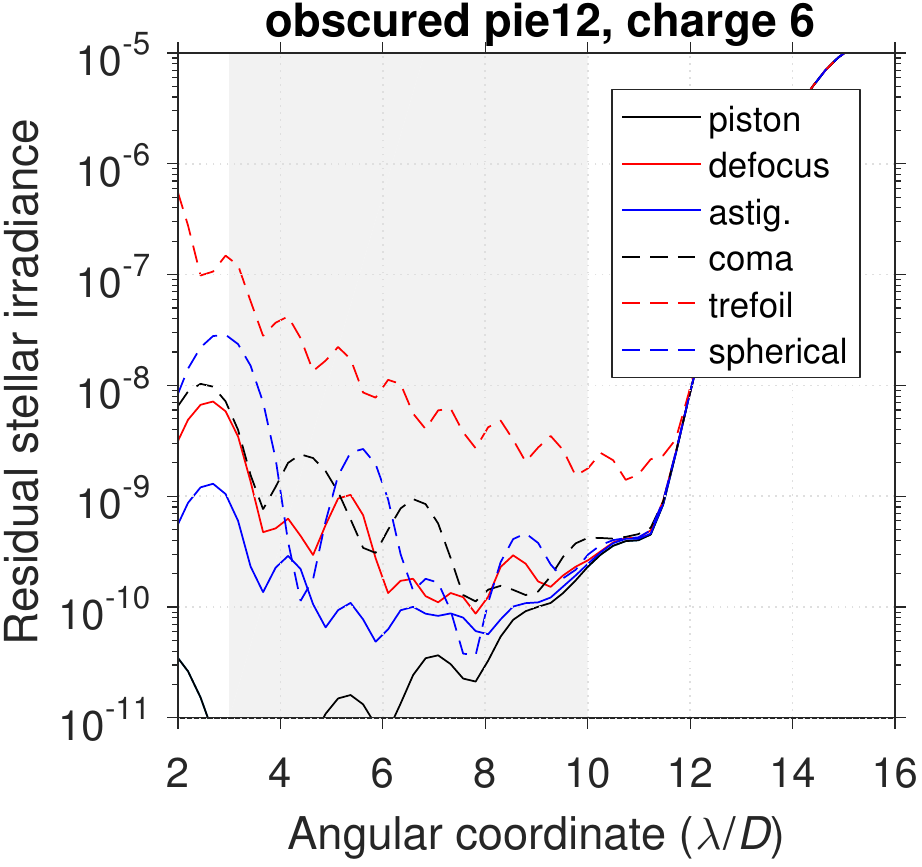}
    \includegraphics[height=0.3\linewidth]{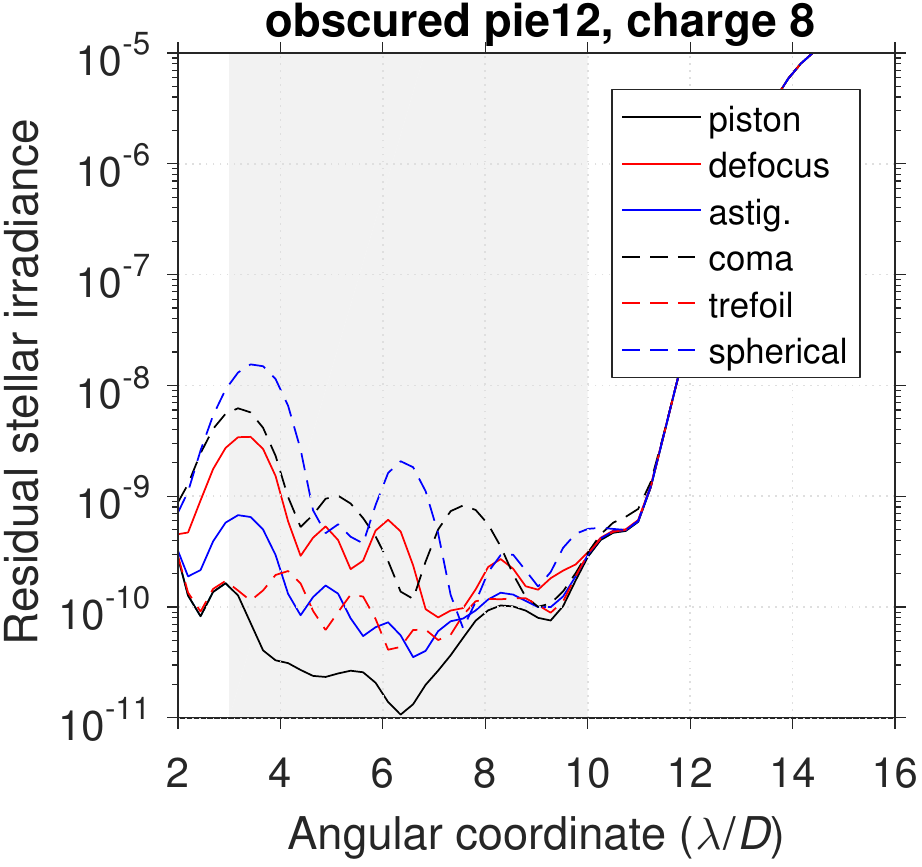}\\
    \includegraphics[height=0.3\linewidth]{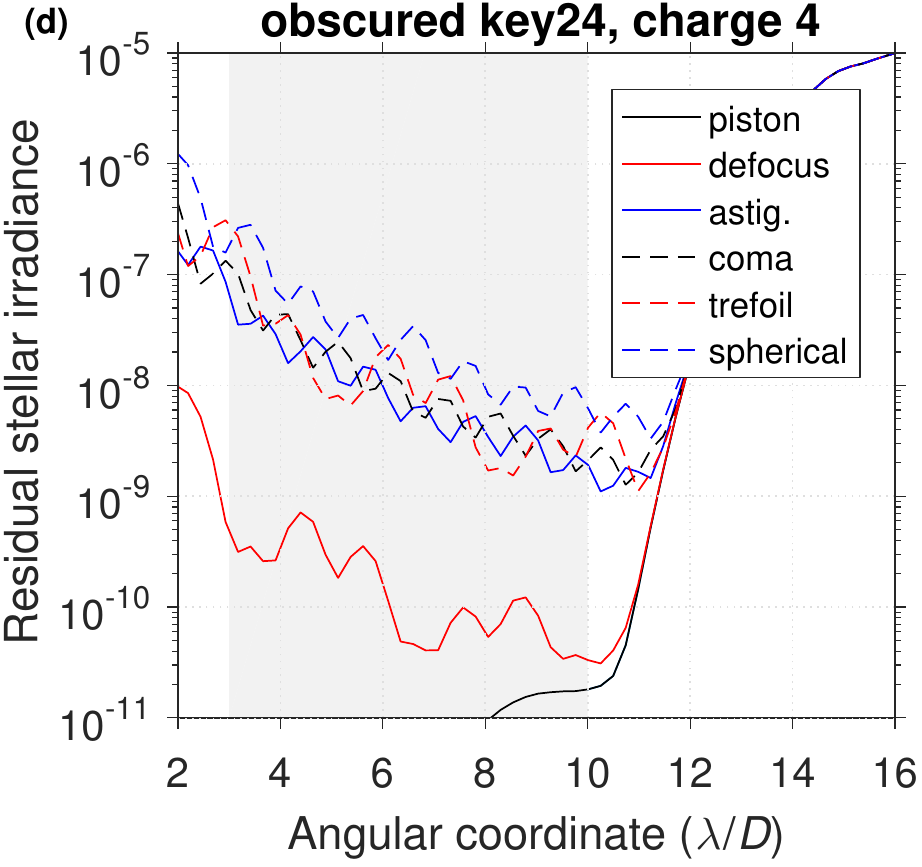}
    \includegraphics[height=0.3\linewidth]{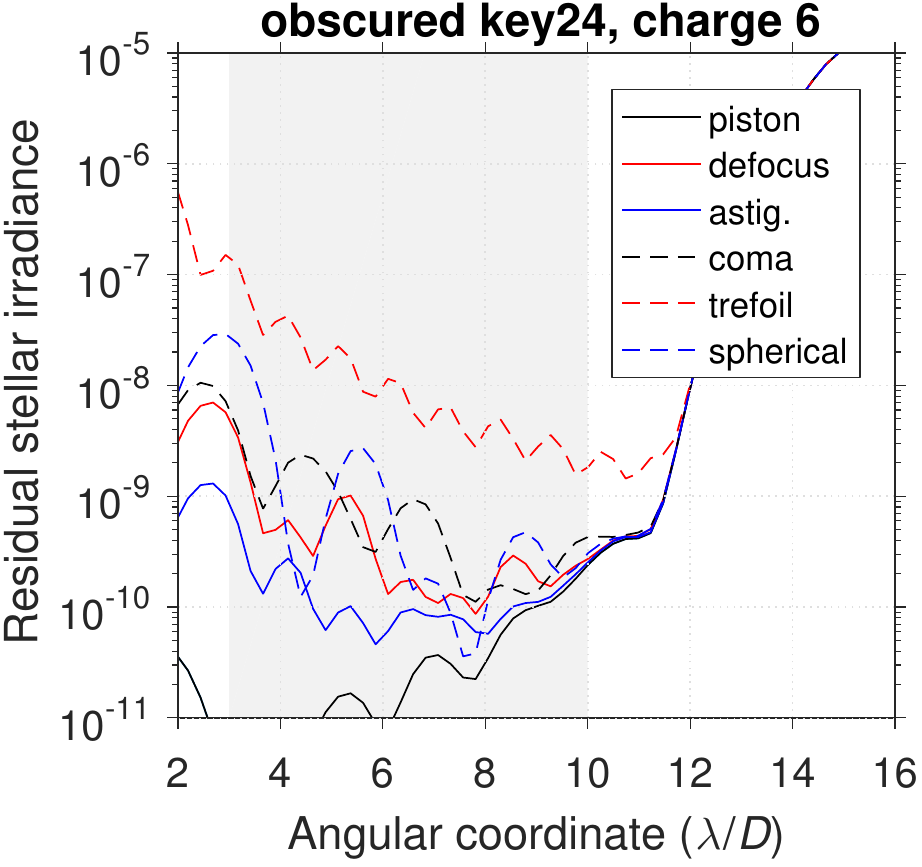}
    \includegraphics[height=0.3\linewidth]{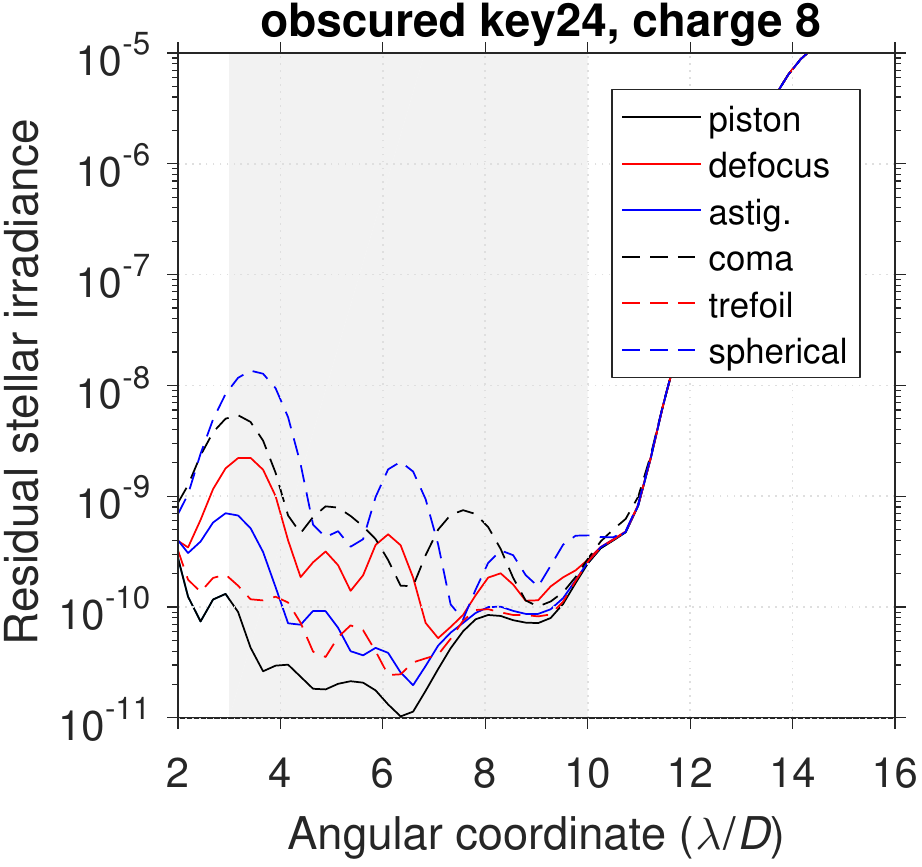}\\
    \caption{Residual starlight with $\lambda/1000$ rms wavefront error in the lowest order Zernike modes.}
    \label{fig:Zsens_obs1}
\end{figure}

\begin{figure}[p]
    \centering
    \includegraphics[height=0.3\linewidth]{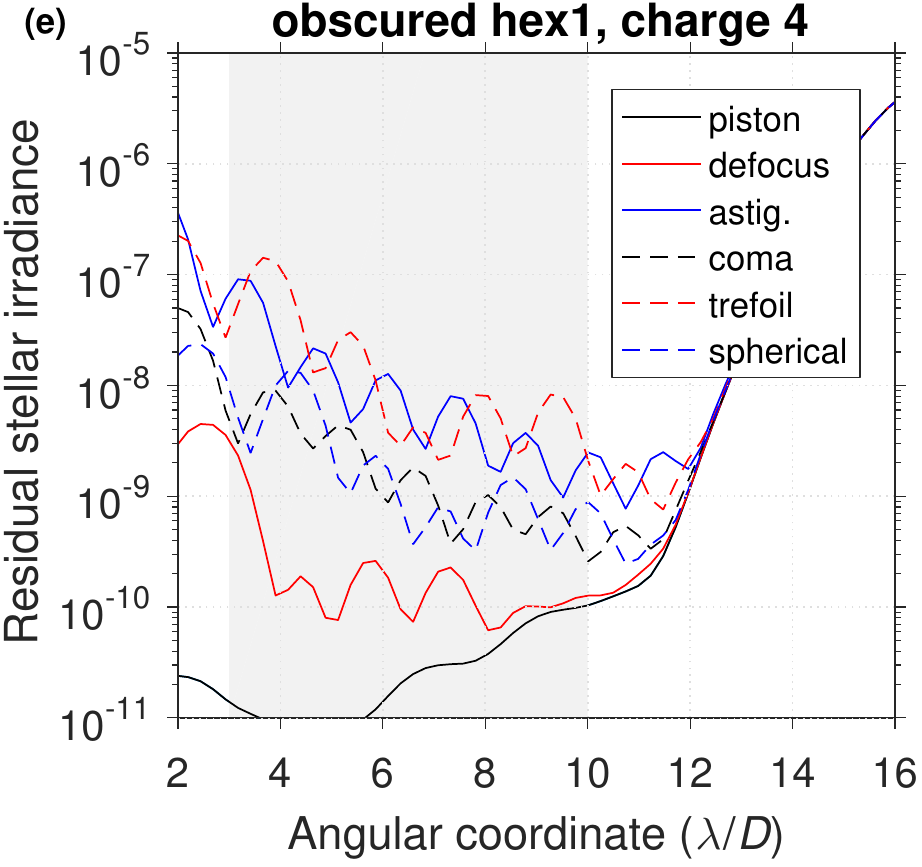}
    \includegraphics[height=0.3\linewidth]{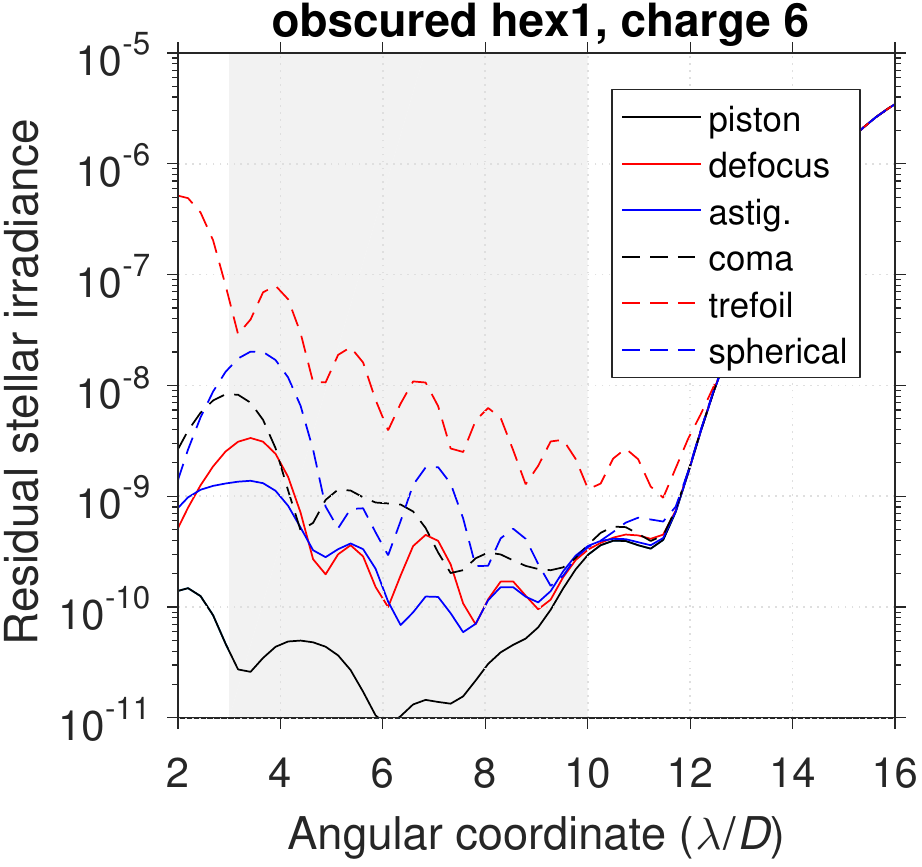}
    \includegraphics[height=0.3\linewidth]{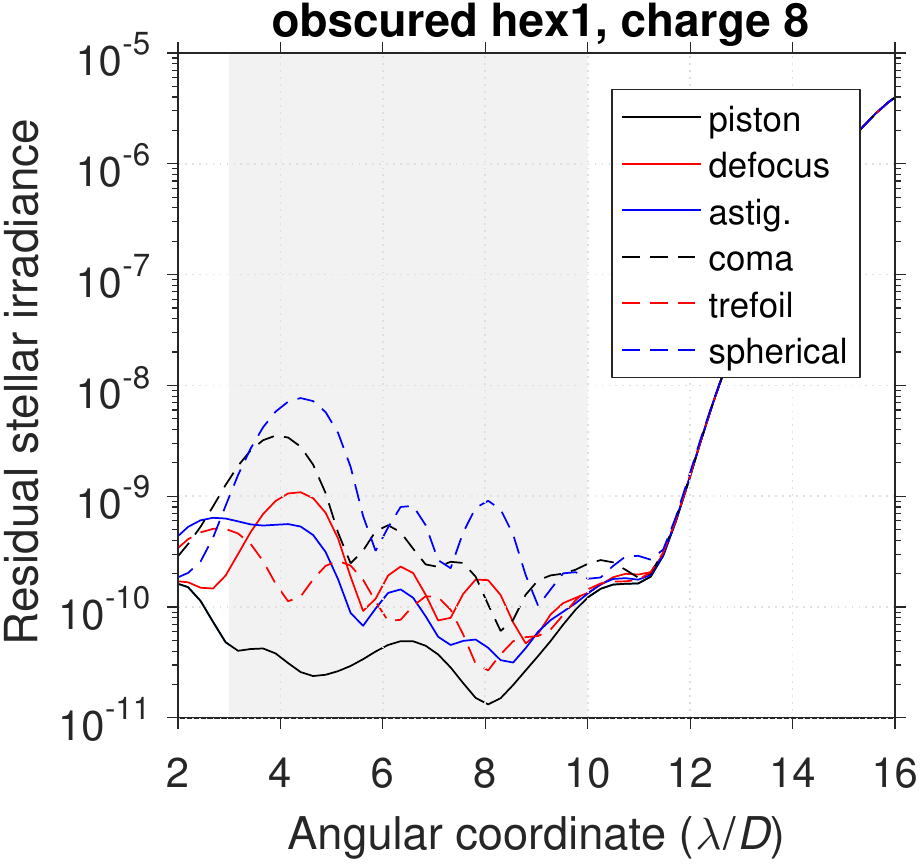}\\
    \includegraphics[height=0.3\linewidth]{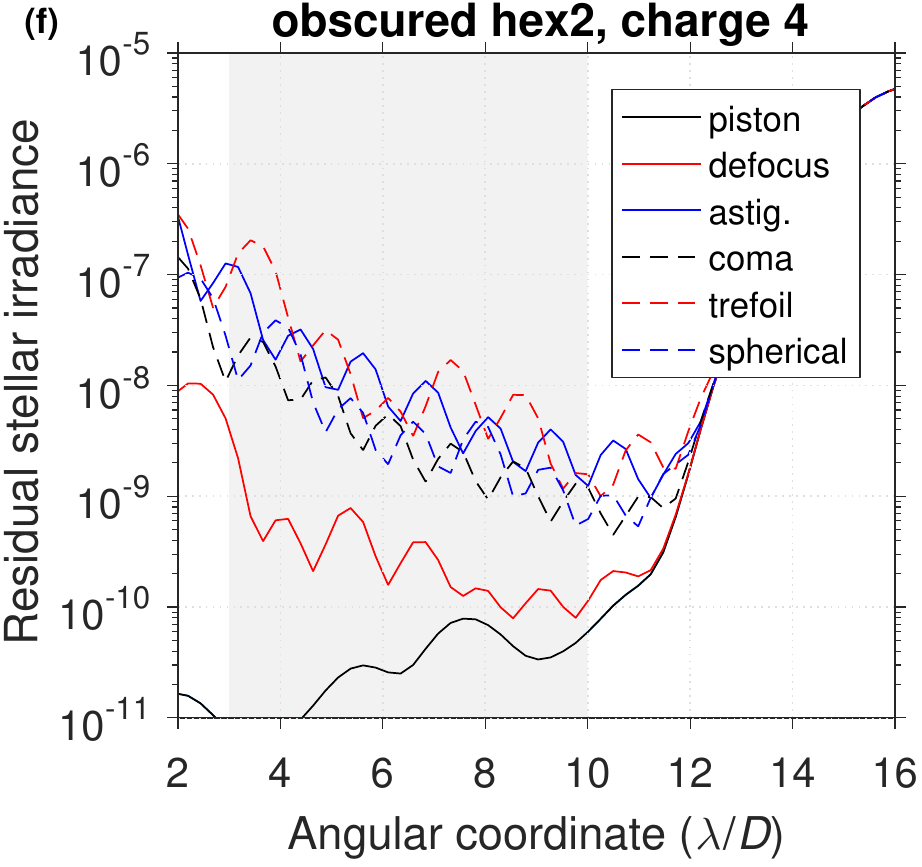}
    \includegraphics[height=0.3\linewidth]{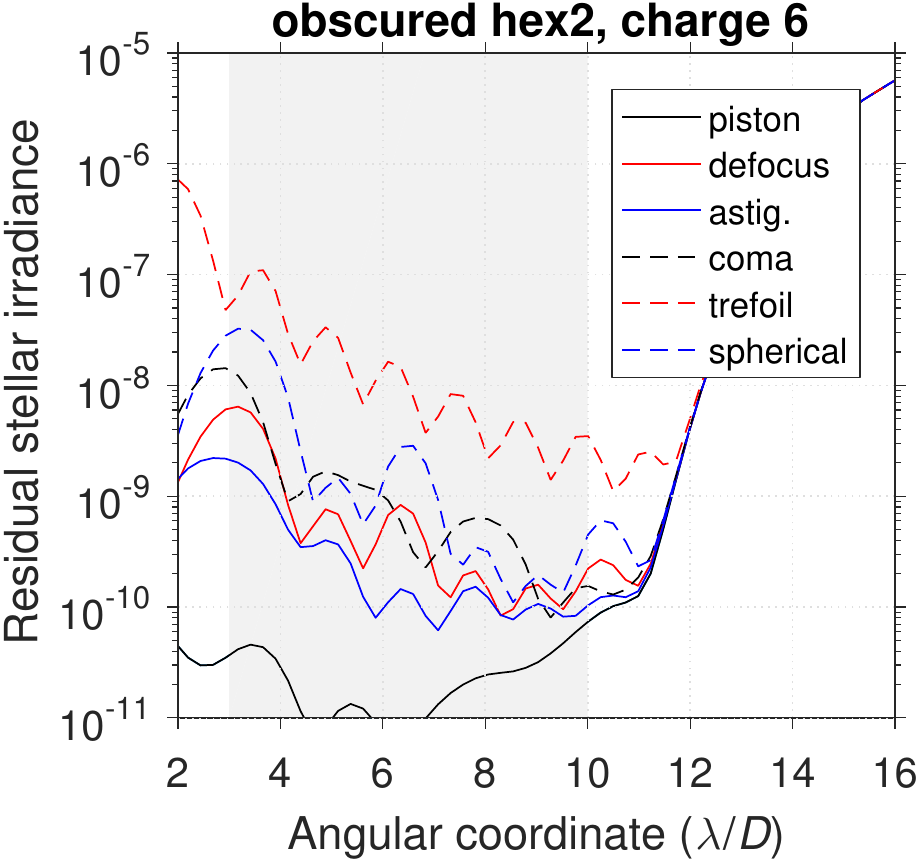}
    \includegraphics[height=0.3\linewidth]{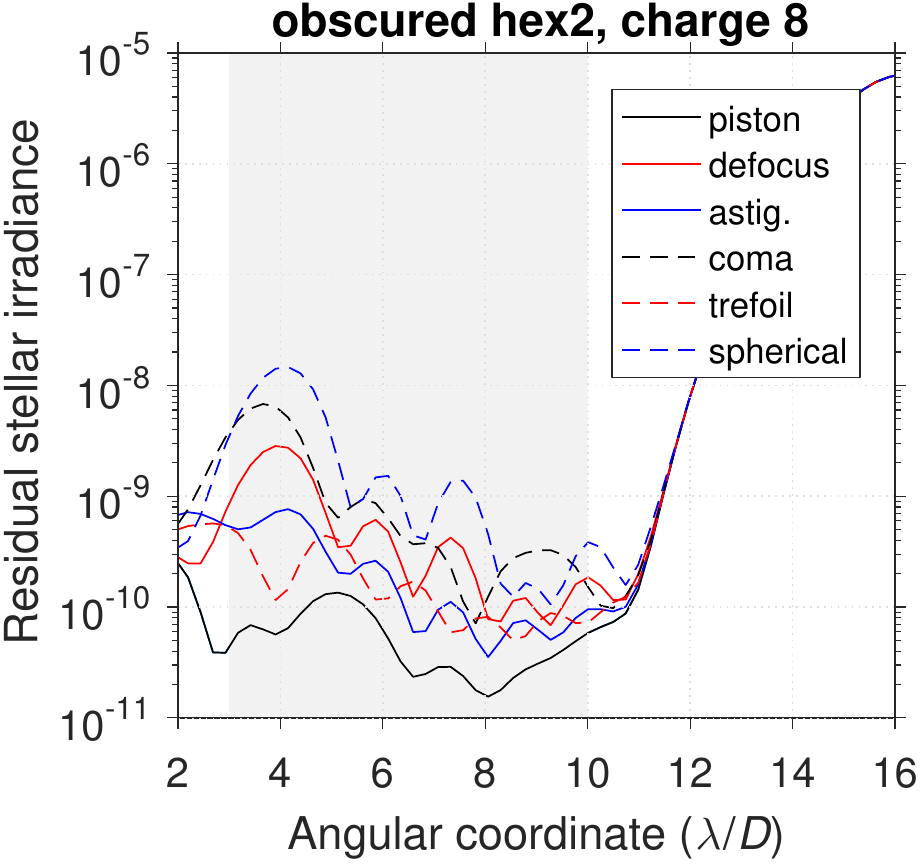}\\
    \includegraphics[height=0.3\linewidth]{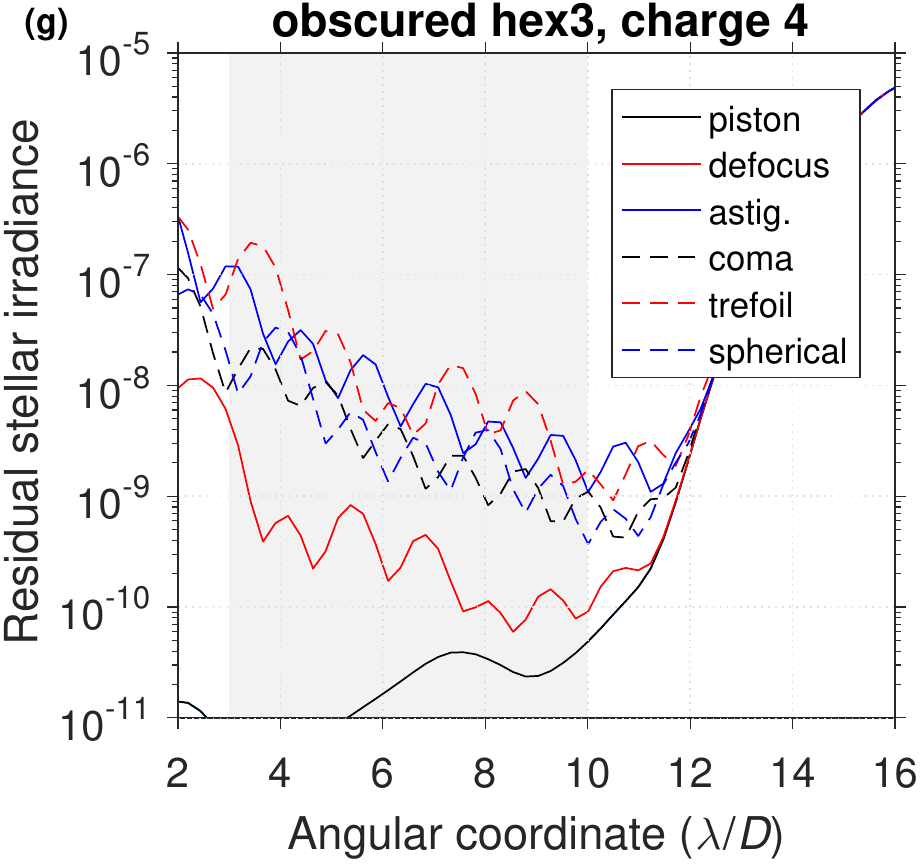}
    \includegraphics[height=0.3\linewidth]{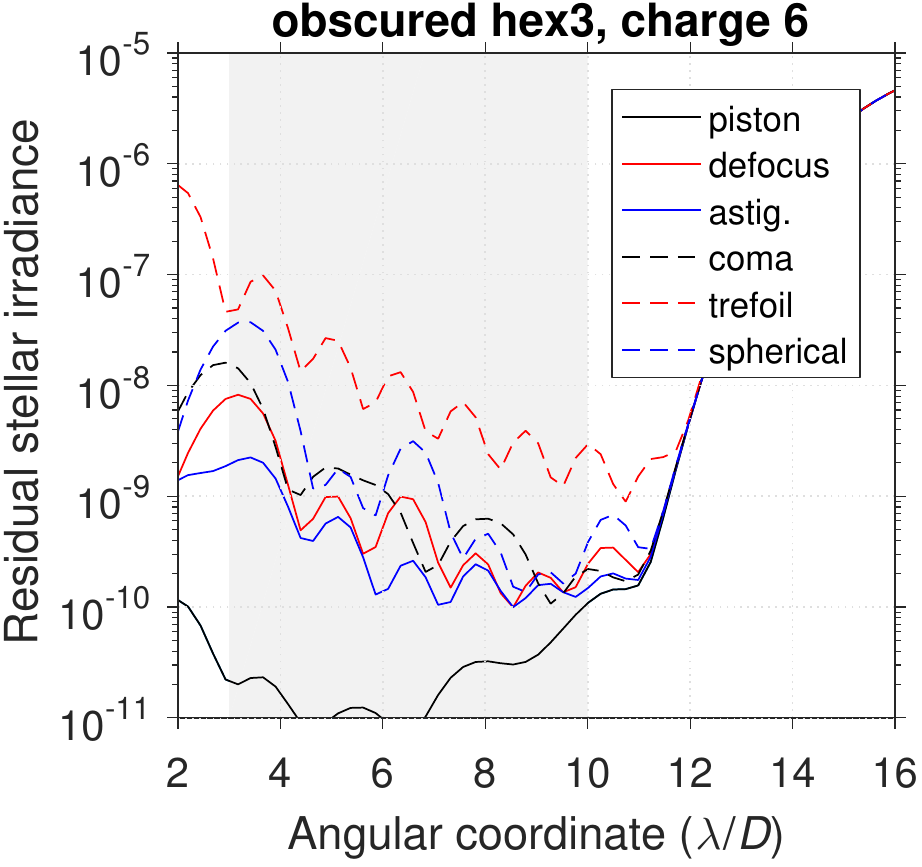}
    \includegraphics[height=0.3\linewidth]{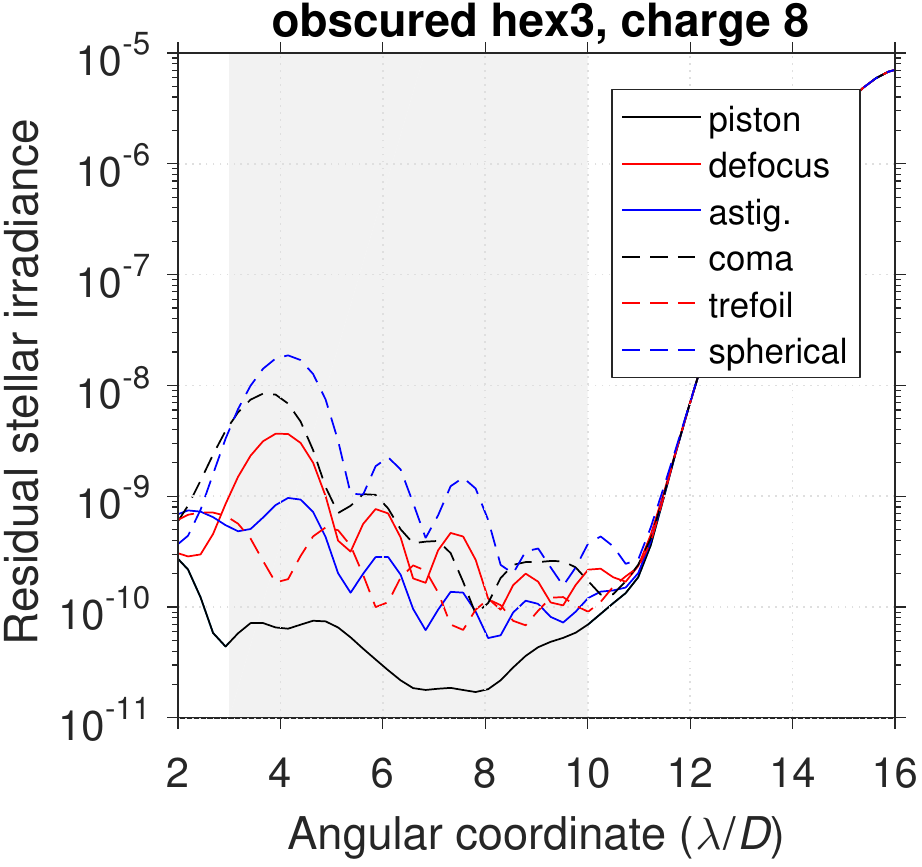}\\
    \includegraphics[height=0.3\linewidth]{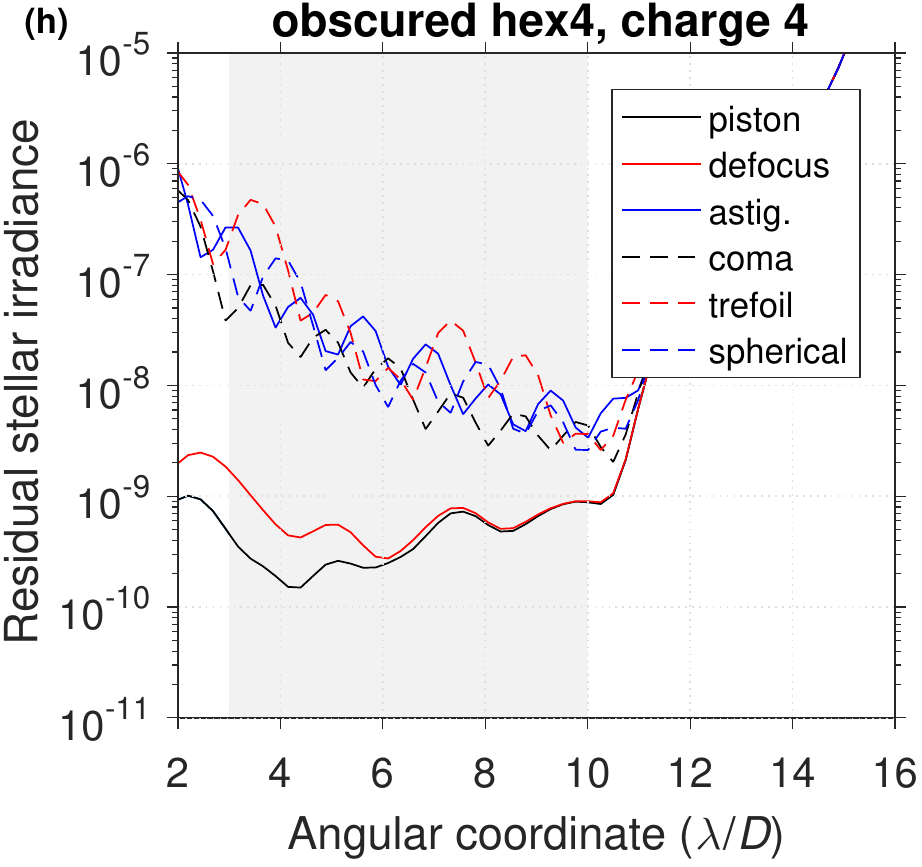}
    \includegraphics[height=0.3\linewidth]{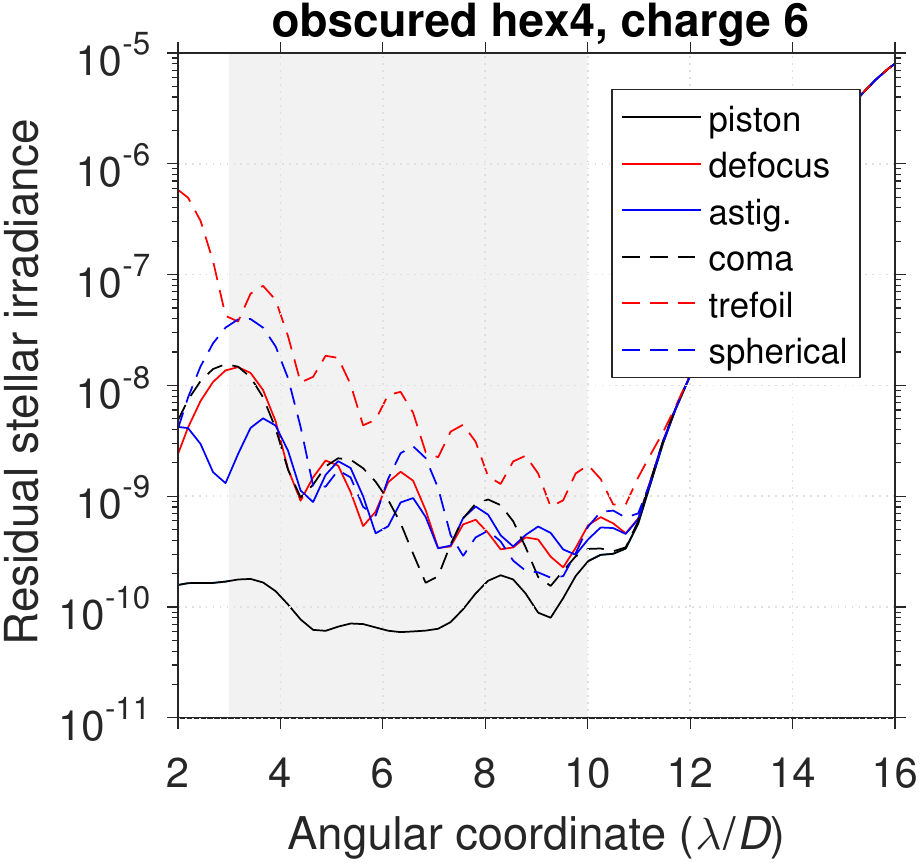}
    \includegraphics[height=0.3\linewidth]{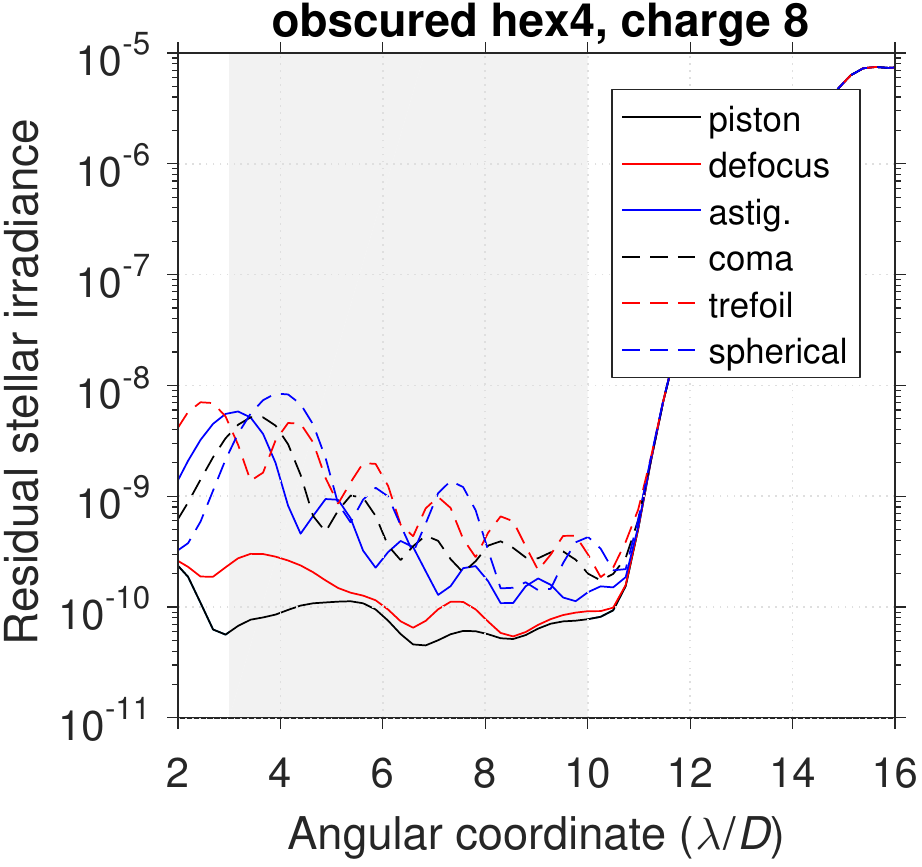}\\
    \caption{Residual starlight with $\lambda/1000$ rms wavefront error in the lowest order Zernike modes.}
    \label{fig:Zsens_obs2}
\end{figure}

\begin{figure}[p]
    \centering
    \includegraphics[height=0.3\linewidth]{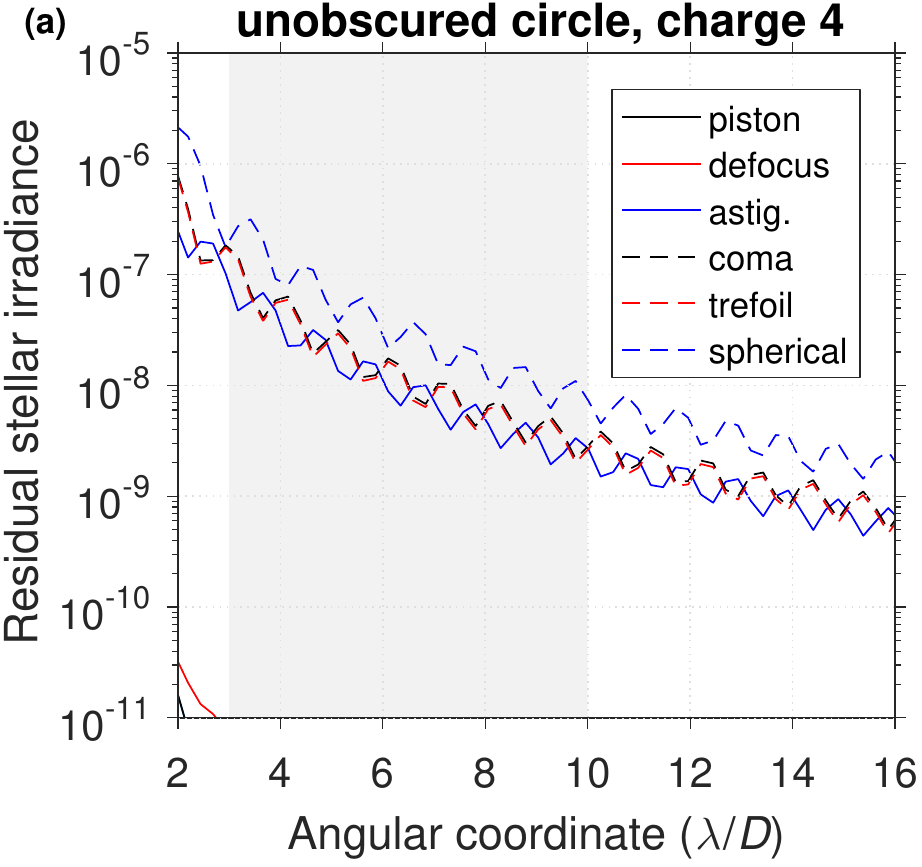}
    \includegraphics[height=0.3\linewidth]{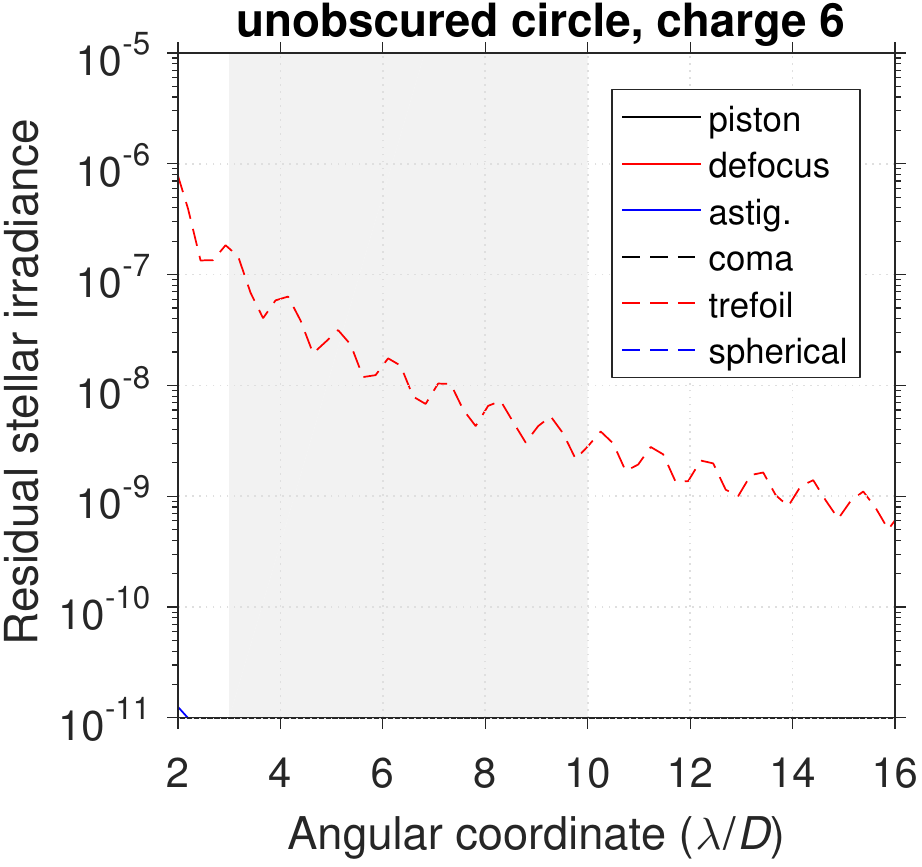}
    \includegraphics[height=0.3\linewidth]{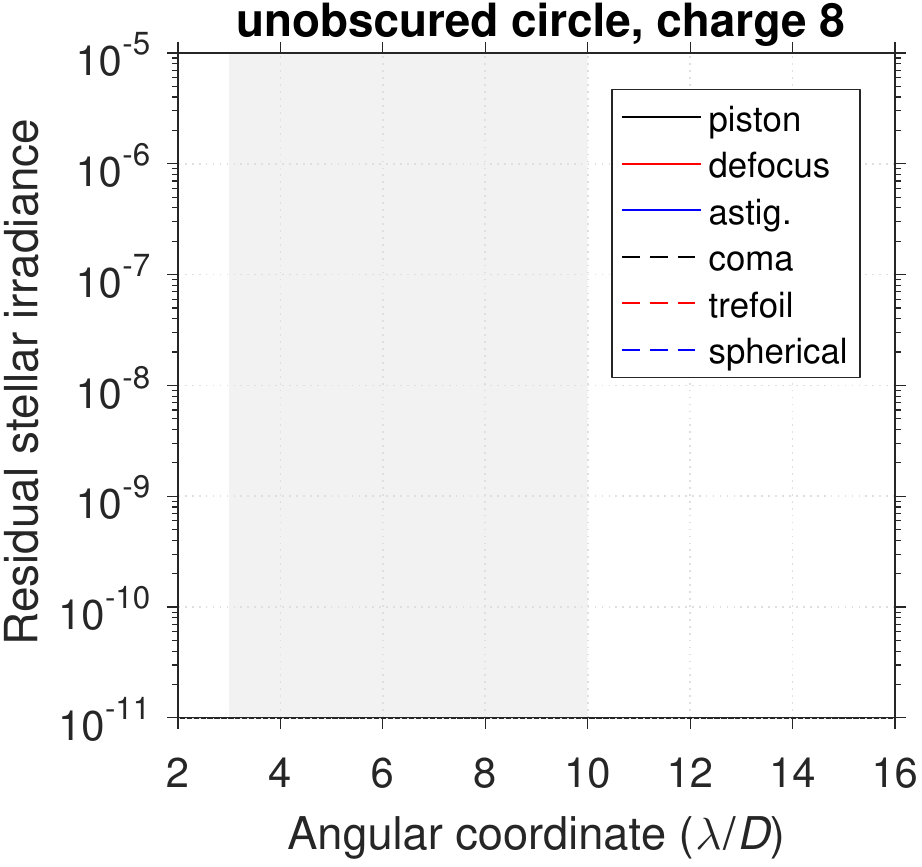}\\
    \includegraphics[height=0.3\linewidth]{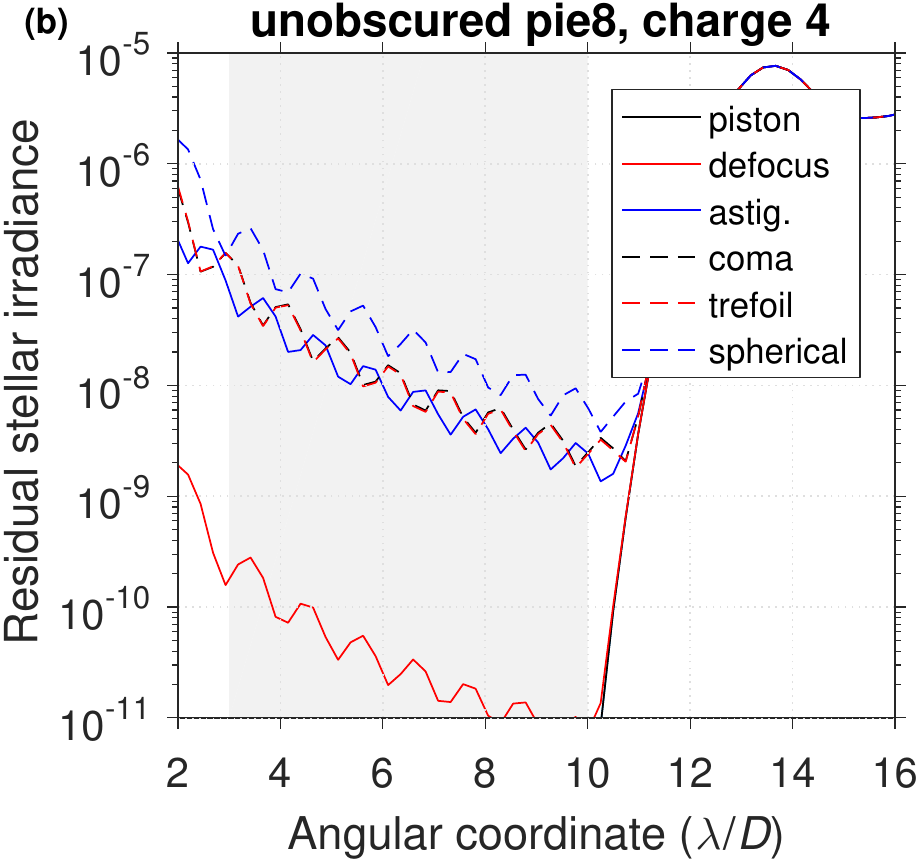}
    \includegraphics[height=0.3\linewidth]{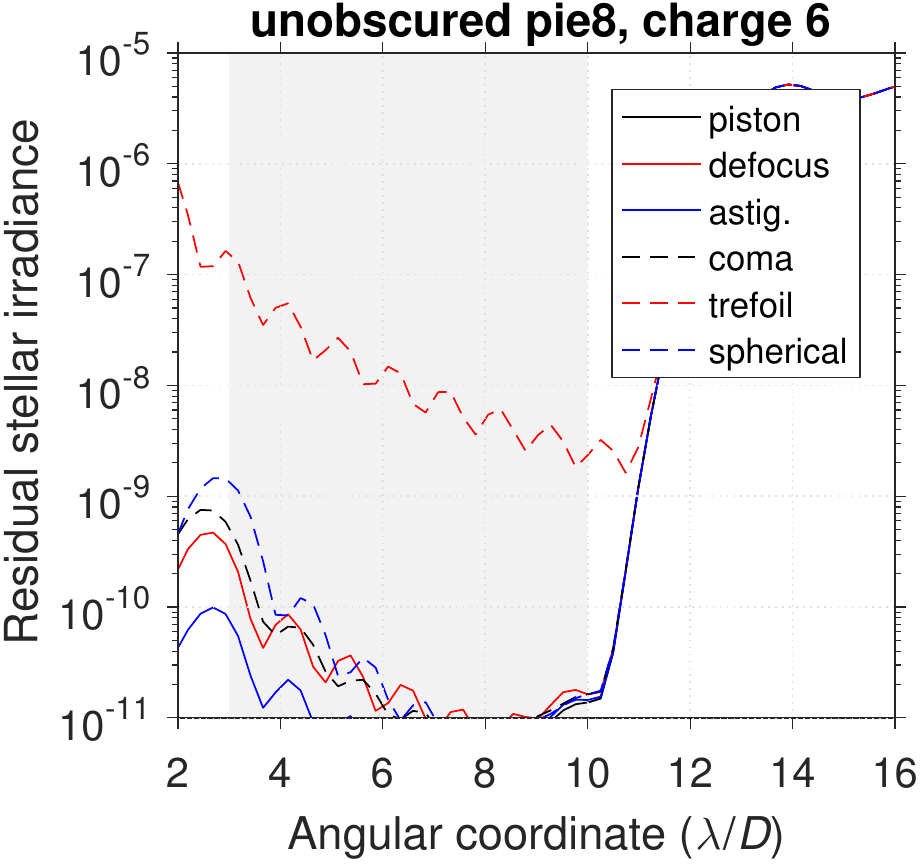}
    \includegraphics[height=0.3\linewidth]{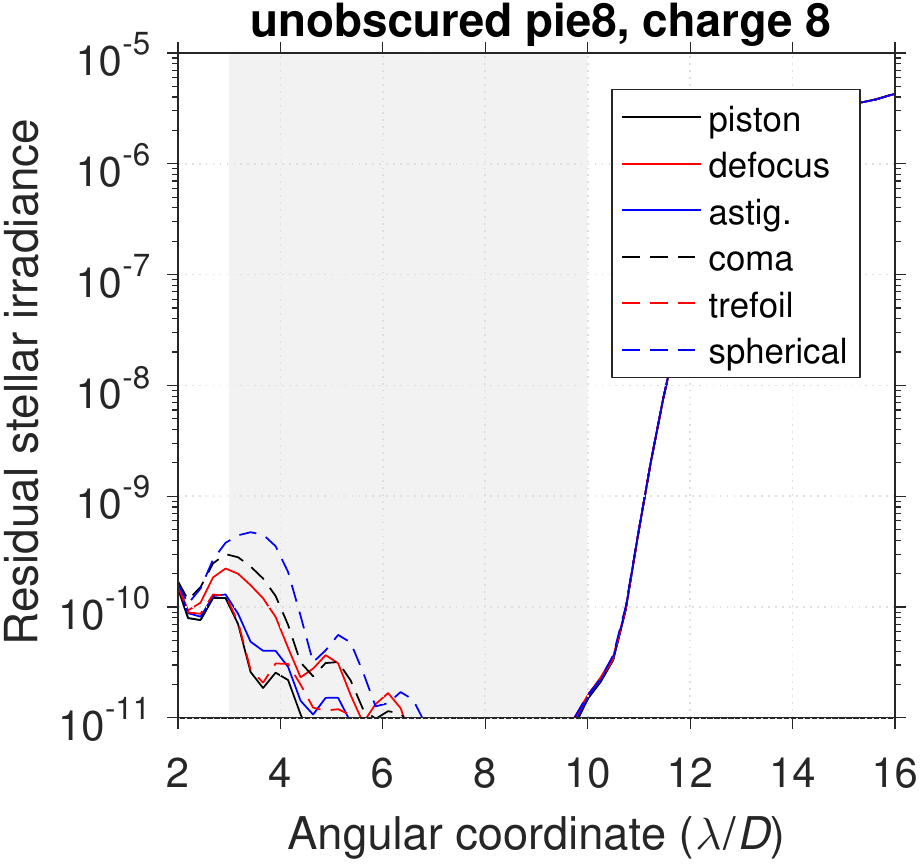}\\
    \includegraphics[height=0.3\linewidth]{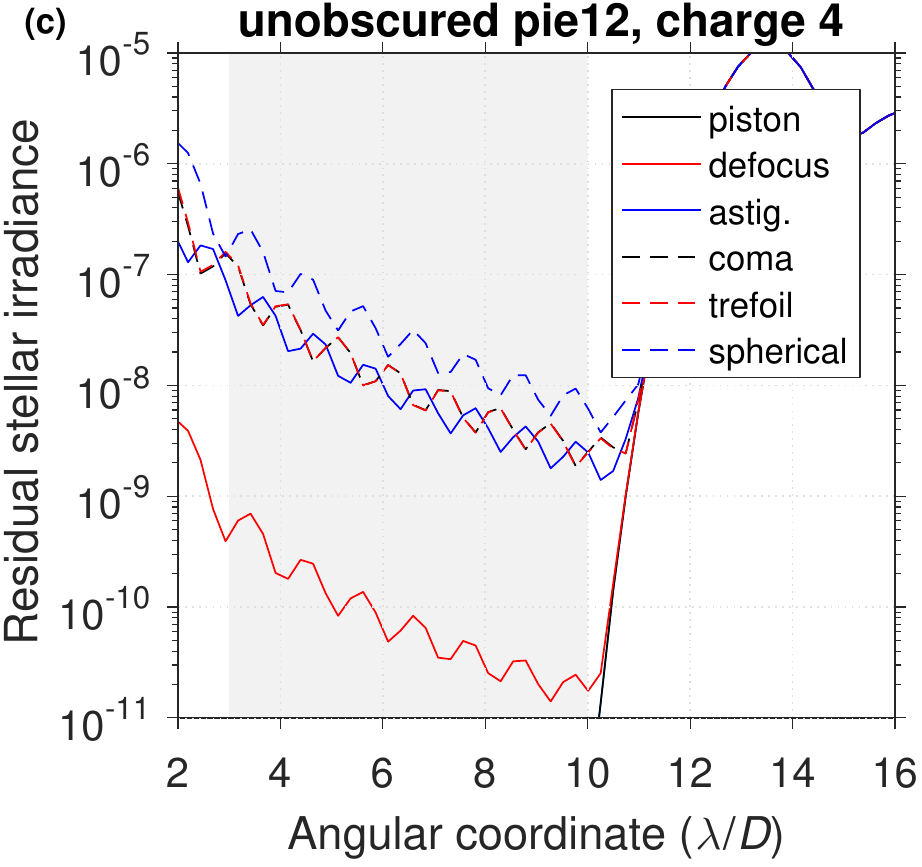}
    \includegraphics[height=0.3\linewidth]{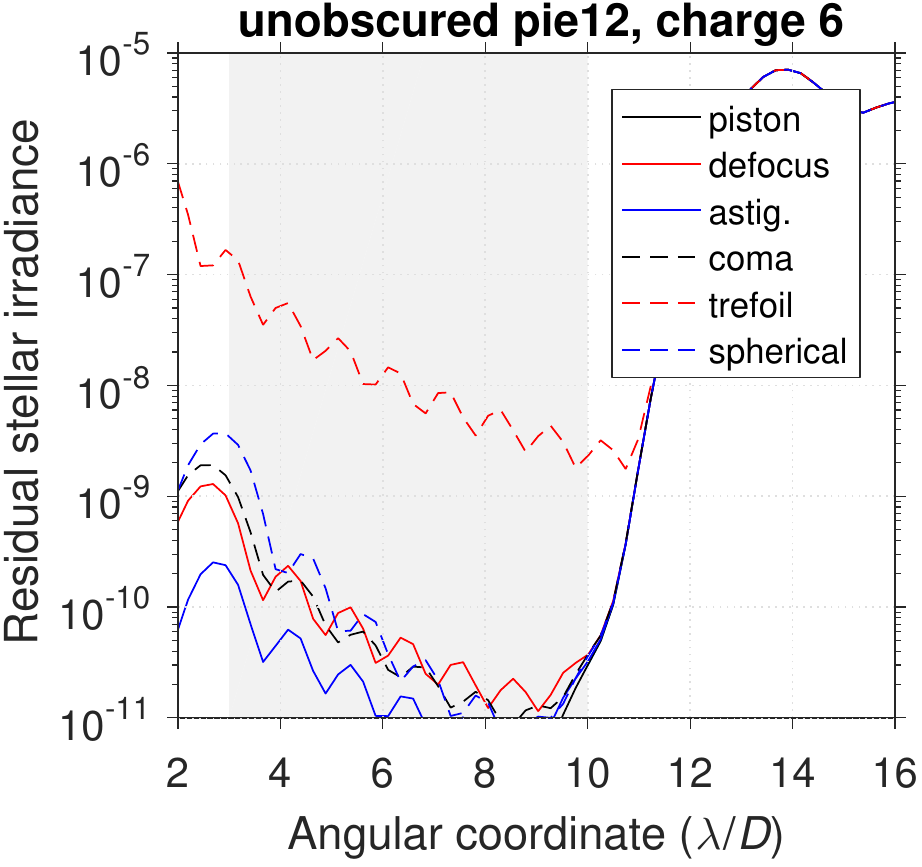}
    \includegraphics[height=0.3\linewidth]{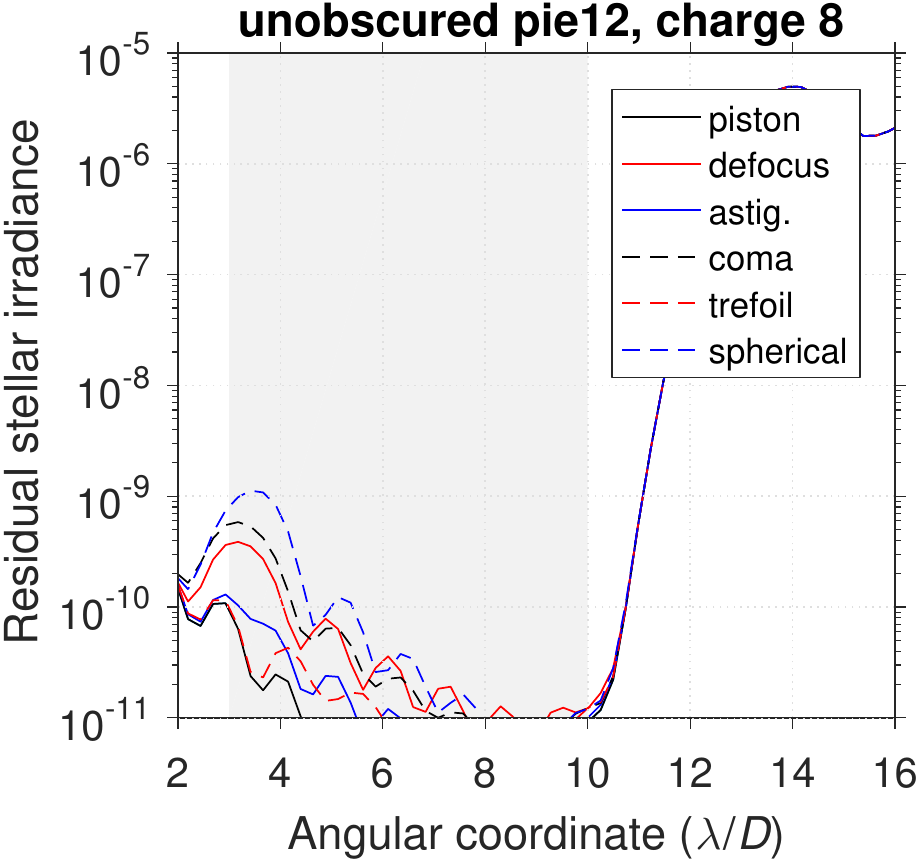}\\
    \includegraphics[height=0.3\linewidth]{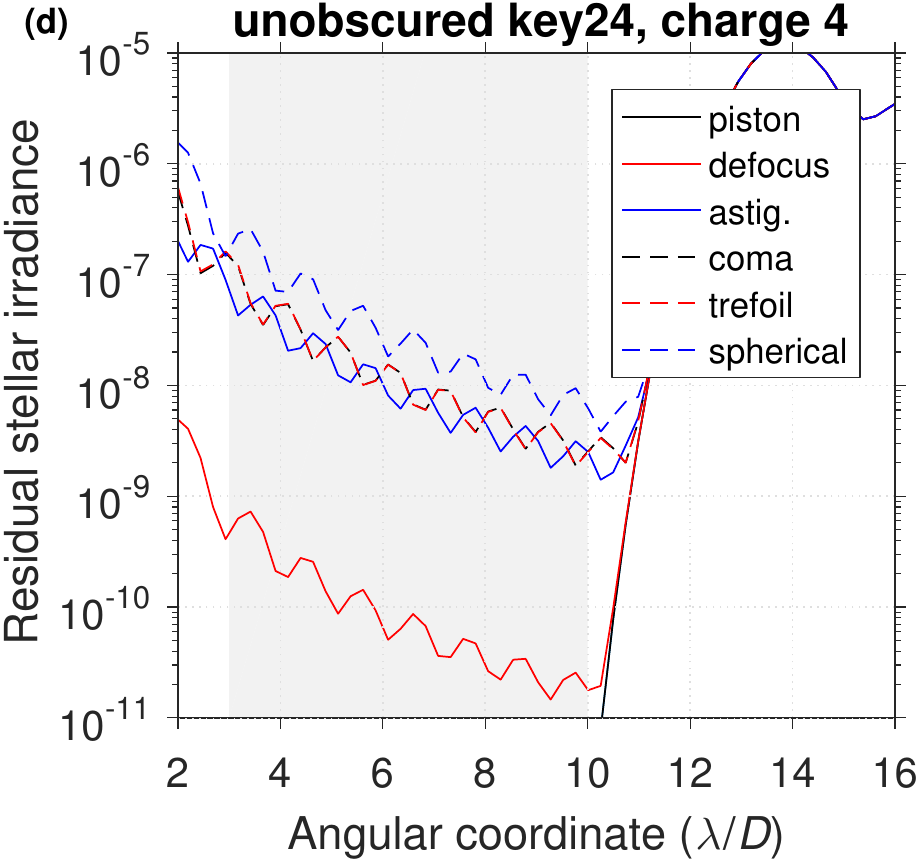}
    \includegraphics[height=0.3\linewidth]{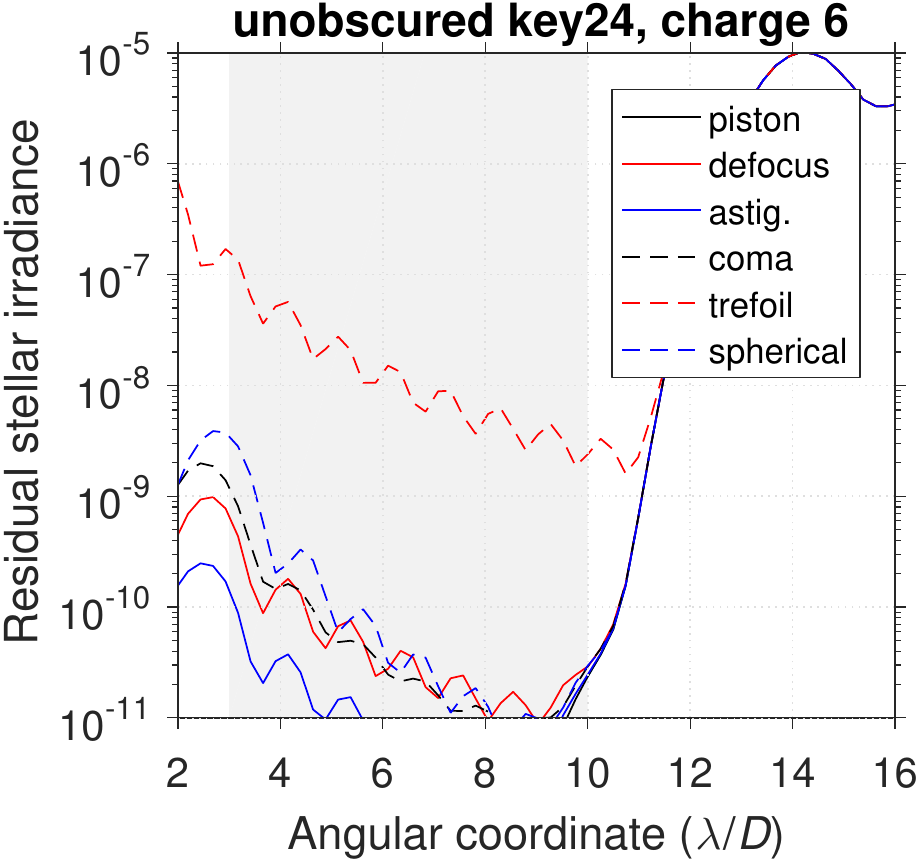}
    \includegraphics[height=0.3\linewidth]{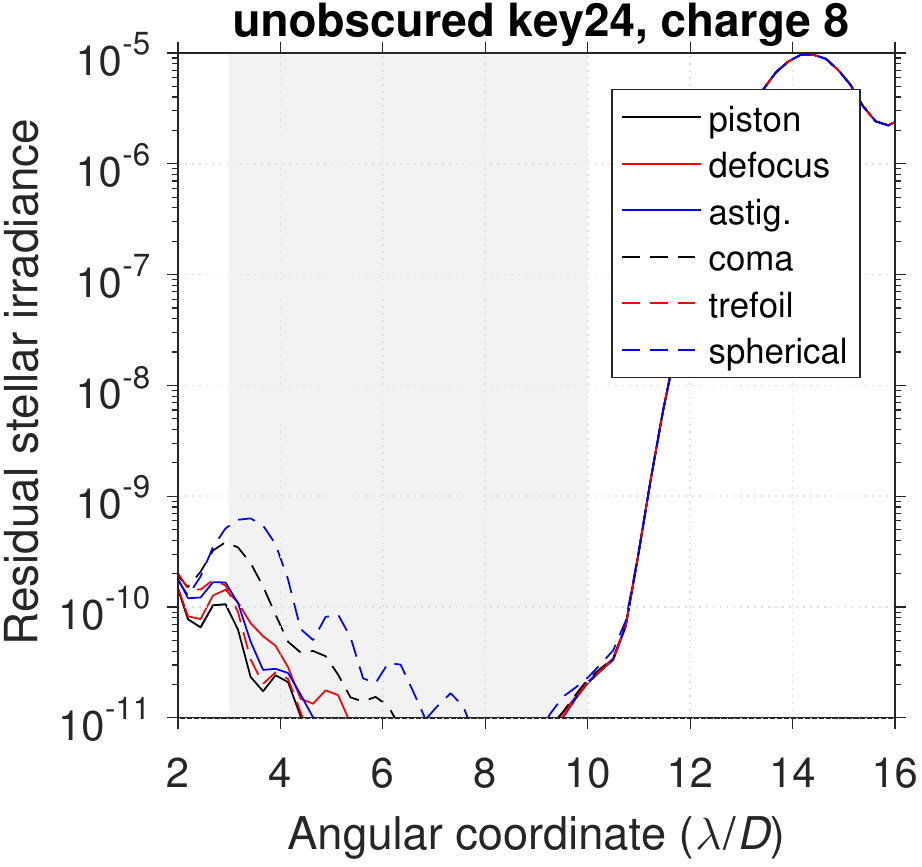}\\
    \caption{Residual starlight with $\lambda/1000$ rms wavefront error in the lowest order Zernike modes.}
    \label{fig:Zsens_unobs1}
\end{figure}

\begin{figure}[p]
    \centering
    \includegraphics[height=0.3\linewidth]{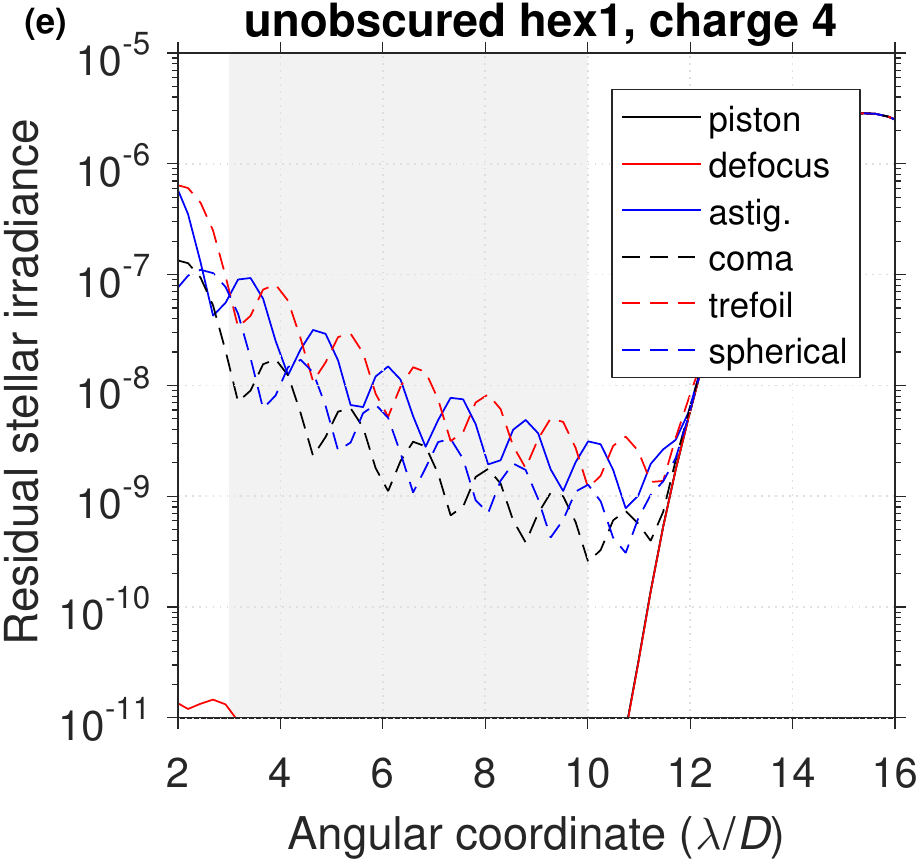}
    \includegraphics[height=0.3\linewidth]{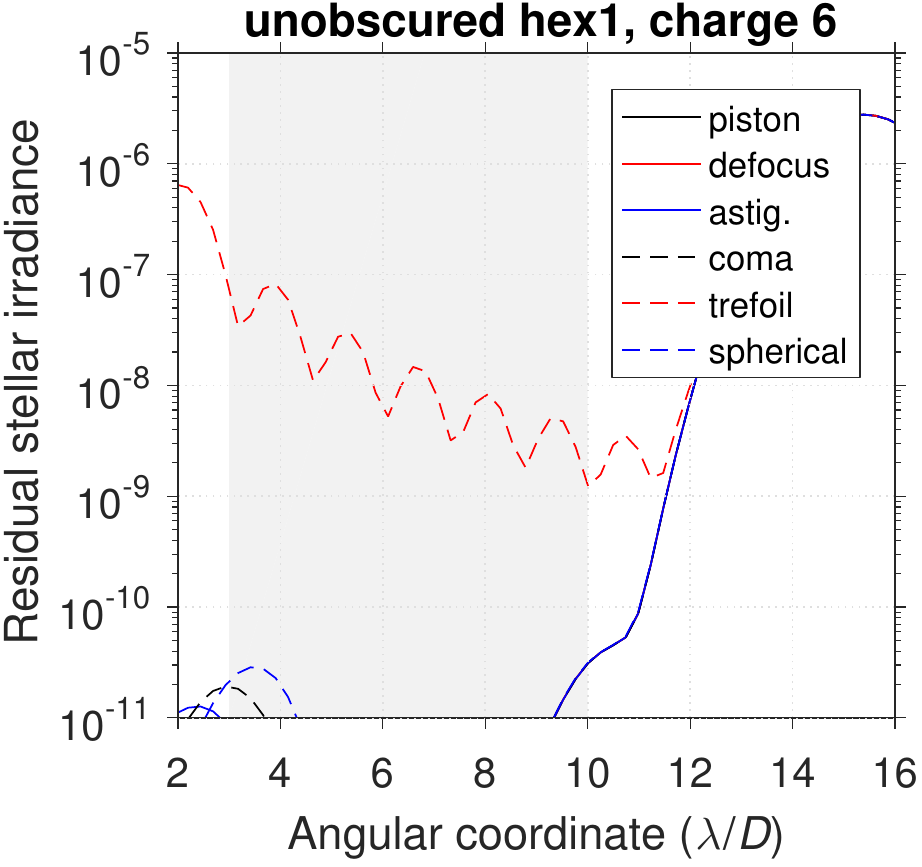}
    \includegraphics[height=0.3\linewidth]{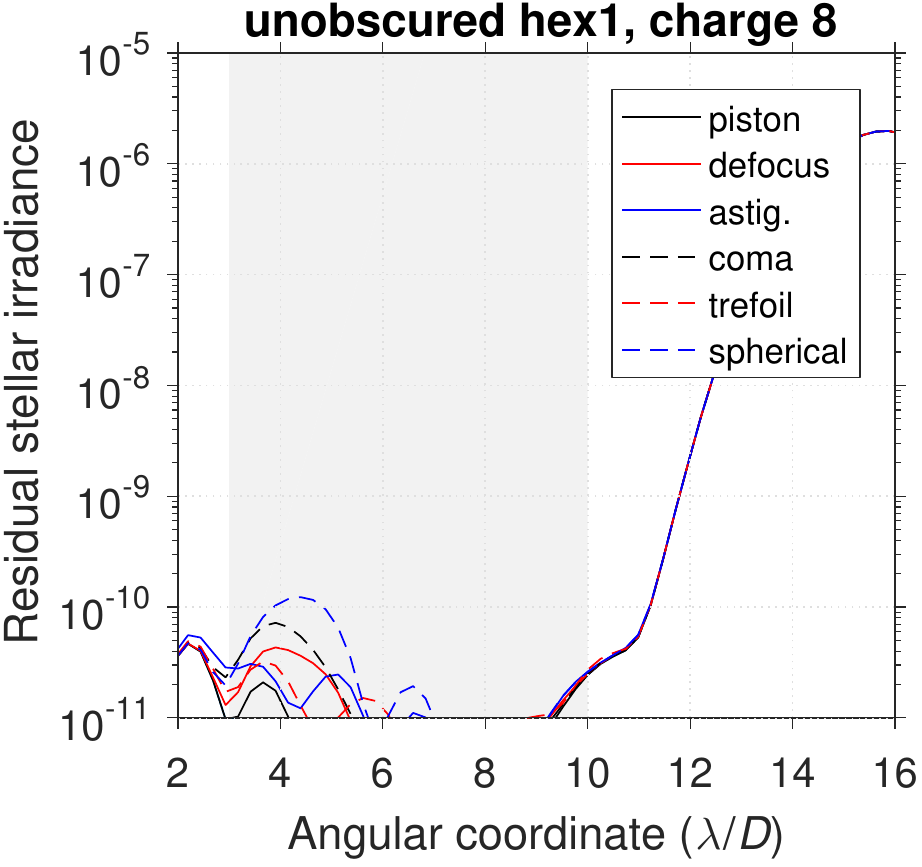}\\
    \includegraphics[height=0.3\linewidth]{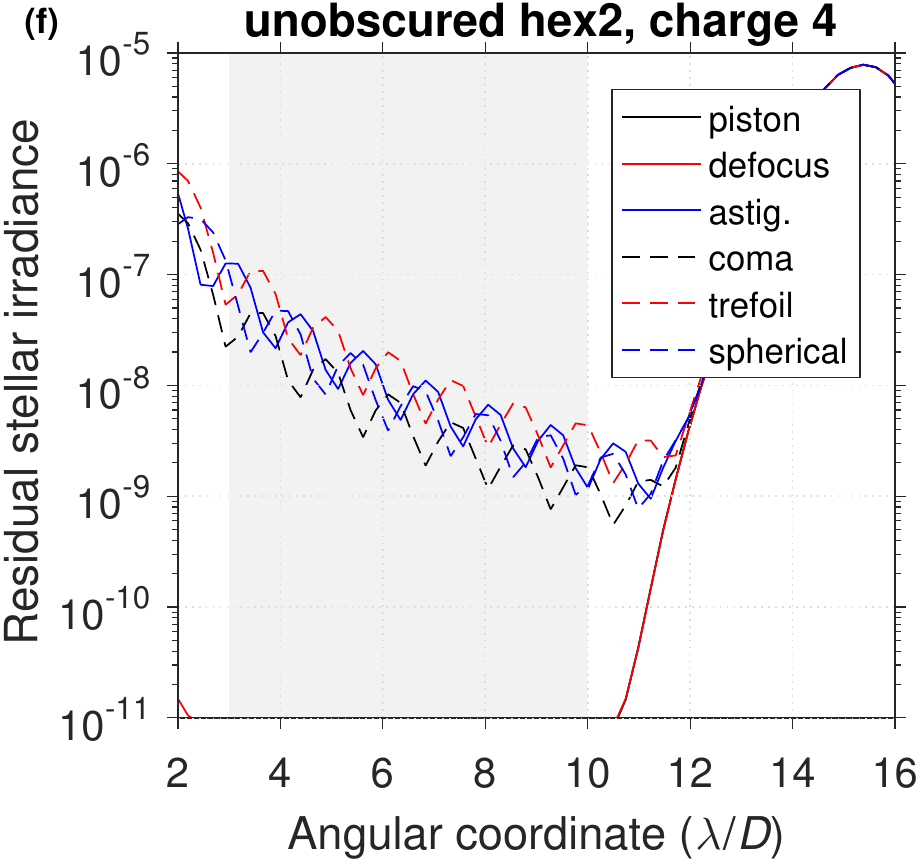}
    \includegraphics[height=0.3\linewidth]{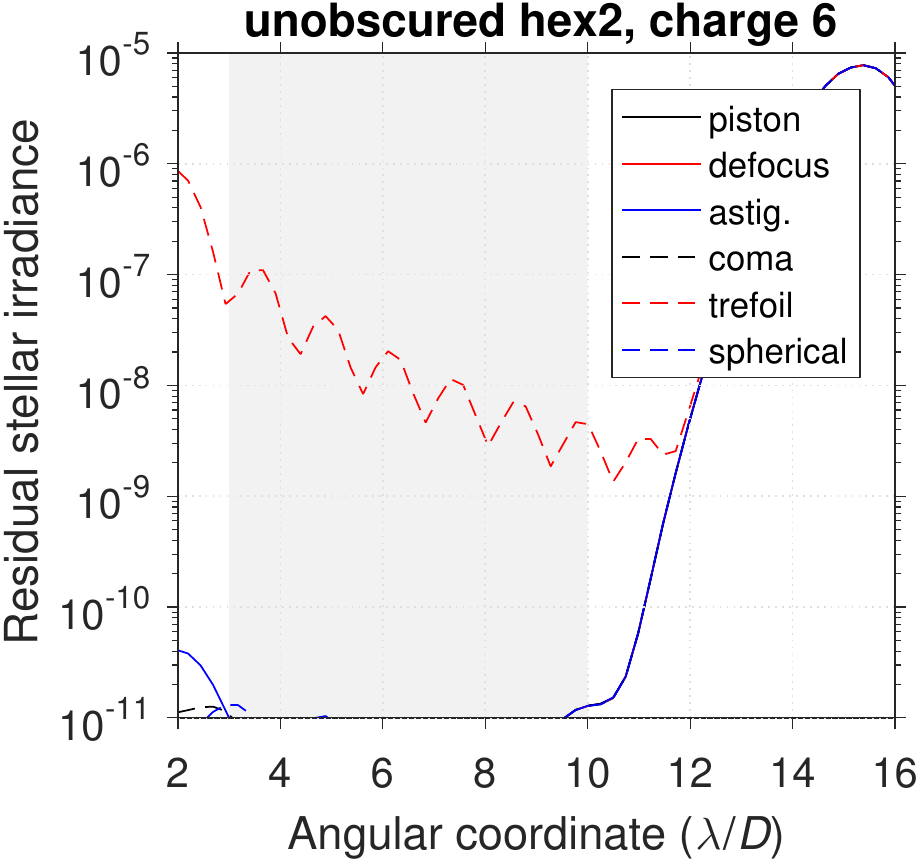}
    \includegraphics[height=0.3\linewidth]{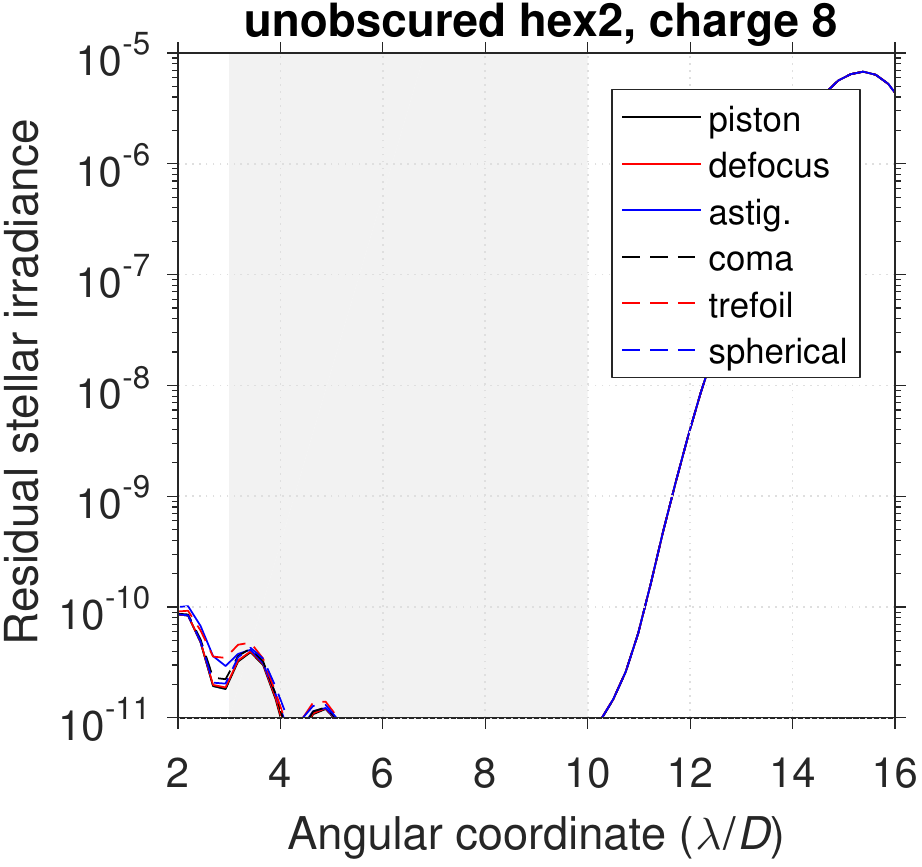}\\
    \includegraphics[height=0.3\linewidth]{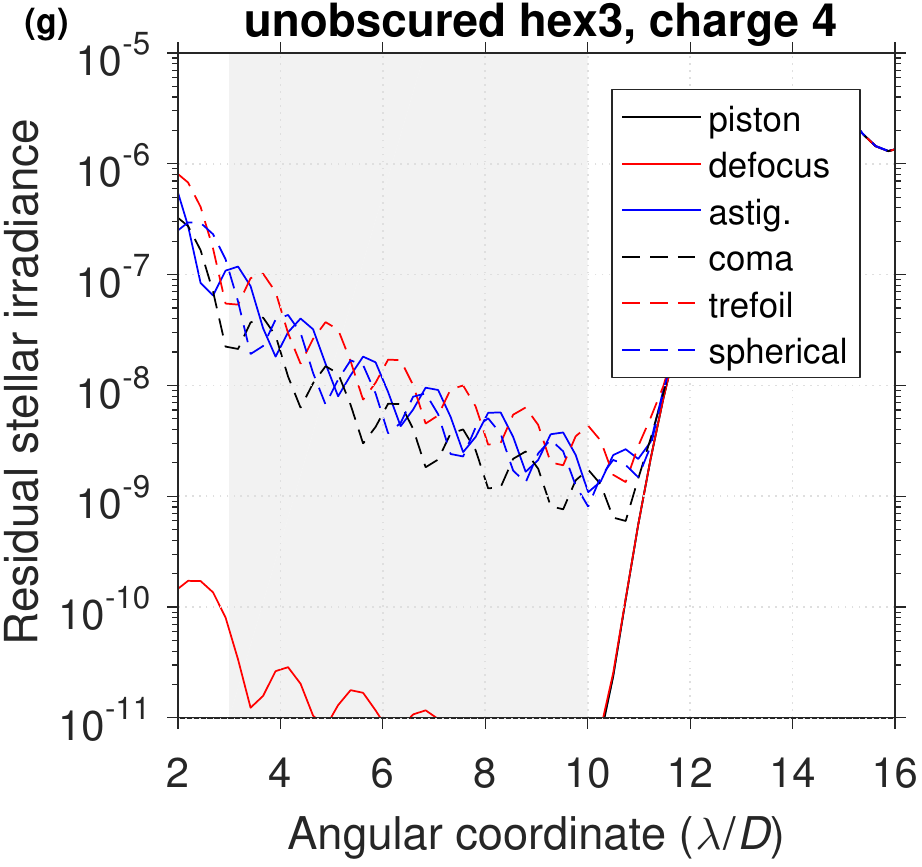}
    \includegraphics[height=0.3\linewidth]{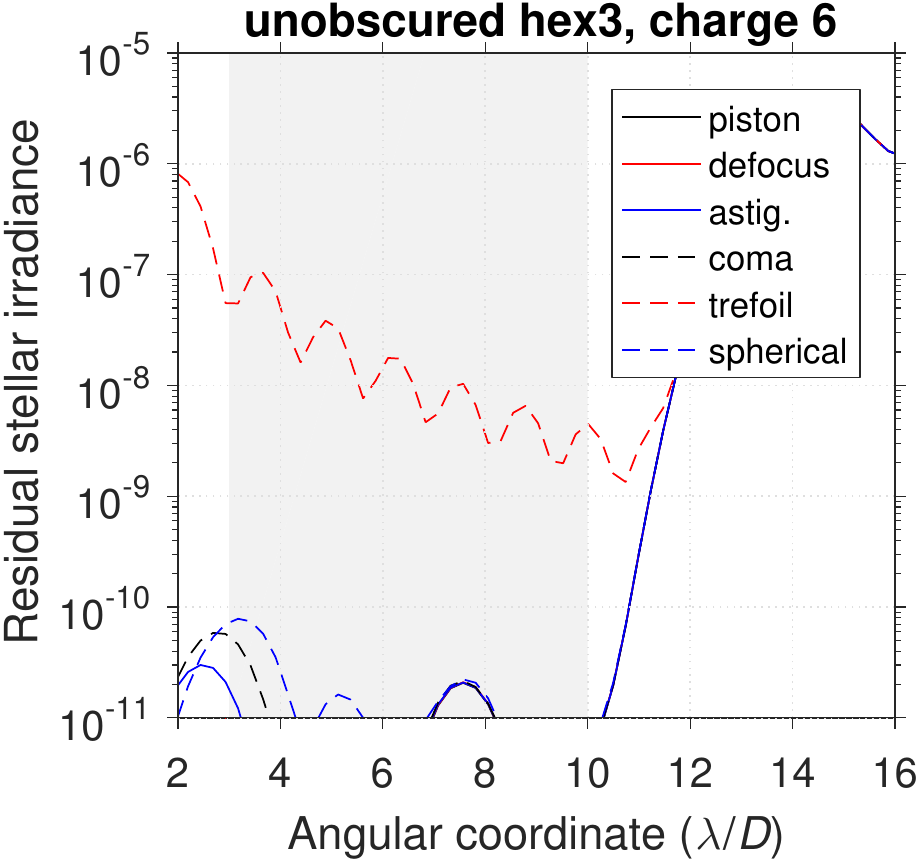}
    \includegraphics[height=0.3\linewidth]{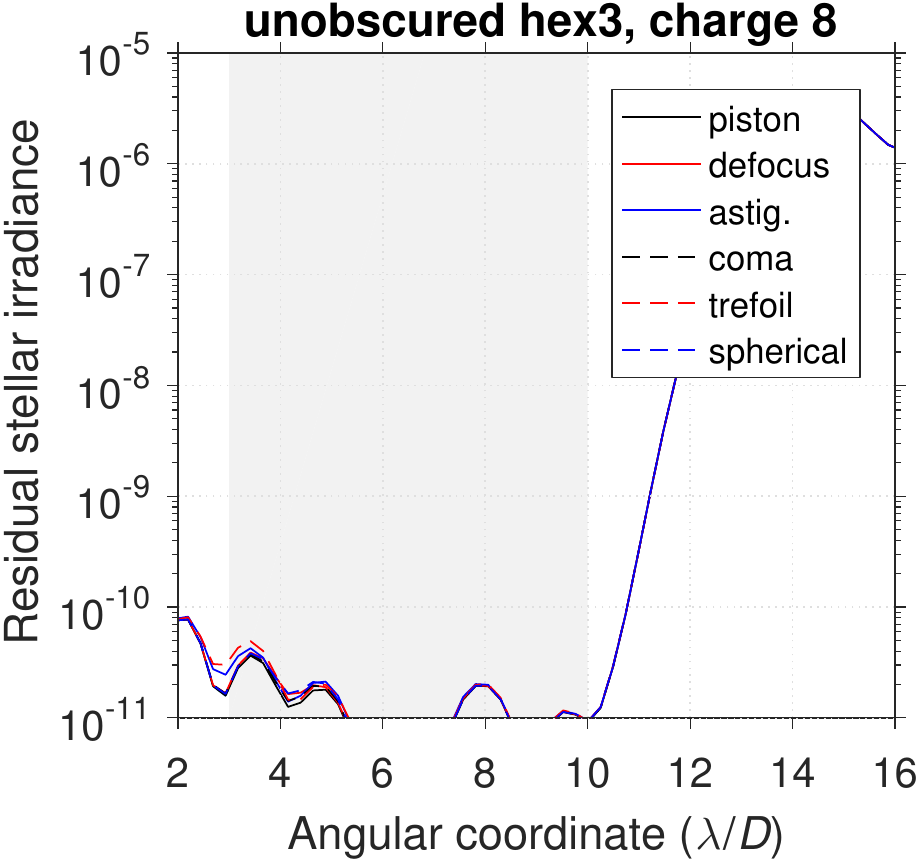}\\
    \includegraphics[height=0.3\linewidth]{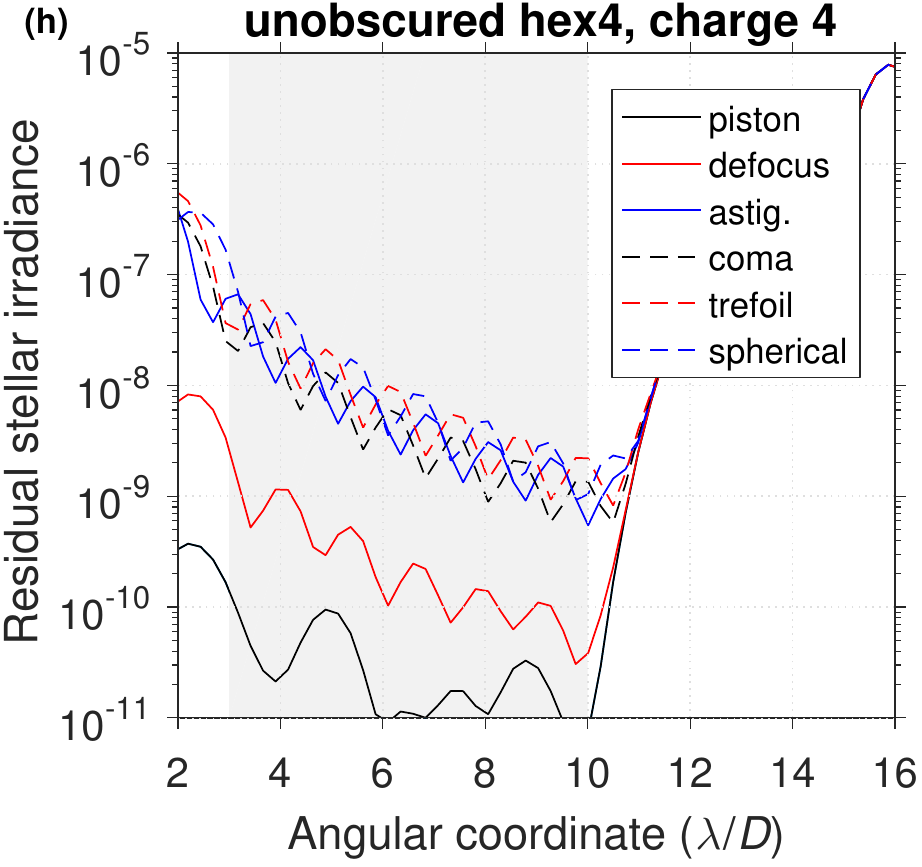}
    \includegraphics[height=0.3\linewidth]{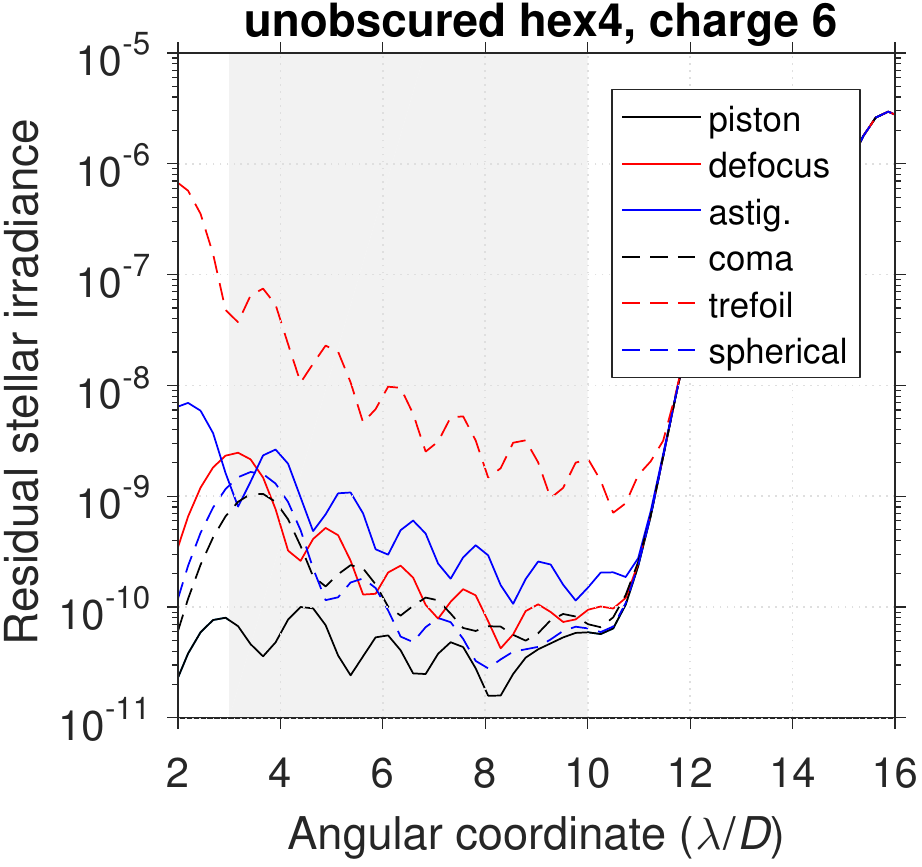}
    \includegraphics[height=0.3\linewidth]{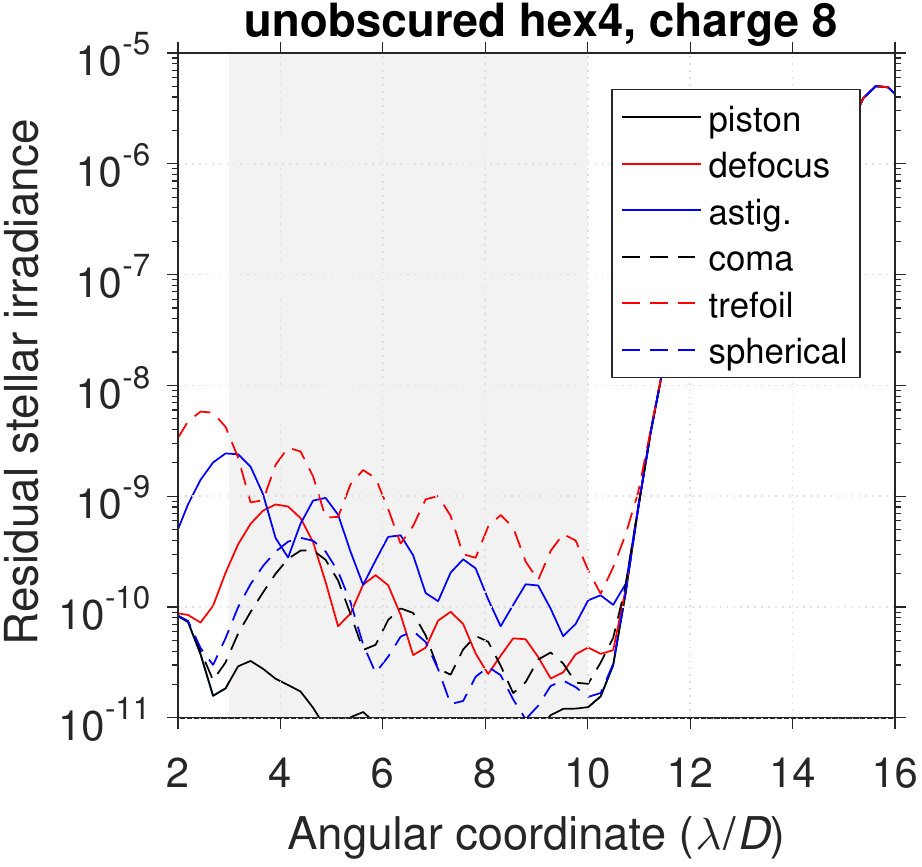}\\
    \caption{Residual starlight with $\lambda/1000$ rms wavefront error in the lowest order Zernike modes.}
    \label{fig:Zsens_unobs2}
\end{figure}

The sensitivity of each numerically optimized coronagraph design to low-order Zernike aberrations is shown in Figs. \ref{fig:Zsens_obs1}-\ref{fig:Zsens_unobs2}. The obscured aperture cases are in Figs. \ref{fig:Zsens_obs1}-\ref{fig:Zsens_obs2} and the unobscured apertures in Figs. \ref{fig:Zsens_unobs1}-\ref{fig:Zsens_unobs2}. The case of a circular pupil (Fig. \ref{fig:Zsens_unobs1}a) confirms the theoretical description above. This trend is also generally followed by the unobscured apertures. However, the obscured pupils tend to be more sensitive to aberrations and deviate from the conventional understanding of vortex coronagraph sensitivities. 

\subsubsection{Sensitivity to Lyot stop alignment}
\begin{figure}[p]
    \begin{center}\textbf{Apertures with central obscurations}\end{center}
    \includegraphics[height=0.31\linewidth]{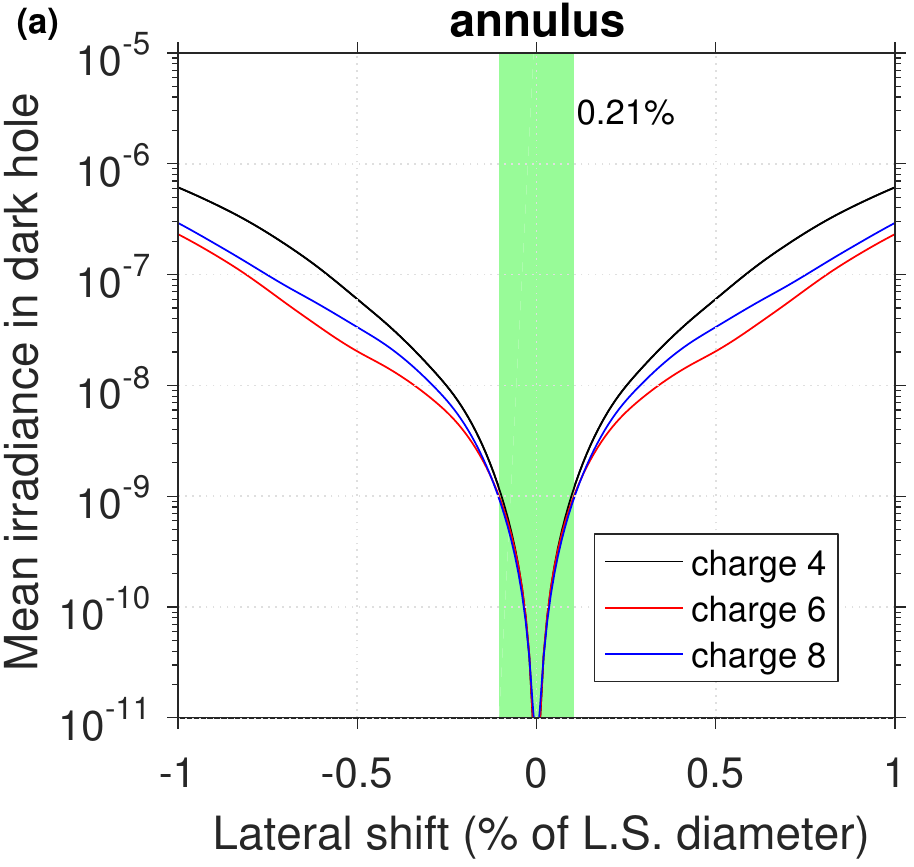}
    \includegraphics[height=0.31\linewidth]{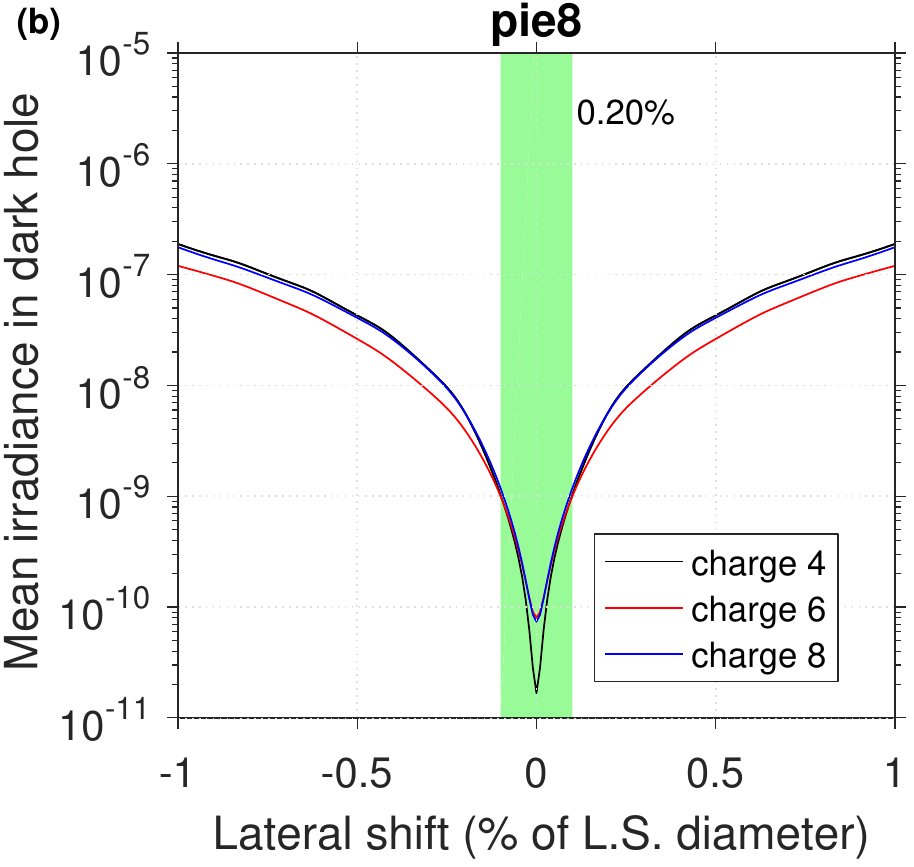}
    \includegraphics[height=0.31\linewidth]{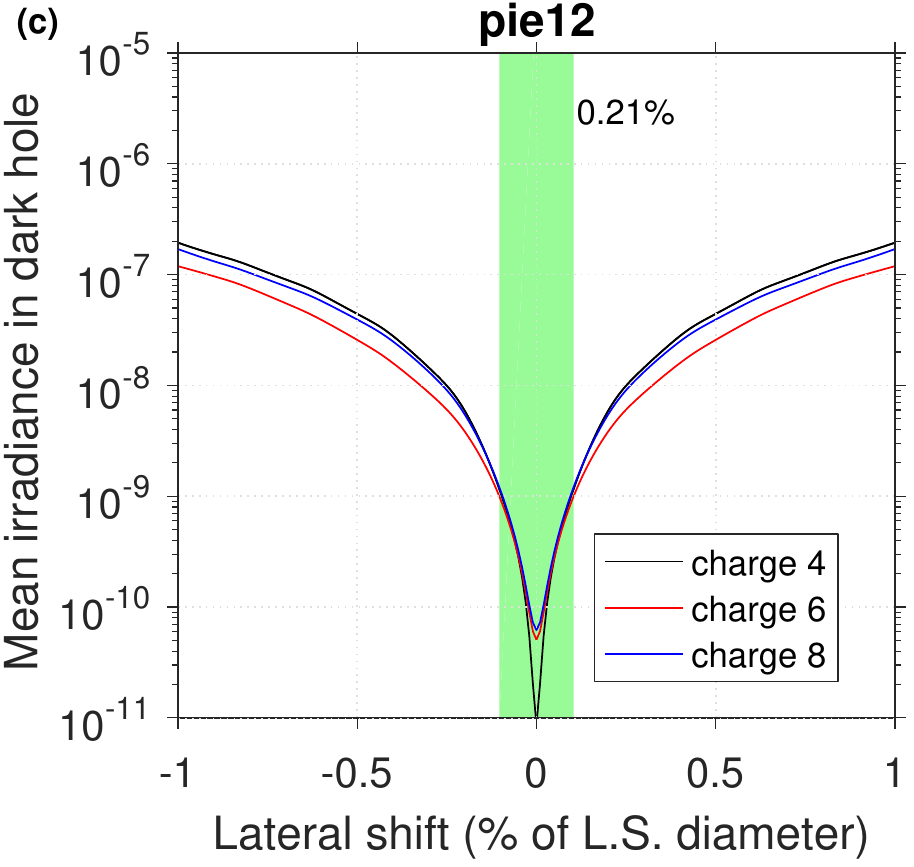}\\
    \includegraphics[height=0.31\linewidth]{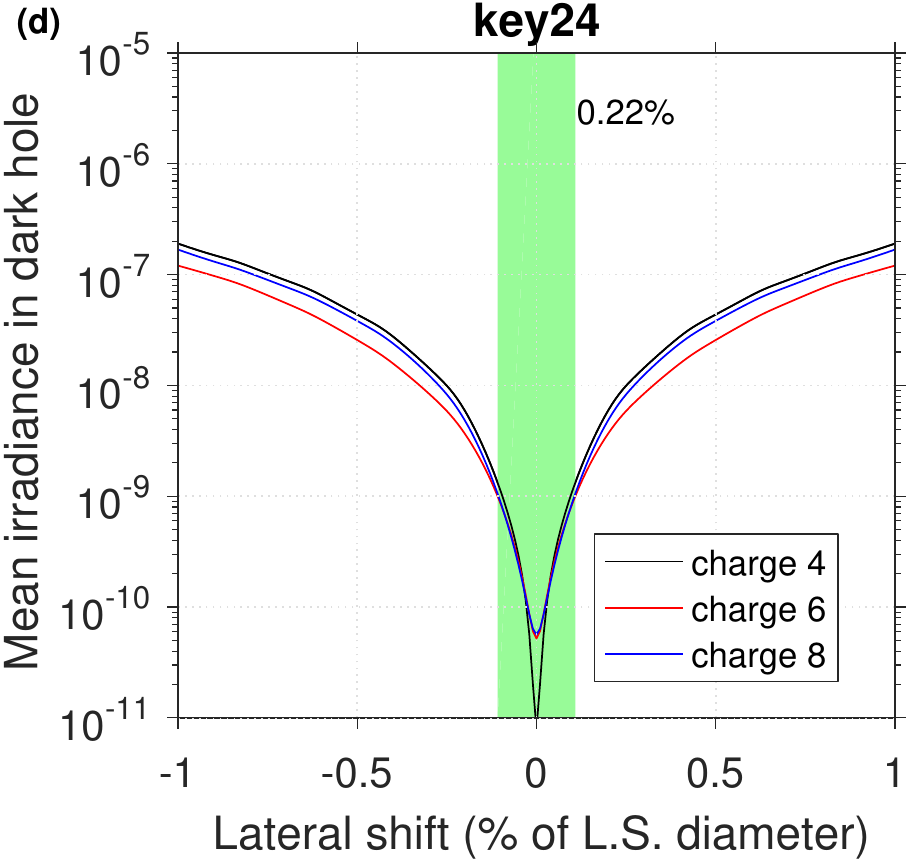}
    \includegraphics[height=0.31\linewidth]{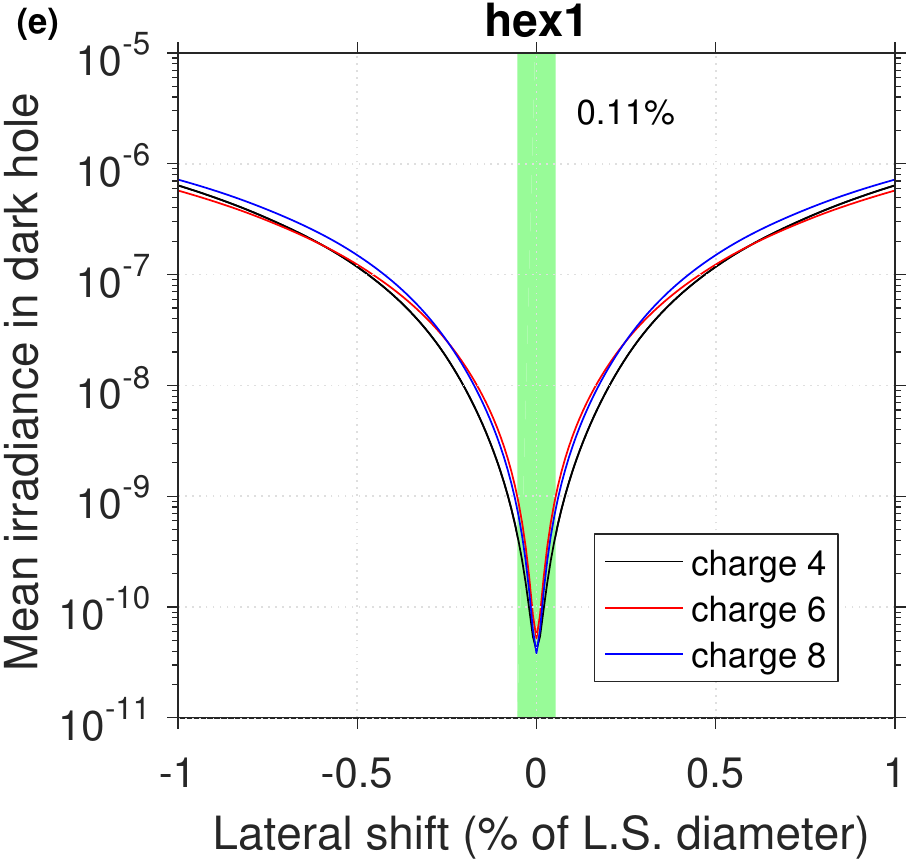}
    \includegraphics[height=0.31\linewidth]{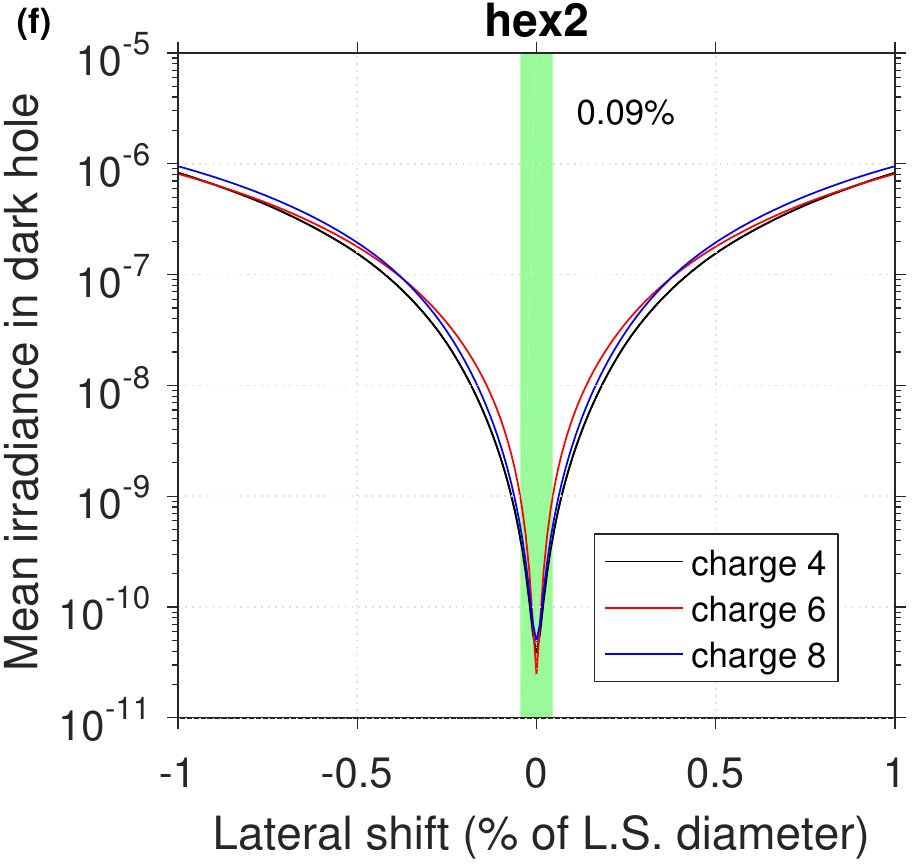}\\
    \includegraphics[height=0.31\linewidth]{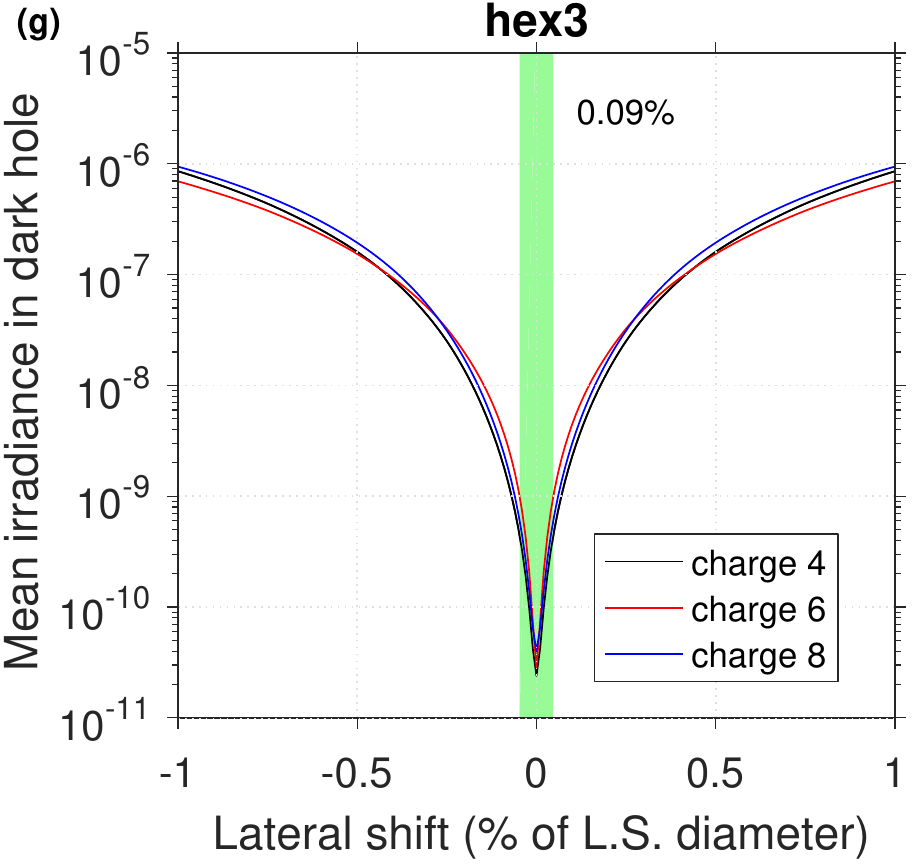}
    \includegraphics[height=0.31\linewidth]{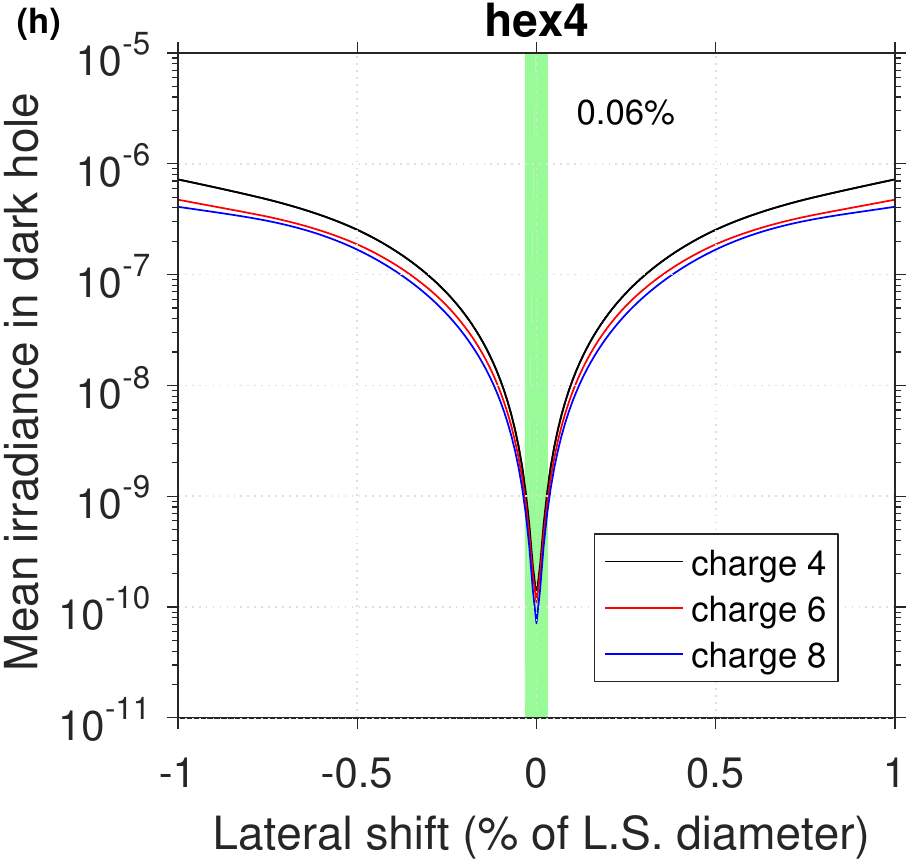}\\
    \caption{Mean irradiance in the dark hole (normalized to the peak of the telescope PSF) as a function of Lyot stop position. The green region shows the range over which $10^{-9}$ is maintained and the annotated number is the region's width (i.e. roughly twice the alignment tolerance).}
    \label{fig:LS_shift_sens_obs}
\end{figure}

\begin{figure}[p]
    \begin{center}\textbf{Apertures without central obscurations}\end{center}
    \includegraphics[height=0.31\linewidth]{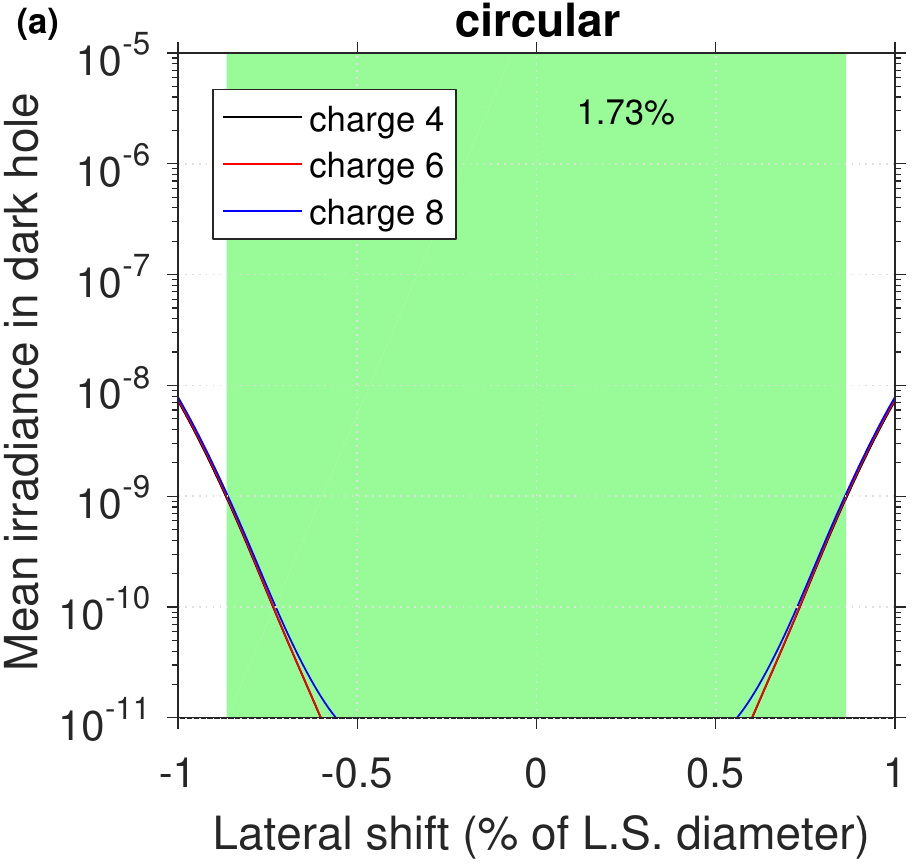}
    \includegraphics[height=0.31\linewidth]{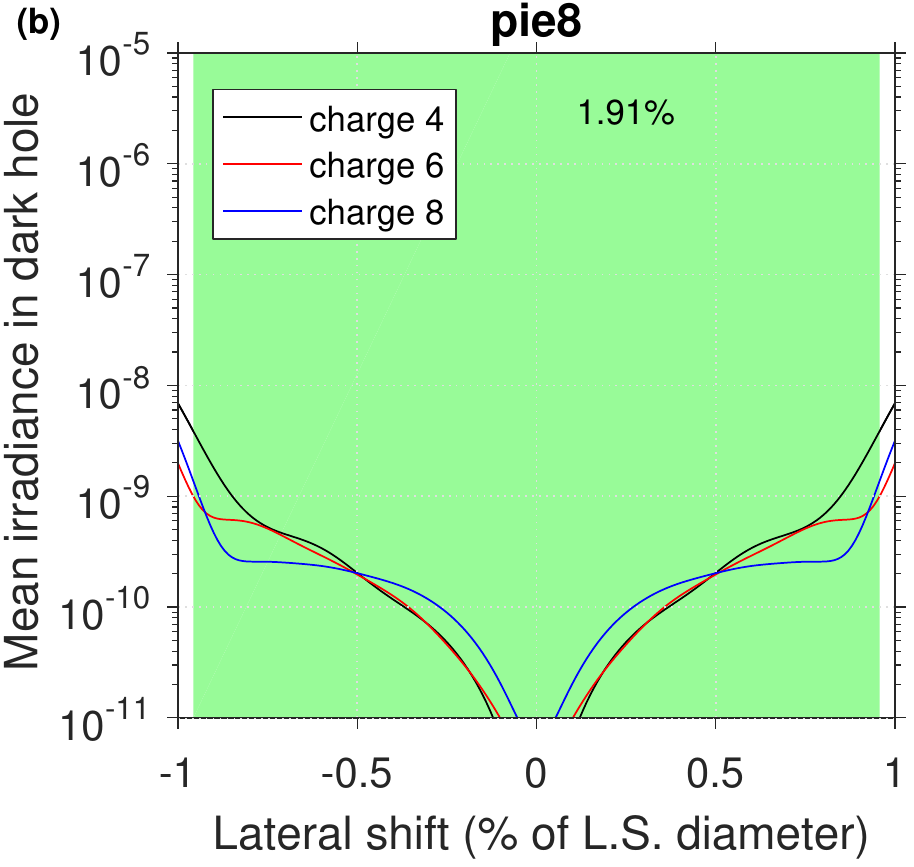}
    \includegraphics[height=0.31\linewidth]{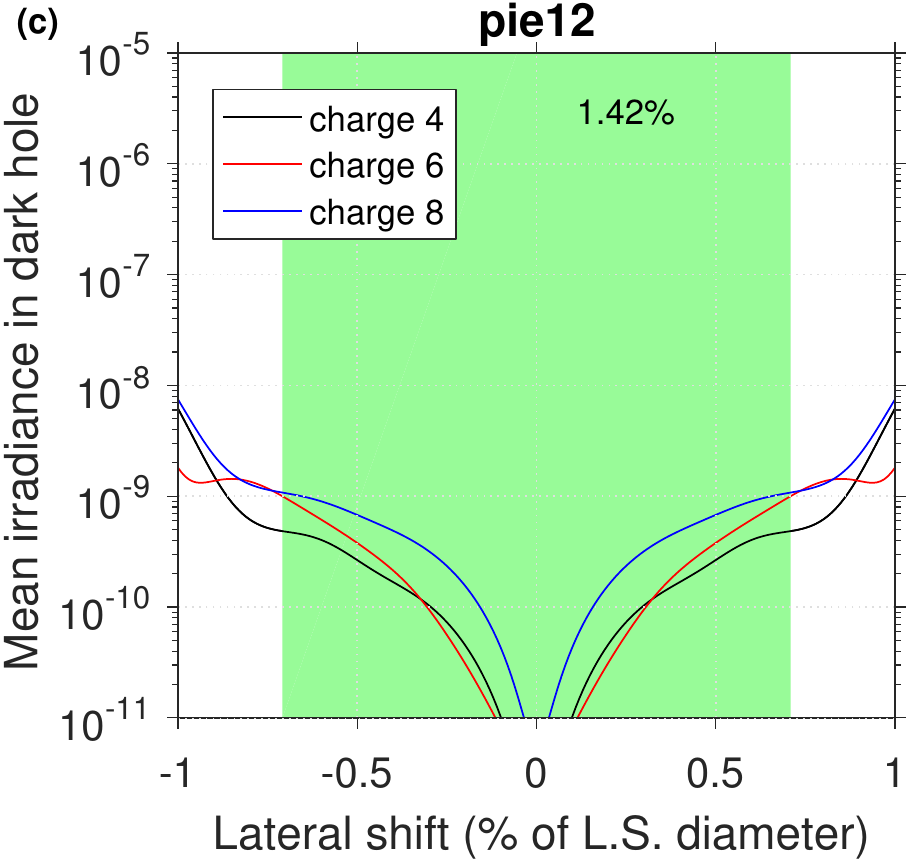}\\
    \includegraphics[height=0.31\linewidth]{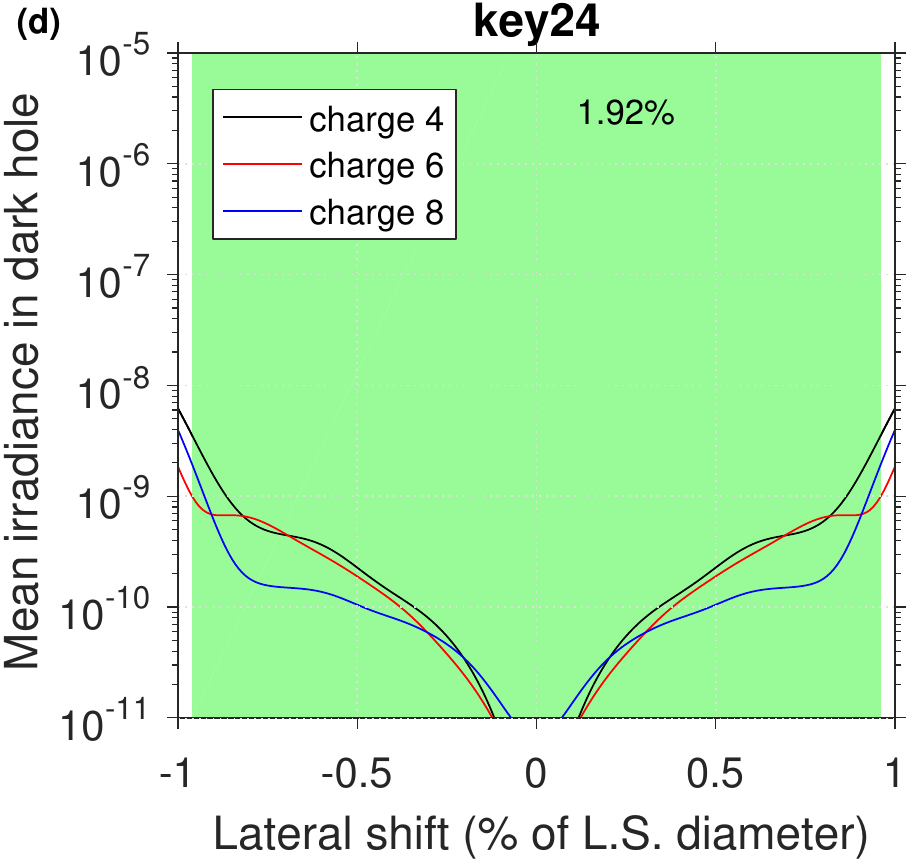}
    \includegraphics[height=0.31\linewidth]{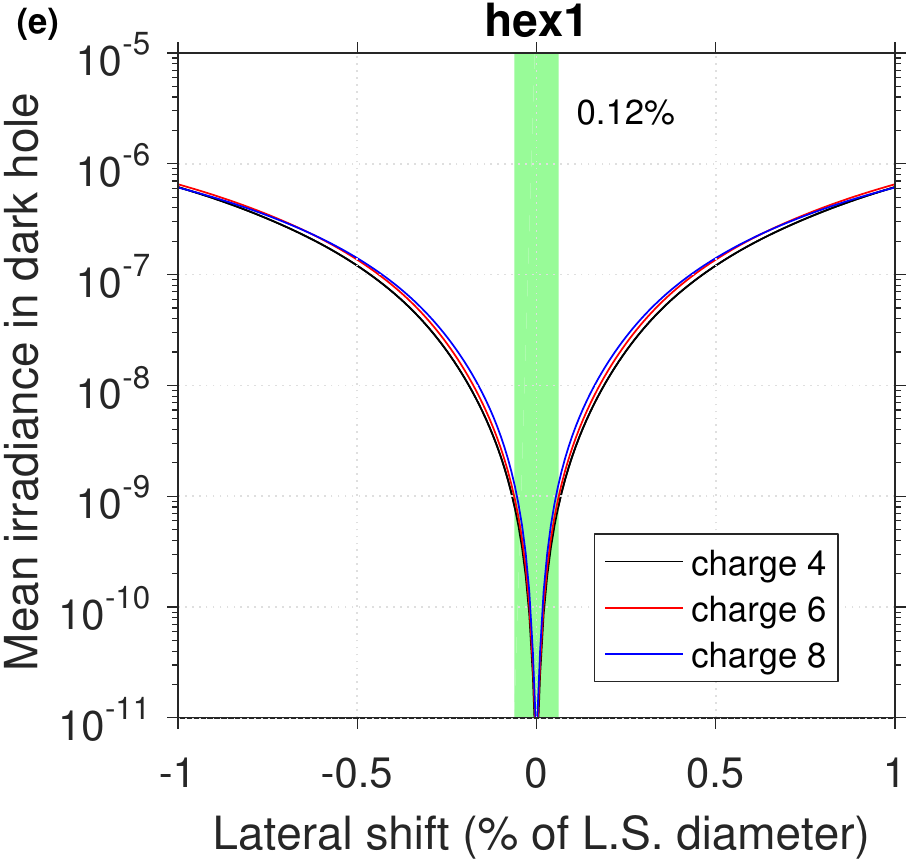}
    \includegraphics[height=0.31\linewidth]{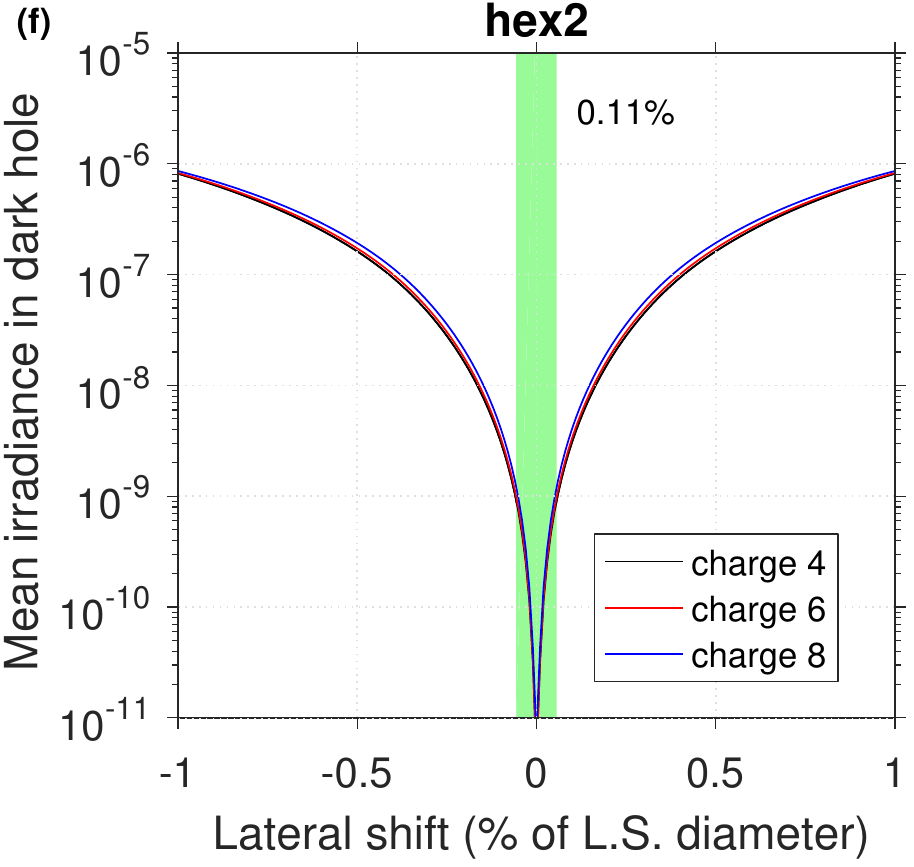}\\
    \includegraphics[height=0.31\linewidth]{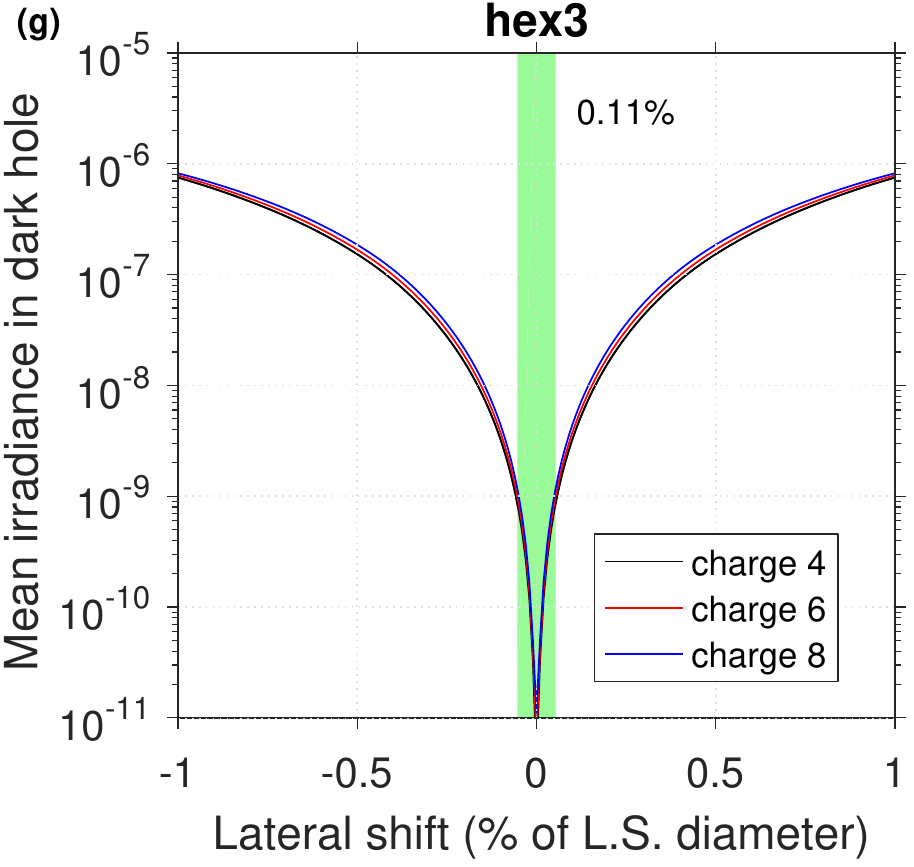}
    \includegraphics[height=0.31\linewidth]{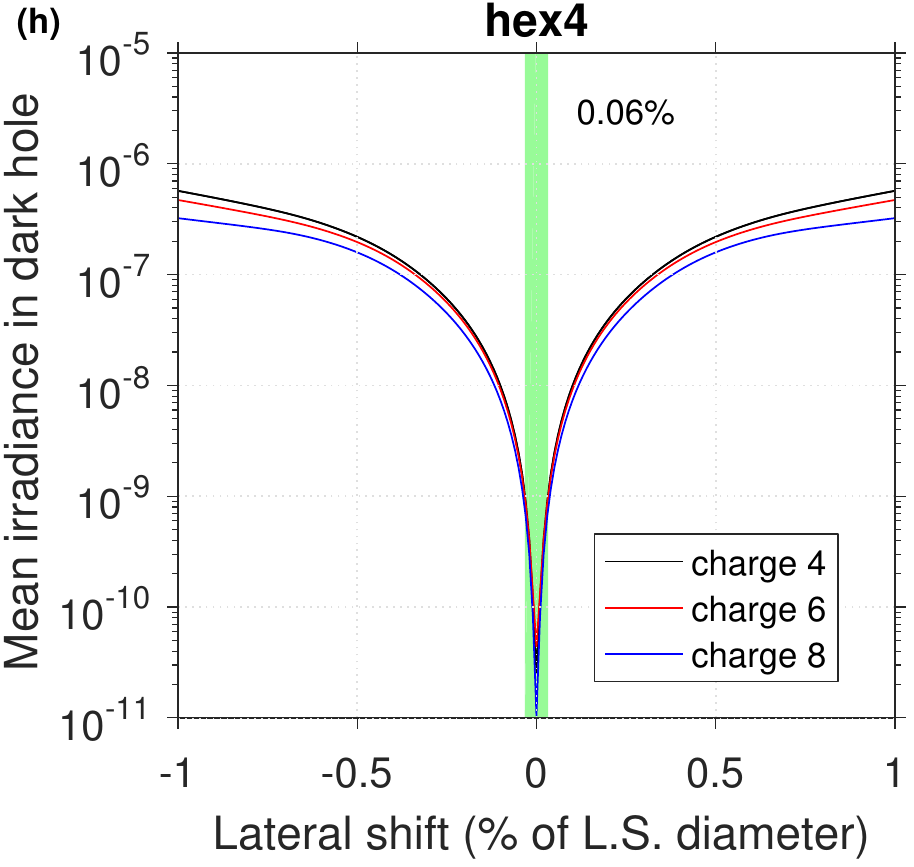}\\
    \caption{Mean irradiance in the dark hole (normalized to the peak of the telescope PSF) as a function of Lyot stop position. The green region shows the range over which $10^{-9}$ is maintained and the annotated number is the region's width (i.e. roughly twice the alignment tolerance).}
    \label{fig:LS_shift_sens_unobs}
\end{figure}

The apodizers presented here are designed assuming the Lyot stop is in a specific position. Shifting the Lyot stop laterally leads to degradation of the starlight suppression level. Figures \ref{fig:LS_shift_sens_obs} and \ref{fig:LS_shift_sens_unobs} show the mean irradiance level in the dark hole as the Lyot stop is shifted horizontally, for obscured and unobscured apertures, respectively. The alignment tolerance ranges from $\sim$0.01\% to $\sim$0.1\% for the obscured apertures and from 0.02\% to 1\% for the unobscured apertures. 

\subsubsection{Sensitivity to coronagraph pupil magnification}
\begin{figure}[p]
    \begin{center}\textbf{Apertures with central obscurations}\end{center}
    \includegraphics[height=0.31\linewidth]{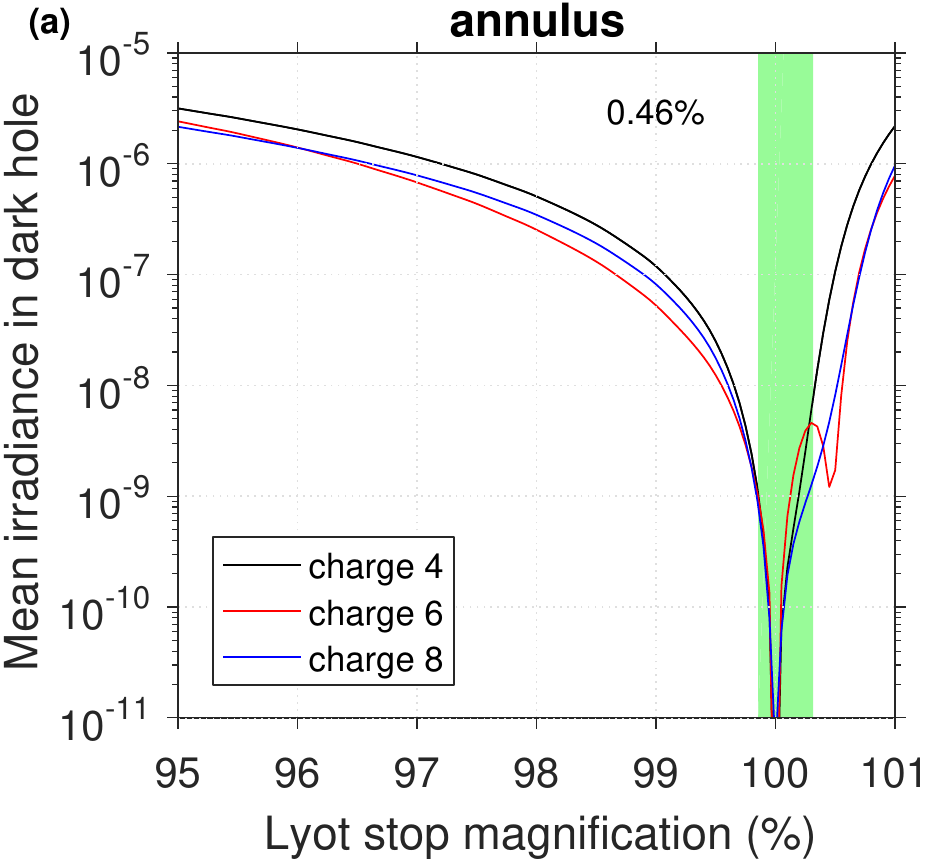}
    \includegraphics[height=0.31\linewidth]{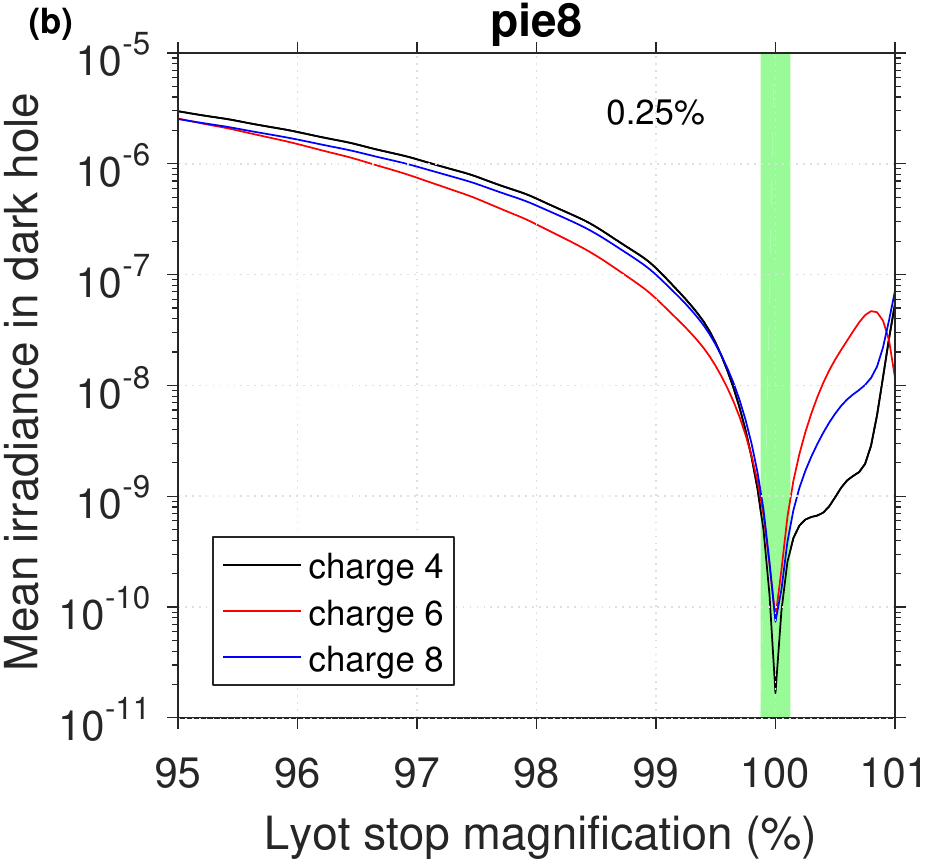}
    \includegraphics[height=0.31\linewidth]{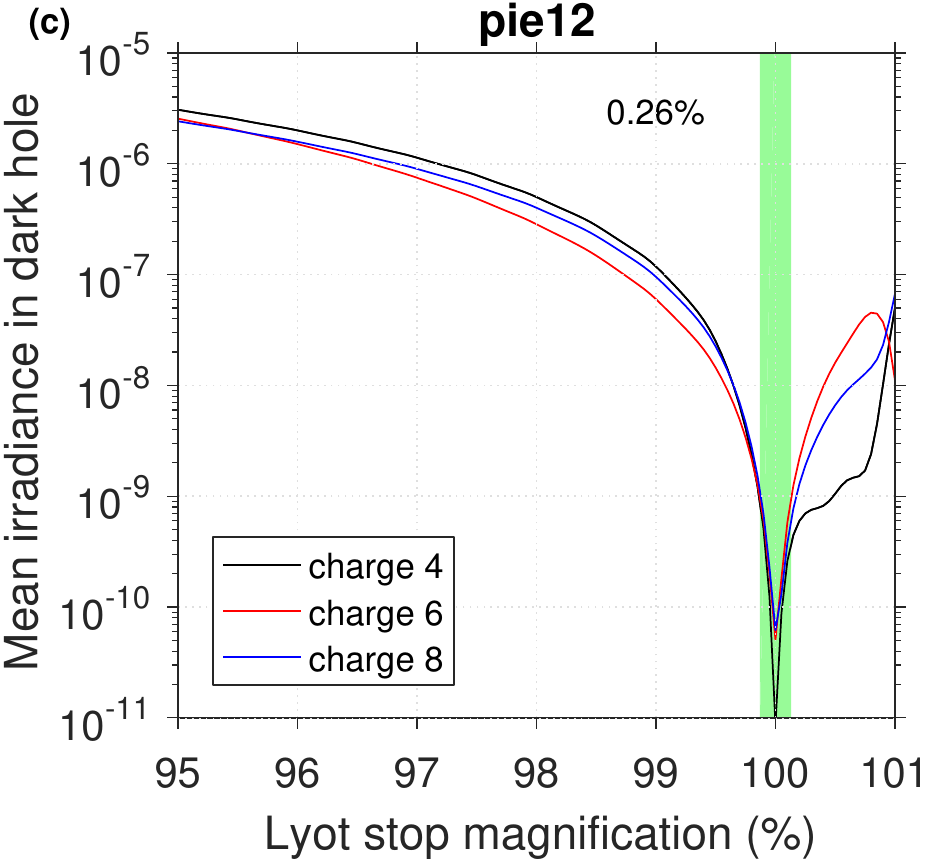}\\
    \includegraphics[height=0.31\linewidth]{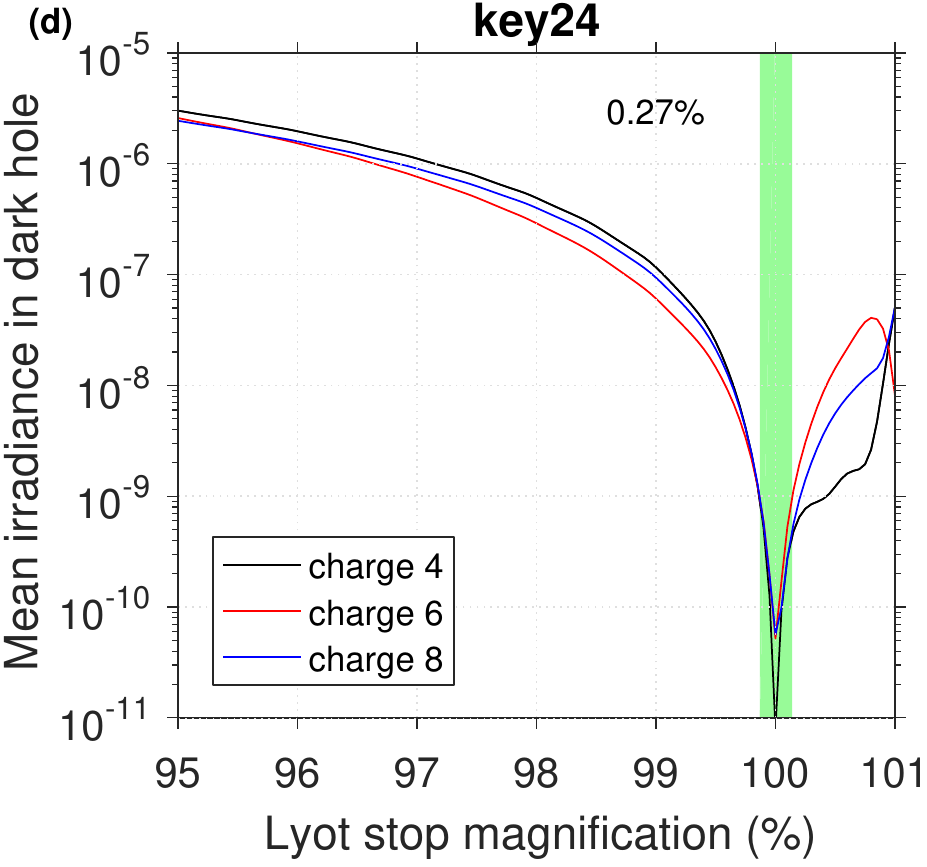}
    \includegraphics[height=0.31\linewidth]{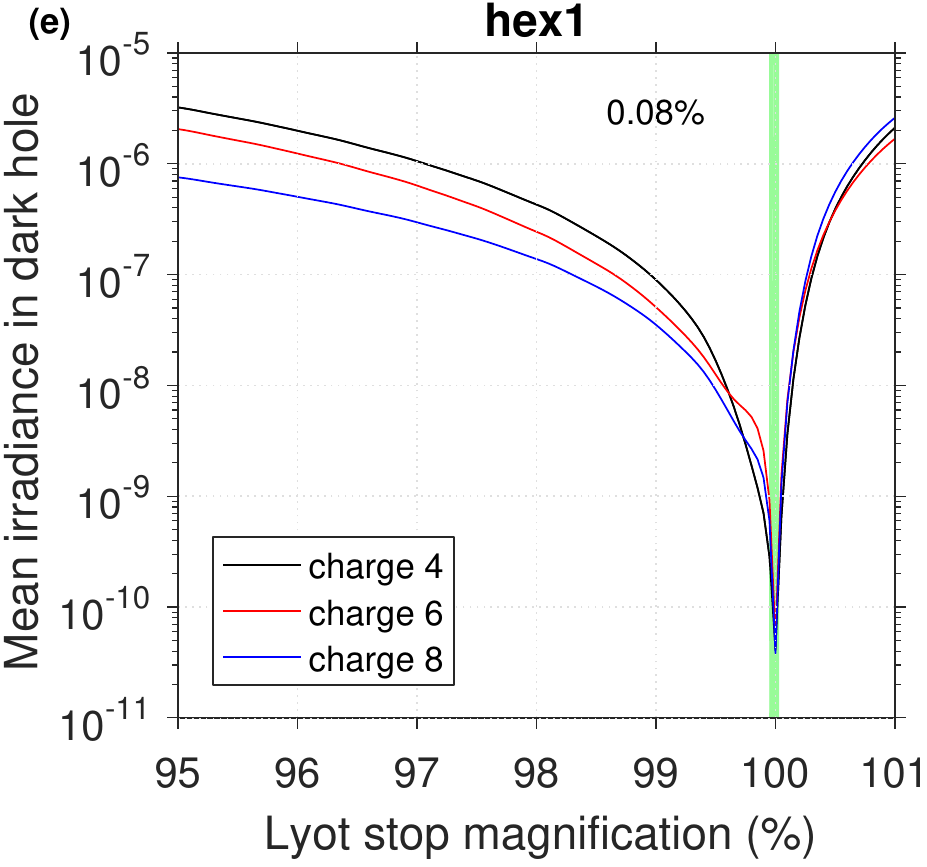}
    \includegraphics[height=0.31\linewidth]{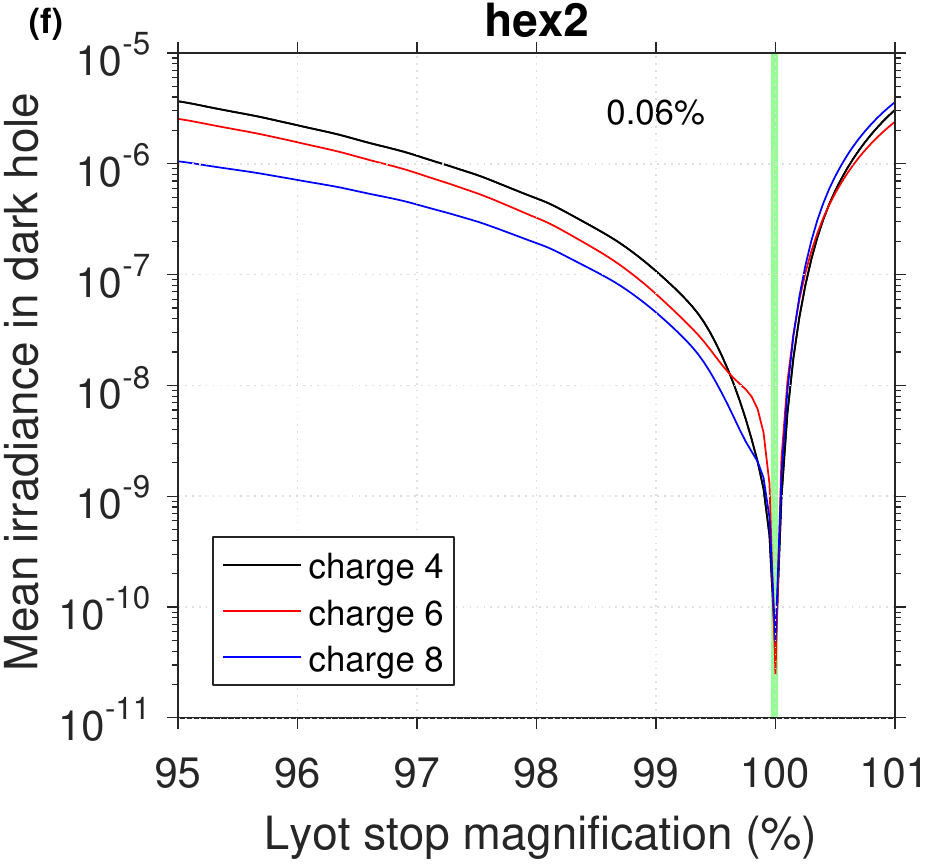}\\
    \includegraphics[height=0.31\linewidth]{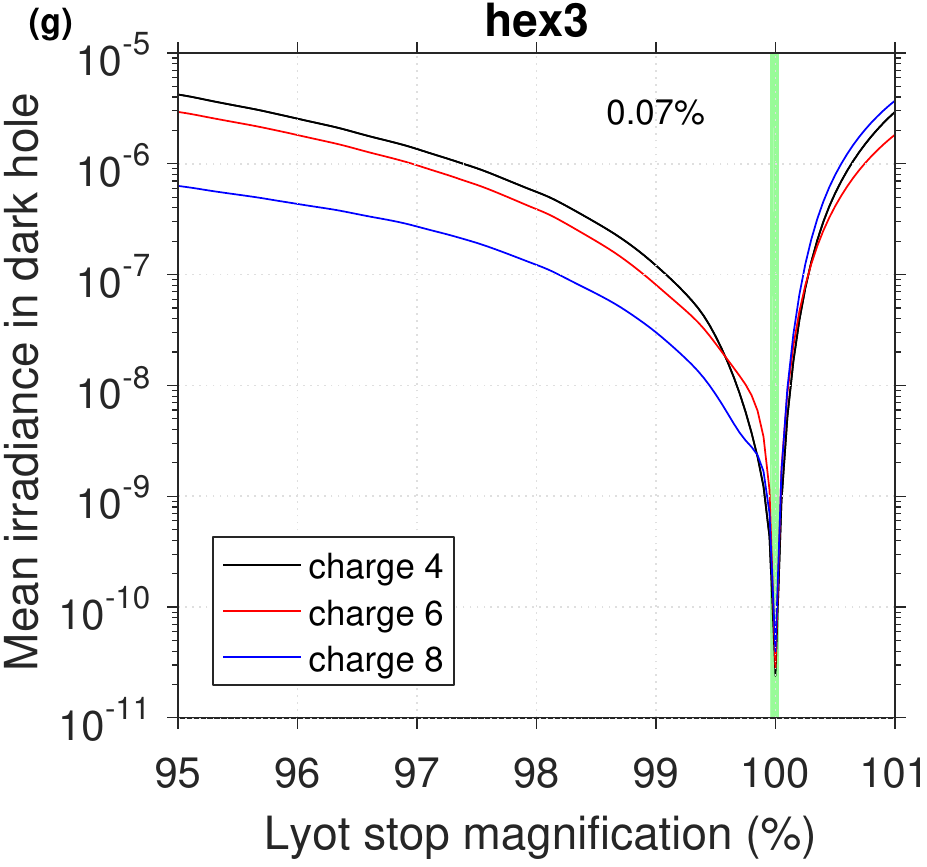}
    \includegraphics[height=0.31\linewidth]{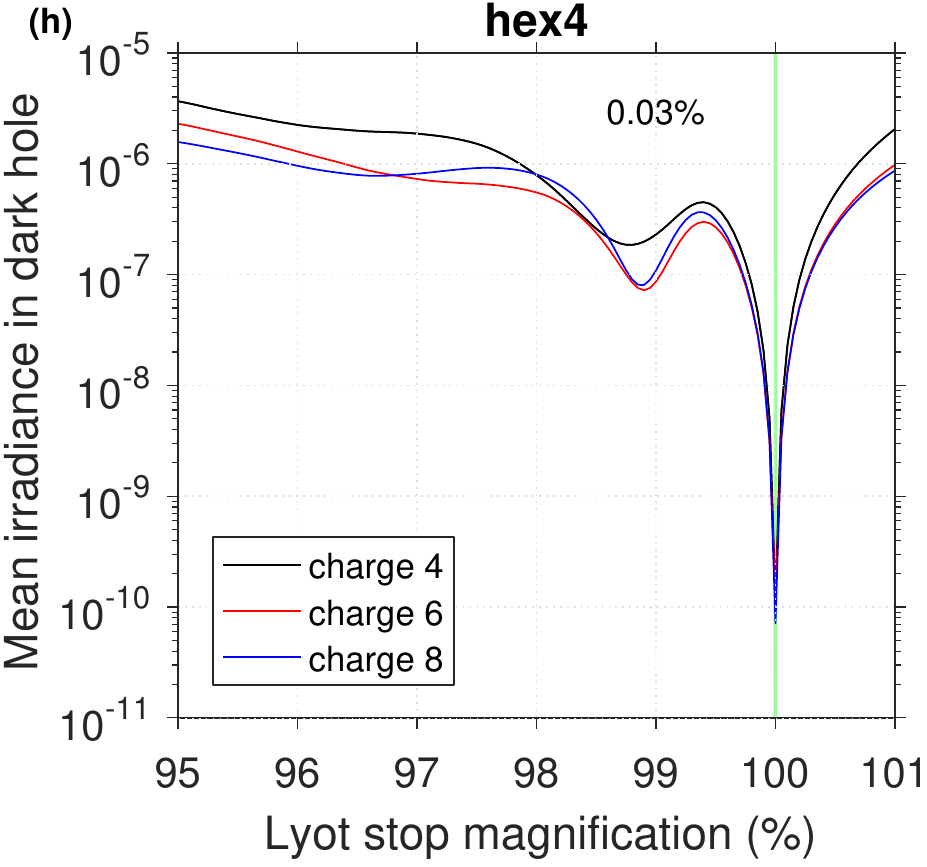}\\
    \caption{Mean irradiance in the dark hole (normalized to the peak of the telescope PSF) as a function of Lyot stop magnification. The green region shows the range over which $10^{-9}$ is maintained and the annotated number is the region's width (i.e. roughly twice the alignment tolerance).}
    \label{fig:LS_mag_sens_obs}
\end{figure}

\begin{figure}[p]
    \begin{center}\textbf{Apertures without central obscurations}\end{center}
    \includegraphics[height=0.31\linewidth]{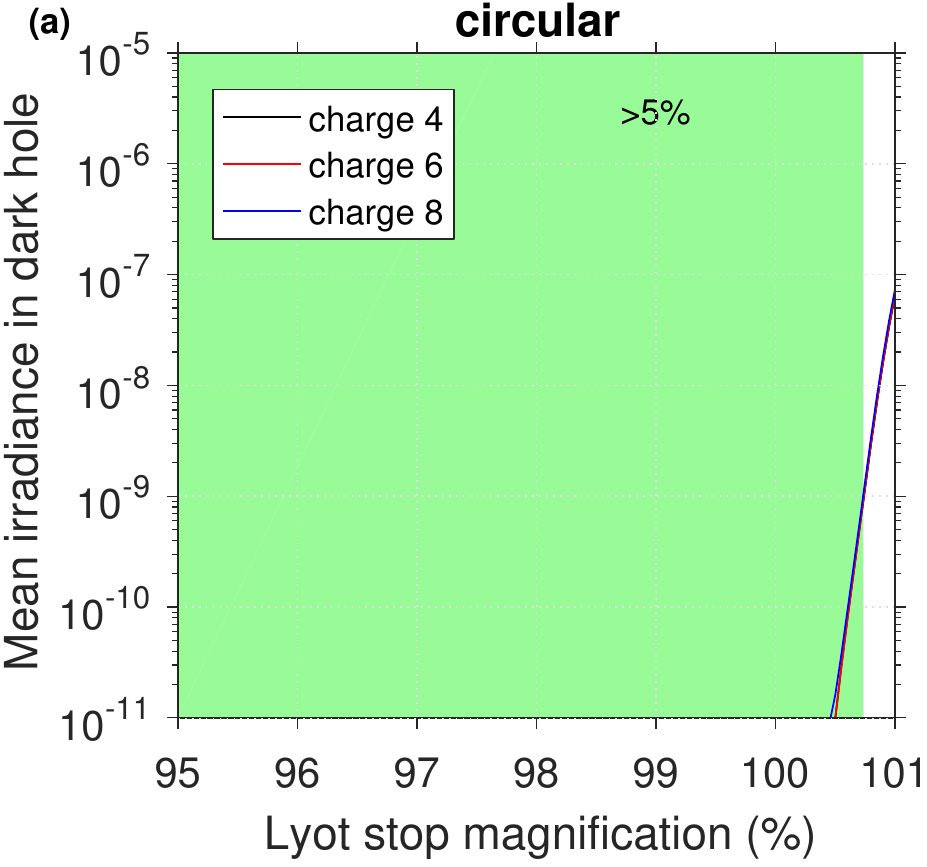}
    \includegraphics[height=0.31\linewidth]{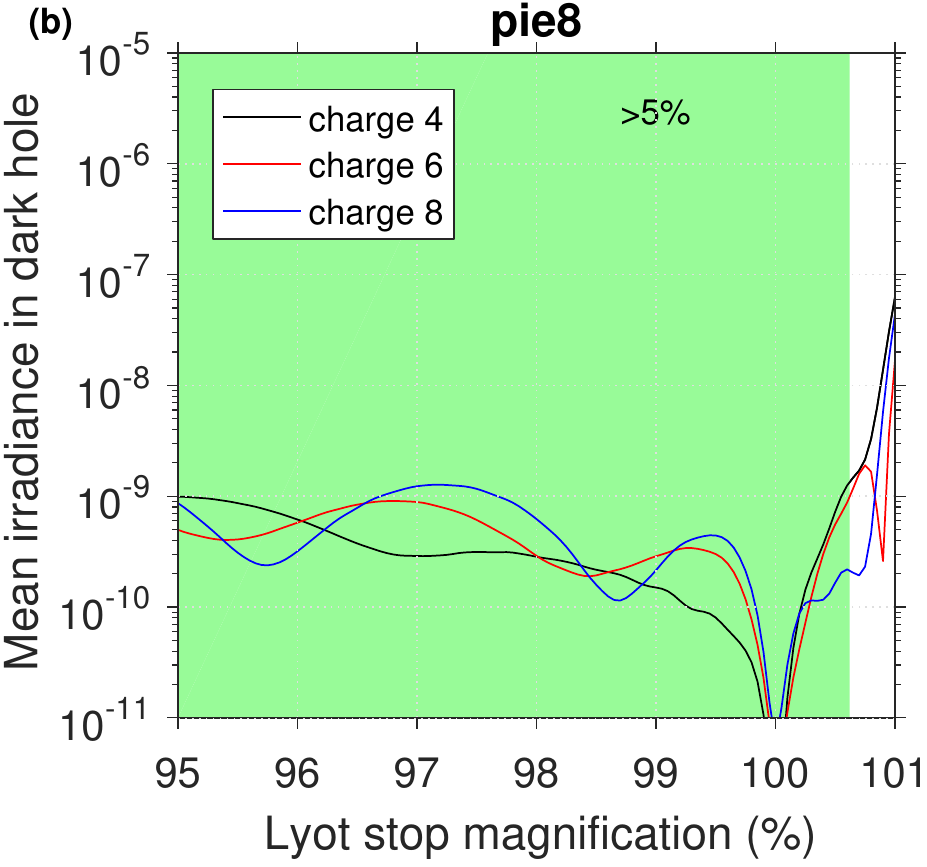}
    \includegraphics[height=0.31\linewidth]{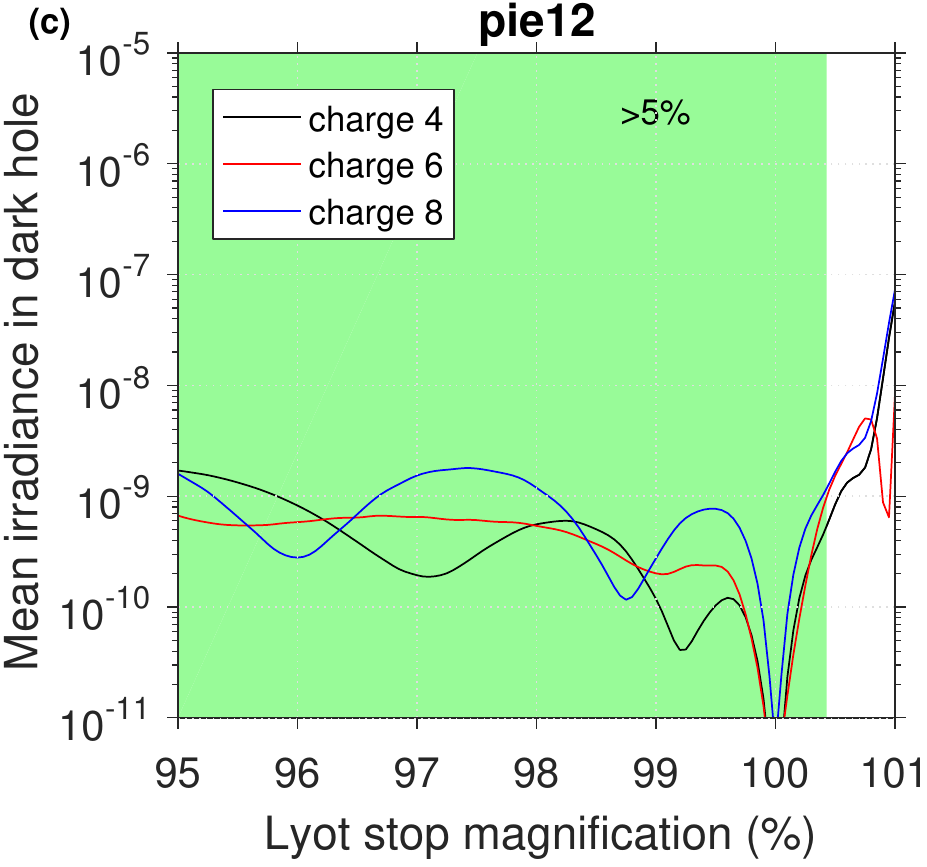}\\
    \includegraphics[height=0.31\linewidth]{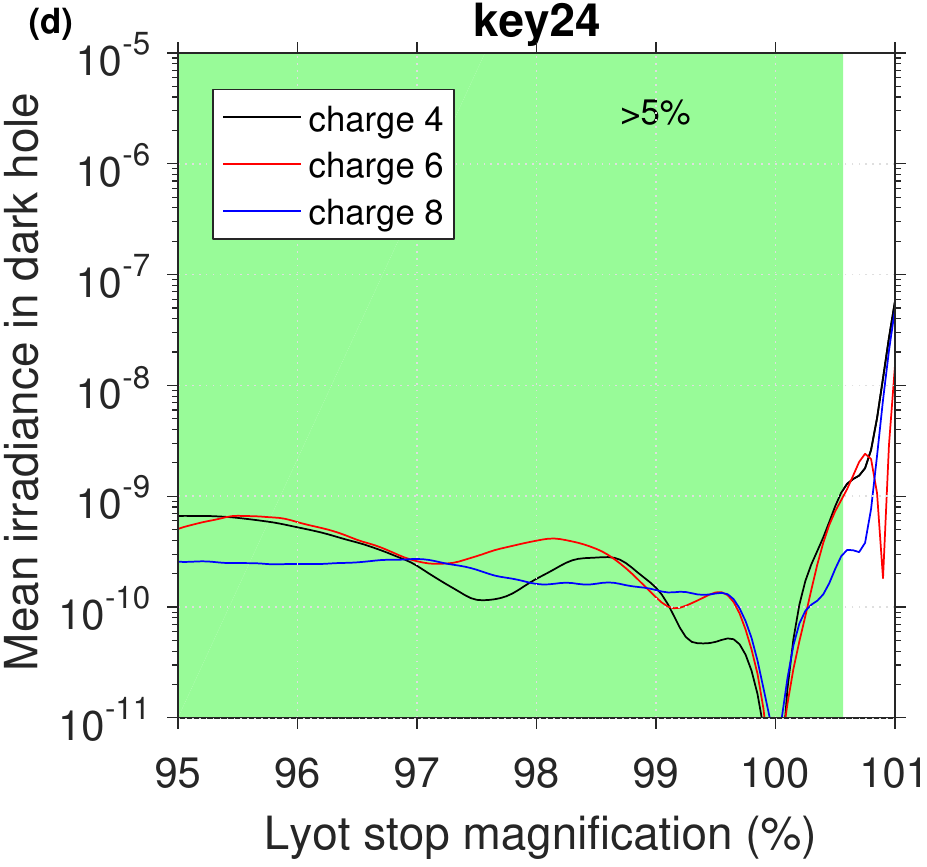}
    \includegraphics[height=0.31\linewidth]{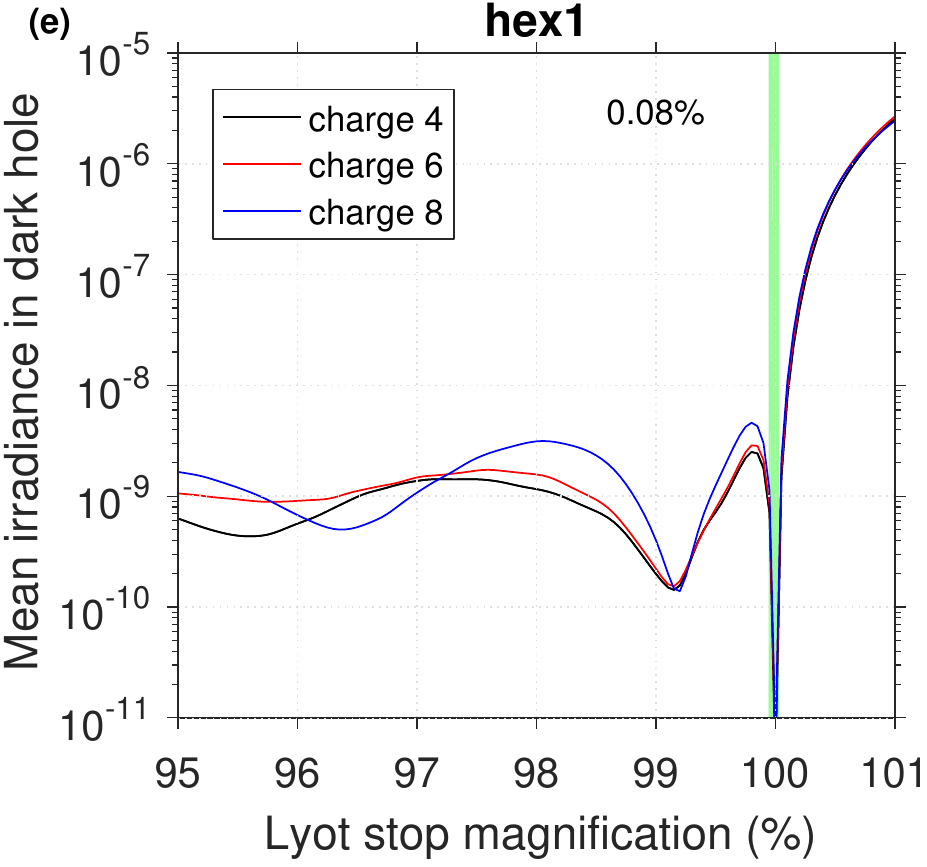}
    \includegraphics[height=0.31\linewidth]{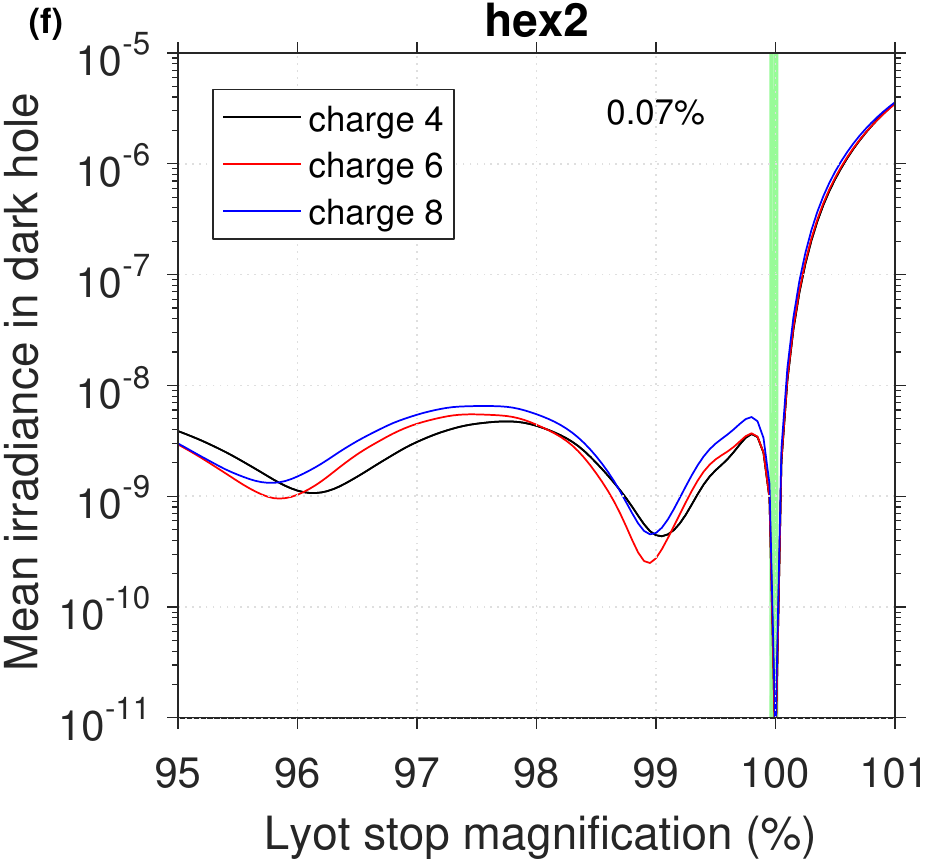}\\
    \includegraphics[height=0.31\linewidth]{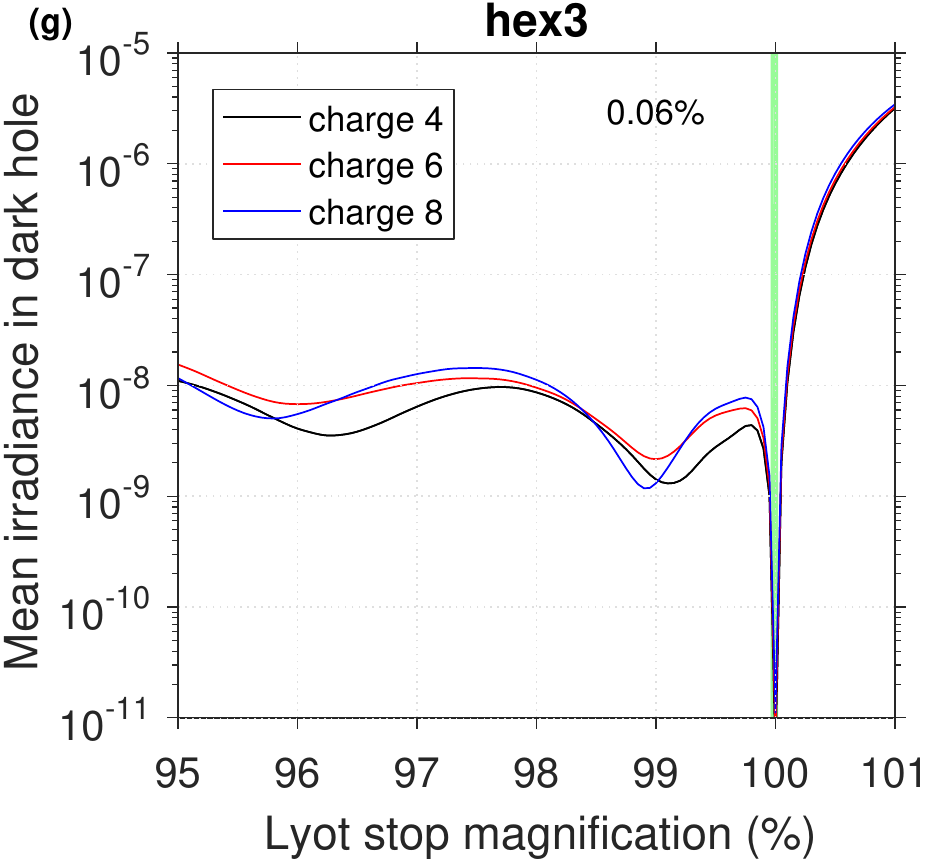}
    \includegraphics[height=0.31\linewidth]{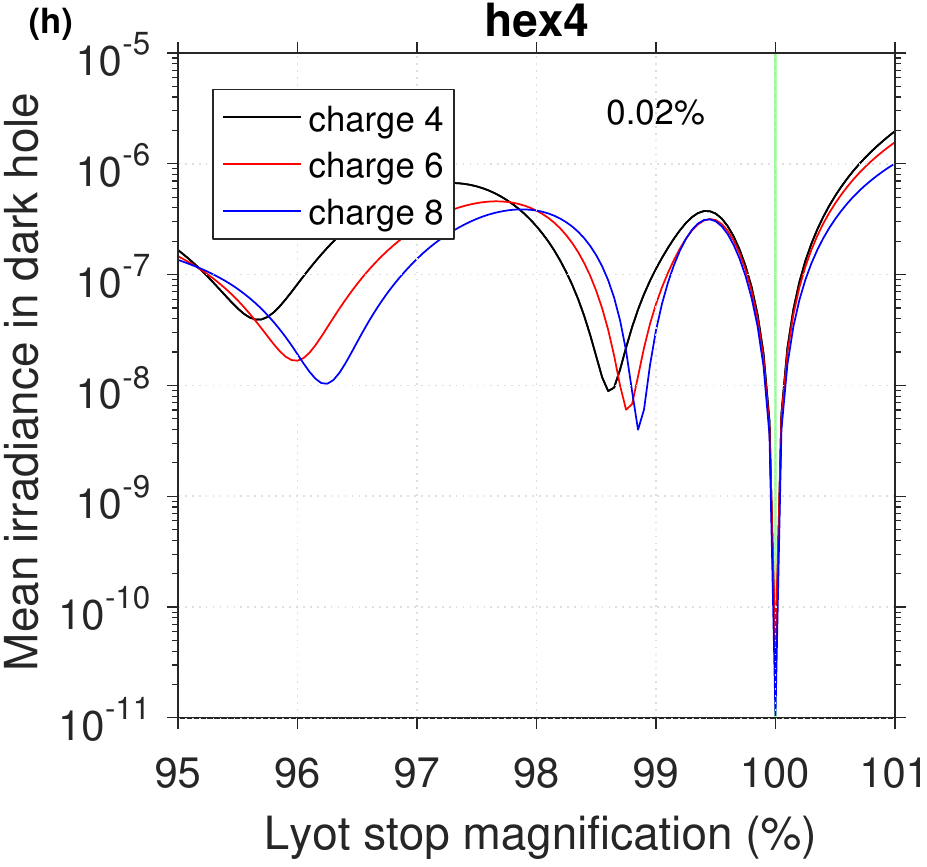}\\
    \caption{Mean irradiance in the dark hole (normalized to the peak of the telescope PSF) as a function of Lyot stop magnification. The green region shows the range over which $10^{-9}$ is maintained and the annotated number is the region's width (i.e. roughly twice the alignment tolerance).}
    \label{fig:LS_mag_sens_unobs}
\end{figure}

Another major concern is that the relative size of the coronagraph exit pupil (field in the plane of the Lyot stop) and the Lyot stop must remain fixed with respect to one another. Any pupil distortions in the system may lead to degraded performance. To quantify this, we calculated the mean irradiance level in the dark hole as the Lyot stop is magnified and the results are shown in Figs. \ref{fig:LS_mag_sens_obs} and \ref{fig:LS_mag_sens_unobs} for obscured and unobscured apertures, respectively. The tolerance can range from a few percent in the best cases to $<$0.01\% in the worst cases. 

\subsubsection{Performance comparison tables}
In order to compare the overall performance and robustness of each of the designs presented above, we have compiled tables of values for the metrics described in section \ref{sec:perfmetrics}. Tables \ref{tab:M1_obs} and \ref{tab:M1_unobs} show the relative integration time for an exoplanet detection $M$ with respect to an idealized coronagraph with $I_{avg}=10^{-10}$ and 100\% throughput. Each table also includes the mean of the performance metric for a stellar angular diameter 1\% $\lambda/D$ and defocus, astigmatism, and coma aberrations, which is representative of realistic performance on future space telescopes. 

Table \ref{tab:Yield_obs} shows the exoEarth candidate yield estimates for an ideal telescope with negligible wavefront errors. These values are meant to be interpreted as the theoretical maximum yield taking into account the flux from the planet, the coronagraph throughput, and the leaked starlight owing to the angular size of nearby stars.

\subsubsection{Discussion}
The optimal aperture for vortex coronagraphs, by all measures, is an unobstructed circle. The pie8, pie12, and key24 segmentation patterns typically lead to slightly better coronagraph performance than hex apertures. The charge 6 versions are found to be optimal in the presence of jitter and low order aberrations. Designs for unobstructed apertures outperform those with a central obscuration in most cases, regardless of aperture segmentation pattern. 

The yield estimates will be drastically degraded in cases where the $M$ metric are $\gg 1$l i.e. once realistic simulations of telescope aberrations are included. On the other hand, with $M\sim1$, the coronagraph is expected to relatively robust to aberrations. Since charge 4 vortex coronagraphs are the particularly sensitive to the expected aberrations, the yields for off-axis telescopes with charge 6 vortex coronagraphs are expected to be the highest in practice. However, future instruments may optimize yield by carrying more than one vortex mask to tailor the coronagraph sensitivities for each observation based on the wavefront stability and/or angular size of the host star. 

\citet{Stark2015} find that the yield depends roughly on $D^2$. If an obstructed design allows for a larger aperture, than the overall yield will likely be greater with a larger telescope, even with the slightly degraded coronagraph performance.

\subsection{Improving upon current designs}

Several avenues to improved designs will be pursued to enhance the performance and robustness of our current designs, especially for more challenging apertures.
\begin{itemize}
\item Apodization using deformable mirrors rather than a gray scale mask will be explored to improve the coronagraph throughput. Our initial simulations suggest that throughput gains are possible with DM-based complex apodization. A detailed analysis of this approach in underway.  
\item Robustness to finite stellar size and low order aberrations will be improved by modifying the optimization to algorithm to build in insensitivity to a given set of pupil modes, including various tilts and Zernike polynomials. 
\item Lyot stop alignment tolerance will be relaxed by creating regions of zero-valued field in the plane of the Lyot stop. The unobstructed circular pupil, for example, may be easily made less sensitive to alignment by reducing the size of the Lyot stop because there is an area of zero field extending throughout the geometric image of the pupil in the plane of the Lyot stop. By the same principle, reducing the intensity near the edges of the Lyot stop may provide a more robust solution. 
\end{itemize}

\section{Conclusions and outlook}\label{sec:outlook}

We have presented a family of apodized vortex coronagraph designs that potentially provide the high-contrast imaging performance needed to directly detect and characterize exoplanets with future segmented aperture space telescopes.

The gray-scale family of designs has the attractive property that the masks suppress starlight without relying on the DMs. Therefore, the usable range of DM surface height, or stroke, may be solely dedicated to compensating for surface errors on the upstream optical surfaces in a closed loop wavefront control scheme. The apodizers are designed to provide most of the amplitude modifications needed to suppress starlight.

The minimum amount of residual starlight present in the dark hole will ultimately depend on the level of wavefront error present in the optical system and manufacturing quality of the coronagraph masks. The extent to which corrections must be applied by the DMs in closed loop to restore the dark hole will be explored in future work. The effect of manufacturing errors will also be incorporated in the design phase. The throughput, however, is not as sensitive to small errors and is expected to maintain the behavior shown in Figs. \ref{fig:Thpts_obs} and \ref{fig:Thpts_unobs} in practical scenarios. 

\section{Acknowledgements}
This work was performed in part at the Jet Propulsion Laboratory, California Institute of Technology, under a contract with the National Aeronautics and Space Administration (NASA). This work was funded by the Exoplanet Exploration Program (ExEP). G. Ruane is supported by an NSF Astronomy and Astrophysics Postdoctoral Fellowship under award AST-1602444. 

\section{References}\label{sec:refs}
\begingroup
\renewcommand{\section}[2]{}%
\bibliographystyle{short}
\def\bibfont{\footnotesize}
\begin{multicols}{2}
\bibliography{library}
\end{multicols}
\endgroup

\newpage
\section{Figures and large tables}

\vspace{2cm}

\begin{table}[p]
\small
\begin{center}\textbf{Apertures with central obscurations}\end{center}
\centering
\begin{tabular}{ *{13}{c} }
 & & \multicolumn{11}{c}{Performance Metric $M$: relative integration time in photon noise limit} \\ \cline{3-13}
 & & \multicolumn{3}{c}{\underline{~\smash{Star diam. ($\lambda/D$)}~}} & \multicolumn{5}{c}{\underline{\smash{~~~~Low order aberration ($\lambda/1000$ rms)~~~~}}} & \multicolumn{2}{c}{\underline{\smash{~~~~~~Lyot stop~~~~~~}}} &  \\
aperture & $l$ & 0 & \textbf{0.01} & 0.1 & \textbf{defoc.} & \textbf{astig.} & \textbf{coma} & tref. & spher. & 0.1\%$\rightarrow$ & 0.99$\times$ & \textbf{mean}\\
 \hline
\multirow{3}{*}{annulus} & 4 & $<1$ & 10 & 798 & 21 & 967 & 1141 & 1342 & 3380 & 88 & 9794 & 534\\
  & 6 & $<1$ & 21 & 1984 & 62 & 20 & 97 & 1658 & 190 & 119 & 6834 & 50\\
  & 8 & 1 & 35 & 3983 & 157 & 32 & 238 & 23 & 599 & 217 & 20521 & 116\\
\hline
\multirow{3}{*}{pie8} & 4 & 2 & 11 & 906 & 18 & 1037 & 1278 & 1421 & 3883 & 120 & 13408 & 586\\
  & 6 & 15 & 37 & 1888 & 69 & 28 & 128 & 2108 & 231 & 203 & 12410 & 66\\
  & 8 & 26 & 61 & 4184 & 164 & 51 & 279 & 42 & 665 & 472 & 39915 & 139\\
\hline
\multirow{3}{*}{pie12} & 4 & 1 & 9 & 1023 & 19 & 1053 & 1291 & 1439 & 3913 & 127 & 13928 & 593\\
  & 6 & 9 & 33 & 2540 & 65 & 22 & 122 & 2108 & 228 & 189 & 12151 & 60\\
  & 8 & 22 & 57 & 3415 & 169 & 49 & 280 & 40 & 674 & 444 & 37737 & 139\\
\hline
\multirow{3}{*}{key24} & 4 & 1 & 9 & 932 & 18 & 1043 & 1279 & 1424 & 3873 & 129 & 13451 & 587\\
  & 6 & 9 & 30 & 2055 & 63 & 21 & 120 & 2106 & 225 & 170 & 12098 & 59\\
  & 8 & 21 & 52 & 2934 & 121 & 38 & 247 & 30 & 614 & 359 & 38363 & 115\\
\hline
\multirow{3}{*}{hex1} & 4 & 34 & 102 & 8223 & 148 & 8796 & 1682 & 16824 & 1902 & 1476 & 79550 & 2682\\
  & 6 & 80 & 346 & 26685 & 809 & 484 & 1492 & 22019 & 4389 & 6238 & 92282 & 783\\
  & 8 & 132 & 438 & 32876 & 959 & 559 & 2440 & 413 & 5708 & 10436 & 132556 & 1099\\
\hline
\multirow{3}{*}{hex2} & 4 & 14 & 57 & 4215 & 99 & 3821 & 1677 & 7816 & 2146 & 821 & 39850 & 1414\\
  & 6 & 14 & 116 & 9407 & 362 & 144 & 594 & 9348 & 1691 & 3073 & 41304 & 304\\
  & 8 & 66 & 286 & 17166 & 719 & 247 & 1405 & 190 & 3237 & 3874 & 61437 & 664\\
\hline
\multirow{3}{*}{hex3} & 4 & 10 & 76 & 6117 & 125 & 4634 & 1759 & 9384 & 2196 & 1022 & 53536 & 1649\\
  & 6 & 21 & 180 & 14146 & 656 & 298 & 930 & 11021 & 2768 & 3763 & 67772 & 516\\
  & 8 & 59 & 334 & 32284 & 1041 & 337 & 1920 & 267 & 4669 & 4578 & 45580 & 908\\
\hline
\multirow{3}{*}{hex4} & 4 & 323 & 437 & 12591 & 404 & 15260 & 9410 & 33355 & 14352 & 7471 & 156098 & 6378\\
  & 6 & 76 & 296 & 20382 & 996 & 737 & 956 & 7144 & 2525 & 6280 & 62927 & 746\\
  & 8 & 134 & 169 & 4317 & 217 & 1015 & 1535 & 1986 & 2816 & 13267 & 203575 & 734\\
\hline
\end{tabular}
\caption{\label{tab:M1_obs} Values of performance metric $M$ (see Eqn. \ref{eqn:m1}) for the obscured apertures, normalized with respect to a coronagraph with $I_{avg}=10^{-10}$ and 100\% throughput. The mean for a stellar angular diameter 1\% $\lambda/D$ and defocus, astigmatism, and coma aberrations (bold terms) represents a probable scenario for future space telescopes.}
\end{table}

\begin{table}[p]
\small
\begin{center}\textbf{Apertures without central obscurations}\end{center}
\centering
\begin{tabular}{ *{13}{c} }
 & & \multicolumn{11}{c}{Performance Metric $M$: relative integration time in photon noise limit} \\ \cline{3-13}
 & & \multicolumn{3}{c}{\underline{~\smash{Star diam. ($\lambda/D$)}~}} & \multicolumn{5}{c}{\underline{\smash{~~~~Low order aberration ($\lambda/1000$ rms)~~~~}}} & \multicolumn{2}{c}{\underline{\smash{~~~~~~Lyot stop~~~~~~}}} &  \\
aperture & $l$ & 0 & \textbf{0.01} & 0.1 & \textbf{defoc.} & \textbf{astig.} & \textbf{coma} & tref. & spher. & 0.1\%$\rightarrow$ & 0.99$\times$ & \textbf{mean}\\
 \hline
\multirow{3}{*}{circular} & 4 & $<1$ & $<1$ & 10 & $<1$ & 212 & 266 & 248 & 760 & $<1$ & $<1$ & 119\\
 & 6 & $<1$ & $<1$ & $<1$ & $<1$ & $<1$ & $<1$ & 368 & $<1$ & $<1$ & $<1$ & $<1$\\
 & 8 & $<1$ & $<1$ & $<1$ & $<1$ & $<1$ & $<1$ & $<1$ & $<1$ & $<1$ & $<1$ & $<1$\\
\hline
\multirow{3}{*}{pie8} & 4 & $<1$ & $<1$ & 34 & 1 & 302 & 359 & 351 & 1011 & $<1$ & 4 & 166\\
 & 6 & $<1$ & $<1$ & 16 & 1 & $<1$ & 1 & 455 & 2 & $<1$ & 10 & 1\\
 & 8 & $<1$ & $<1$ & 19 & 1 & $<1$ & 1 & $<1$ & 3 & 1 & 12 & 1\\
\hline
\multirow{3}{*}{pie12} & 4 & $<1$ & 1 & 88 & 3 & 347 & 400 & 401 & 1114 & $<1$ & 4 & 188\\
 & 6 & $<1$ & 1 & 49 & 2 & 1 & 2 & 473 & 4 & $<1$ & 7 & 1\\
 & 8 & $<1$ & 1 & 43 & 2 & 1 & 3 & 1 & 6 & 2 & 16 & 2\\
\hline
\multirow{3}{*}{key24} & 4 & $<1$ & 1 & 84 & 3 & 346 & 398 & 399 & 1106 & $<1$ & 5 & 187\\
 & 6 & $<1$ & $<1$ & 37 & 1 & $<1$ & 2 & 486 & 4 & $<1$ & 4 & 1\\
 & 8 & $<1$ & $<1$ & 7 & 1 & $<1$ & 1 & $<1$ & 4 & 1 & 8 & 1\\
\hline
\multirow{3}{*}{hex1} & 4 & $<1$ & $<1$ & 20 & $<1$ & 1196 & 306 & 1440 & 399 & 236 & 20 & 376\\
 & 6 & 1 & 1 & 3 & 1 & 1 & 1 & 2428 & 1 & 468 & 38 & 1\\
 & 8 & 2 & 2 & 88 & 4 & 3 & 5 & 3 & 8 & 1055 & 120 & 3\\
\hline
\multirow{3}{*}{hex2} & 4 & $<1$ & $<1$ & 17 & $<1$ & 614 & 357 & 856 & 468 & 159 & 23 & 243\\
 & 6 & $<1$ & $<1$ & 1 & $<1$ & $<1$ & $<1$ & 1348 & $<1$ & 267 & 22 & $<1$\\
 & 8 & 1 & 1 & 1 & 1 & 1 & 1 & 1 & 1 & 564 & 64 & 1\\
\hline
\multirow{3}{*}{hex3} & 4 & $<1$ & $<1$ & 24 & $<1$ & 643 & 349 & 871 & 466 & 181 & 79 & 248\\
 & 6 & 1 & 1 & 2 & 1 & 1 & 1 & 1405 & 1 & 328 & 189 & 1\\
 & 8 & 2 & 2 & 5 & 2 & 2 & 2 & 2 & 2 & 665 & 198 & 2\\
\hline
\multirow{3}{*}{hex4} & 4 & 2 & 8 & 658 & 20 & 518 & 452 & 727 & 703 & 812 & 9312 & 249\\
 & 6 & 4 & 10 & 446 & 26 & 54 & 19 & 1027 & 21 & 921 & 5480 & 27\\
 & 8 & 2 & 9 & 702 & 26 & 54 & 14 & 133 & 14 & 1221 & 5263 & 26\\
\hline
\end{tabular}
\caption{\label{tab:M1_unobs} Values of performance metric $M$ (see Eqn. \ref{eqn:m1}) for the unobscured apertures, normalized with respect to a coronagraph with $I_{avg}=10^{-10}$ and 100\% throughput. The mean for a stellar angular diameter 1\% $\lambda/D$ and defocus, astigmatism, and coma aberrations (bold terms) represents a probable scenario for future space telescopes.}
\end{table}

\begin{table}[p]
\small
\centering
\begin{tabular}{ *{8}{c} }
 & & \multicolumn{6}{c}{ExoEarth yield} \\ \cline{3-8}
 & & \multicolumn{3}{c}{\underline{~~~~~~~~~\smash{Obscured}~~~~~~~~~}} & \multicolumn{3}{c}{\underline{~~~~~~~\smash{Unobscured}~~~~~~~}}\\
aperture & $l$ & 4~m & 6.5~m & 12~m & 4~m & 6.5~m & 12~m\\
 \hline
\multirow{3}{*}{circular} & 4 & 1 & 3 & 8 & 8 & 19 & 55\\
 & 6 & 1 & 2 & 5 & 7 & 17 & 53\\
 & 8 & 1 & 2 & 5 & 5 & 14 & 45\\
\hline
\multirow{3}{*}{pie8} & 4 & 1 & 3 & 7 & 5 & 11 & 29\\
 & 6 & 1 & 2 & 5 & 4 & 11 & 30\\
 & 8 & 1 & 1 & 4 & 4 & 10 & 28\\
\hline
\multirow{3}{*}{pie12} & 4 & 1 & 3 & 7 & 3 & 8 & 21\\
 & 6 & 1 & 2 & 5 & 3 & 8 & 22\\
 & 8 & 1 & 1 & 4 & 3 & 8 & 22\\
\hline
\multirow{3}{*}{key24} & 4 & 1 & 3 & 7 & 3 & 8 & 20\\
 & 6 & 1 & 2 & 5 & 4 & 9 & 25\\
 & 8 & 1 & 2 & 4 & 4 & 11 & 31\\
\hline
\multirow{3}{*}{hex1} & 4 & 0 & 1 & 3 & 4 & 10 & 27\\
 & 6 & 0 & 1 & 2 & 3 & 8 & 24\\
 & 8 & 0 & 1 & 2 & 2 & 5 & 16\\
\hline
\multirow{3}{*}{hex2} & 4 & 1 & 1 & 3 & 5 & 12 & 33\\
 & 6 & 0 & 1 & 2 & 4 & 10 & 30\\
 & 8 & 0 & 1 & 2 & 3 & 8 & 24\\
\hline
\multirow{3}{*}{hex3} & 4 & 0 & 1 & 3 & 4 & 10 & 28\\
 & 6 & 0 & 1 & 2 & 4 & 9 & 28\\
 & 8 & 0 & 1 & 2 & 3 & 7 & 22\\
\hline
\multirow{3}{*}{hex4} & 4 & 1 & 2 & 4 & 1 & 3 & 8\\
 & 6 & 0 & 1 & 1 & 1 & 3 & 8\\
 & 8 & 1 & 1 & 4 & 1 & 3 & 8\\
\hline
\end{tabular}
\label{tab:Yield_obs}
\caption{\label{tab:Yield_obs} ExoEarth yield for ideal vortex coronagraphs }
\end{table}
\end{document}